\documentclass[12pt]{article}
\usepackage{graphicx} 
\usepackage{amsmath, amssymb} 
\usepackage{fancyhdr} 
\usepackage{geometry} 
\usepackage{caption}
\usepackage{natbib}

\usepackage{accents}

\usepackage{mathabx}
\usepackage{xcolor}
\usepackage{subfig}
\usepackage{authblk}
\usepackage{url}
\usepackage{algorithm, algorithmic}
\newtheorem{proposition}{Proposition}

\usepackage{hyperref}
\usepackage{enumitem}

\newcommand{\eidx}{{n, p_1 \ldots p_L}}
\newcommand{\Sumn}{{\sum_{n=1}^{N}}}
\newcommand{\Sumps}{\sum_{p_1 \cdots p_L}^{P_1 \cdots P_L}}
\newcommand{\scell}{\hat{\sigma}_{1, p_1 \ldots p_L}}
\newcommand{\Sumks}{\sum_{k_1 \cdots k_L}^{K_1 \cdots K_L}}
\newcommand{\Jidx}{{n, k_1 \ldots k_L}}
\newcommand{\idx}{{p_1 \ldots p_L}}
\newcommand{\Wcell}{\mathcal{W}_{n}^{\text{cell}}}
\newcommand{\Wcase}{\mathcal{W}_{n}^{\text{case}}}
\newcommand{\wcell}{w^{\text{cell}}_{\eidx}}
\newcommand{\wcase}{w^{\text{case}}_{n}}

\newcommand{\sbigotimes}{%
  \mathop{\mathchoice{\textstyle\bigotimes}{\bigotimes}{\bigotimes}{\bigotimes}}%
}

\usepackage[english]{babel}
\usepackage{csquotes}

\begin{document}
\def\spacingset#1{\renewcommand{\baselinestretch}%
{#1}\small\normalsize} \spacingset{1}


\title{\bf Casewise and Cellwise Robust Multilinear Principal Component Analysis}
\author[1]{Mehdi Hirari}
\author[1]{Fabio Centofanti}
\author[1]{Mia Hubert\thanks{Corresponding author. e-mail: \texttt{Mia.Hubert@kuleuven.be}}}
\author[1]{Stefan Van Aelst}
\affil[1]{Section of Statistics and Data Science, Department 
          of Mathematics, KU Leuven, Belgium}

\setcounter{Maxaffil}{0}
\renewcommand\Affilfont{\itshape\small}
\date{December 31, 2025}         
  \maketitle

\bigskip
\begin{abstract}
Multilinear Principal Component Analysis (MPCA) is an important tool for analyzing tensor data. It performs dimension reduction similar to PCA for multivariate data. However, standard MPCA is sensitive to outliers. It is highly influenced by  observations deviating from the bulk of the data, called casewise outliers, as well as by individual outlying cells in the tensors, so-called cellwise outliers. This latter type of outlier is highly likely to occur in tensor data, as tensors typically consist of many cells. This paper introduces a novel robust MPCA method that can handle both types of outliers simultaneously, and can cope with missing values as well. This method uses a single loss function to reduce the influence of both casewise and cellwise outliers. The solution that minimizes this loss function is computed using an iteratively reweighted least squares algorithm with a robust initialization. Graphical diagnostic tools are also proposed to identify the different types of outliers that have been found by the new robust MPCA method.  
The performance of the method and associated graphical displays is assessed through simulations and illustrated on two real datasets.
\end{abstract}

\noindent%

\bigskip \bigskip
The Version of Record of this manuscript has been published and is available in \textbf{Journal of Computational and Graphical Statistics,} \url{https://doi.org/10.1080/10618600.2026.2637632}.
\vfill

\newpage
\spacingset{1.2} 

\section{Introduction}

In many real-world scenarios, data are naturally represented as  tensors, i.e., multiway arrays where each dimension or mode encodes specific characteristics of the data objects. For example, video data consist of a sequence of frames where each frame can be represented by a three-dimensional tensor where the first two modes characterize pixel location and the third mode is a color encoding of each pixel. See the Dog Walker video data in Section~\ref{Ex:Dog walker} for an example.

A common practice to analyze such data is to reshape or vectorize it so that matrix-based models and methods can be applied. However, this approach often breaks the natural structure and correlations of the data, removing higher-order dependencies, for instance. This can lead to the loss of more compact and meaningful representations that could be preserved in the original multidimensional form \citep{Ye:GPCA}. On the other hand, tensor-based multilinear data analysis has shown that tensor models can effectively leverage the multilinear structure of the data, providing better insights and precision. An important aspect of multilinear data analysis is tensor decomposition, typically performed using either a CANDECOMP/PARAFAC (CP) decomposition \citep{Carroll:PARanalysis,Harshman:PARAFAC} or a Tucker decomposition \citep{Tucker:Tucker}. Originally developed in the fields of psychometrics and chemometrics \citep{Smilde:Bookmultiway, Kroonenberg:BookMulti}, these decomposition techniques have gained visibility in various area, including computer vision \citep{Vasilescu:MultiliProj, Zhao:BayesianCP, Shi:FeatureExt}, medical imaging \citep{Goldfarb:L1, Liu:GenHOOI}, and signal processing \citep{Lathauwer:MSVD}.

\citet{Lu:MPCA} introduced a new method for tensor decomposition, Multilinear Principal Component Analysis (MPCA), with an application in gait recognition \citep{Han:IndividualGait,Sarkar:Humanid}. MPCA extends the classical PCA paradigm and performs dimensionality reduction across all tensor modes by identifying a basis for each mode such that the projected tensors capture the maximum amount of variation present in the data.

\begin{figure}[!ht]
\centering
\includegraphics[trim=0cm 0cm 0cm 3cm, clip=TRUE,width=0.6\textwidth]{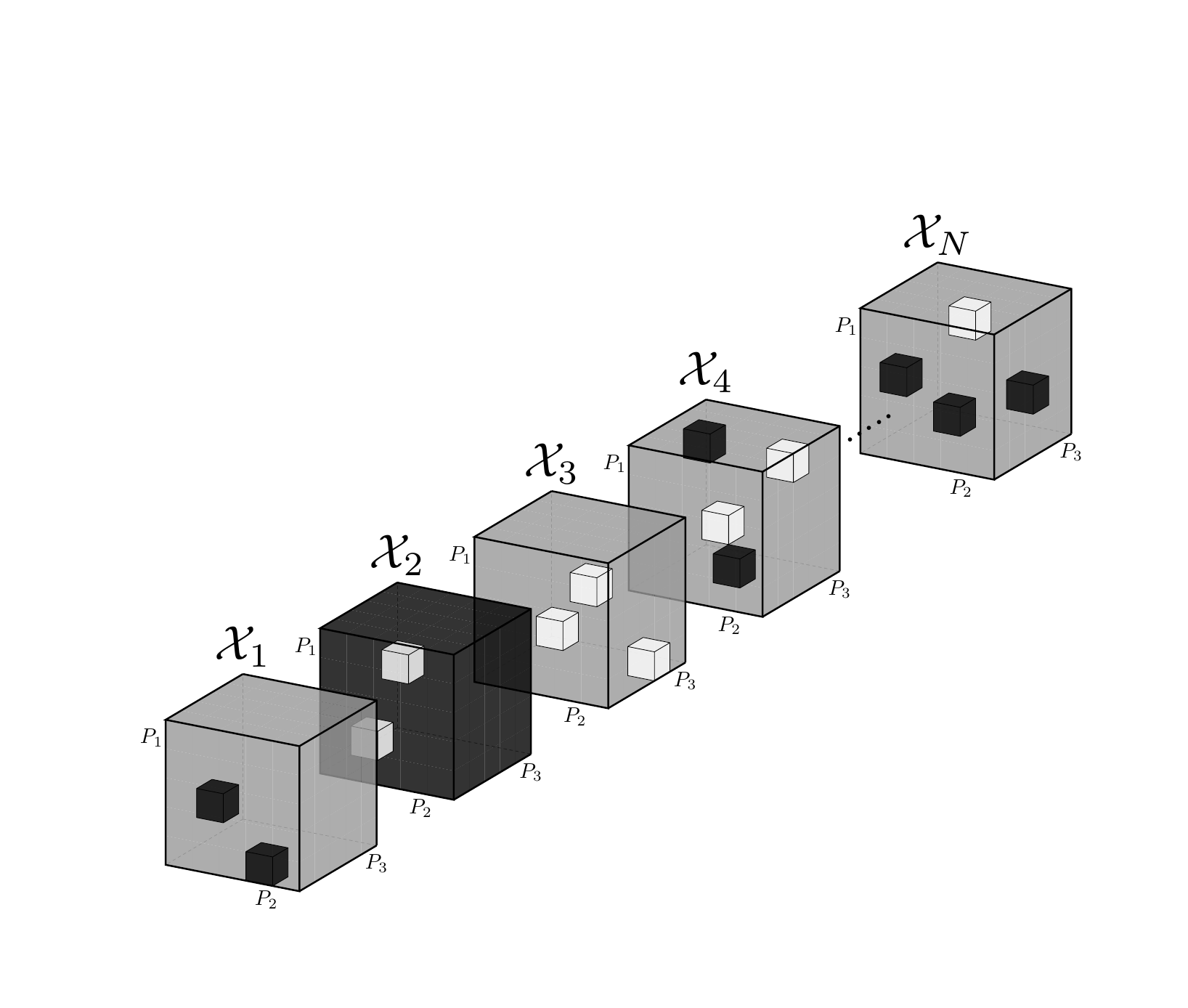}
\vspace{-5mm}
\caption{Illustration of a casewise outlier (black cube $\mathcal{X}_2$), cellwise outliers (black cells), along with missing values (white cells) in tensor data.}
\label{fig:casecellmixed}
\end{figure}

As for other data types, also for tensor data a common issue that may arise is the presence of outliers. Robust methods address this issue by constructing fits that are little affected by anomalies. These robust methods then make it possible to identify outliers based on their deviation from the robust fit. Since its inception in the 1960s, research in robust statistics mainly focused on casewise outliers, which is the setting where a minority of entire observations can deviate from the bulk of the data. 
However, the advent of high-dimensional data instigated a paradigm shift in robust statistics towards cellwise outliers \citep{Alqallaf:cell}, where the focus is on individual cells that deviate from the bulk of the data.
Cellwise robust methods on the one hand aim to mitigate the effect of the deviating cells while on the other hand they want to preserve the valuable information in the other components of these  observations \citep{Raymaekers:Challenges}. Ideally, modern robust methods can handle both cellwise and casewise outliers \citep{Centofanti:RODESSA, Centofanti:cellRCOV}. Another common issue usually encountered in practice is the presence of missing entries, so a versatile method should be able to handle missing data as well. Figure~\ref{fig:casecellmixed} provides a simplified illustration of casewise and cellwise outliers, along with missing values in tensor data.

In the multivariate setting, it is well-known that classical PCA is not robust. Several methods have been developed that are robust in the presence of casewise outliers \citep{Locantore:RobPCA,Croux:Proj,Hubert:ROBPCA}.  \cite{Serneels:RobPCA} proposed a casewise robust PCA method that can also cope with missing data. \cite{Torre:Robframework} and \cite{Maronna:RobElement} proposed robust methods for data containing cellwise outliers and missing values. The more recently proposed MacroPCA \citep{Hubert:MacroPCA} and cellPCA \citep{Centofanti:cellpca} methods can handle missing values, cellwise and casewise outliers simultaneously.

Similarly to PCA, the standard PARAFAC, Tucker and MPCA methods for tensor decomposition are sensitive to both casewise and cellwise outliers.
Robust methods for PARAFAC decomposition include the casewise robust method of \cite{Engelen:RPARAFAC} and \cite{Todorov:PARAFAC}, its extension for tensors with missing data \citep{Hubert:RPARAFAC-SI} and the MacroPARAFAC method \citep{Hubert:MacroPARAFAC} that can handle cellwise outliers as well.
\cite{Inoue:RMPCA} introduced robust adaptations of MPCA that can either deal with casewise or cellwise outliers in the data. However, there does not yet exist an MPCA method that can simultaneously handle both types of outliers as well as missing values.

We introduce a robust MPCA method (ROMPCA) capable of managing both types of outliers and missing values. ROMPCA minimizes a single loss function in which the effect of both casewise and cellwise outliers is appropriately dampened, while missing cells are discarded to avoid the need for a robust imputation strategy. Section~\ref{sec:meth} introduces basic multilinear concepts and our newly proposed ROMPCA method. We also provide a formal model for cellwise and casewise outliers, as well as missing values, in tensor data.
To solve the ROMPCA minimization problem, we introduce an iteratively reweighted least squares algorithm which uses alternating least squares in each iteration step. Section~\ref{sec:simu} evaluates its performance through extensive simulations while Section~\ref{sec:outdetect} presents several diagnostics to detect and visualize outlying cells and cases. Finally, we apply our method to two real datasets in Section~\ref{sec:realdata} and Section~\ref{sec:disc} concludes.

\section{Methodology}
\label{sec:meth}
\subsection{Preliminaries and Notation}

This section reviews some of the basic multilinear concepts and notations that will be used throughout the paper. More details can be found in \cite{Lu:MPCA}.
An $L$-order tensor is denoted as $\mathcal{A}=\left[ a_{p_1\ldots p_L}\right] \in \mathbb{R}^{P_1  \times \cdots \times P_L}$  where each of the $L$ indices $p_\ell=1, \ldots, P_\ell$ addresses the mode-$\ell$ of $\mathcal{A}$.
The tensor $\mathcal{A}$ resides in the tensor space $\mathbb{R}^{P_1}\otimes \ldots \otimes \mathbb{R}^{P_L}$, which is the tensor product (outer product) of $L$ vector spaces $\mathbb{R}^{P_1}, \ldots, \mathbb{R}^{P_L}$. 
The mode-$\ell$ vectors of $\mathcal{A}$ are defined as the $P_\ell$-dimensional vectors obtained from $\mathcal{A}$ by varying the index $p_\ell$ while keeping all the other indices fixed. 
For instance, $\mathcal{A}_{:, p_2,\ldots,p_L}$ is a mode-1 vector, where the colon notation $:$ indicates the full range of a given index.
The $p_\ell$th mode-$\ell$ slice of $\mathcal{A}$ is obtained by fixing the mode-$\ell$ index of $\mathcal{A}$ to be $p_\ell$, i.e., $\mathcal{A}_{:, \ldots,:,p_\ell,:,\ldots,:}$.
The mode-$\ell$ unfolding of $\mathcal{A}$ is denoted as $\mathbf{A}_{(\ell)} \in \mathbb{R}^{P_\ell \times (P_1  \cdots P_{\ell-1}  P_{\ell+1} \cdots  P_L)}$ whose  column vectors are the mode-$\ell$ vectors of $\mathcal{A}$. The order of the columns of the mode-$\ell$ vectors in $\mathbf{A}_{(\ell)}$ is not important as long as it is consistent throughout the computation. The operation $\text{vec}(\mathcal{A})$ vectorizes any tensor $\mathcal{A}$ into a column vector $\mathbf{a} \in \mathbb{R}^{\prod_{\ell}^{L} P_\ell}$.

The mode-$\ell$ product of a tensor $\mathcal{A}$ by a matrix $\mathbf{V}=\left[ v_{p_\ell k_\ell}\right] \in \mathbb{R}^{P_\ell \times K_\ell}$ is a tensor $\mathcal{B}=\left[ b_{p_1\ldots p_L}\right] \in \mathbb{R}^{P_1  \times \cdots \times P_{\ell-1} \times K_{\ell} \times P_{\ell+1} \times \cdots \times P_L}$, denoted by $  \mathcal{B} = \mathcal{A} \times_\ell \mathbf{V}^T$,
where each entry of $\mathcal{B}$ is defined as the sum of products of corresponding entries in $\mathcal{A}$ and $\mathbf{V}$, that is $b_{p_1 \ldots p_{\ell-1} k_{\ell} p_{\ell+1} \ldots p_L} = \sum_{p_\ell=1}^{P_\ell} a_{p_1 \ldots p_L} v_{p_\ell k_\ell}$.

Any $L$-order tensor $\mathcal{A}$ can be decomposed  as the mode-$\ell$ product of an $L$-order core tensor $\mathcal{B}$ and a collection of $L$ matrices $\mathbf{V}^{(\ell)}\in \mathbb{R}^{P_\ell \times P_\ell}$ with orthonormal columns, as follows
\begin{equation}
    \mathcal{A} = \mathcal{B} \times_1 \mathbf{V}^{(1)} \times_2  \cdots \times_L \mathbf{V}^{(L)} := \mathcal{B} \times \lbrace\mathbf{V}\rbrace,
\label{eq:multdecomp}
\end{equation}
where $\mathcal{B} = \mathcal{A} \times_1  \mathbf{V}^{(1)T} \times_2  \cdots \times_L {\mathbf{V}^{(L)T}} := \mathcal{A} \times \lbrace{\mathbf{V}}^T\rbrace$ \citep{Tucker:Tucker}. 
This decomposition can also be expressed using either vector or matrix formulation \citep{BallardKolda:bookTensor} as follows
\begin{align*}
    &\text{vec}\left( \mathcal{A} \right) = \left(\mathbf{V}^{(L)} \otimes \cdots \otimes \mathbf{V}^{(1)} \right)\text{vec}\left(\mathcal{B} \right):= \left(\sbigotimes_{\ell=1}^L \mathbf{V}^{(\ell)} \right) \text{vec}\left(\mathcal{B} \right) , \\
    &\mathbf{A}_{(\ell)} = \mathbf{V}^{(\ell)}\mathbf{B}_{(\ell)} \left(\mathbf{V}^{(L)} \otimes \cdots \otimes \mathbf{V}^{(\ell+1)} \otimes \mathbf{V}^{(\ell-1)} \otimes \cdots \otimes \mathbf{V}^{(1)} \right)^{T},
\end{align*}
where $\otimes$ denotes the Kronecker product. The mode-$\ell$ product of $\mathcal{A}$ over \textit{all but one} of the collection of  matrices $\mathbf{V}^{(\ell)}$ is defined as
\begin{equation*}
    \mathcal{A}^{(-\ell)} = \mathcal{A} \times_1 \mathbf{V}^{(1)T} \times_2 \cdots \times_{\ell-1} \mathbf{V}^{(\ell - 1)T} \times_{\ell+1} \mathbf{V}^{(\ell + 1)T} \times_{\ell+2} \cdots \times_L \mathbf{V}^{(L)T} : = \mathcal{A} \times_{-\ell} \lbrace\mathbf{V}^T\rbrace.
\end{equation*}

The inner product between two tensors of the same dimension $\mathcal{A}$ and $\mathcal{B}$ is defined as $$\langle \mathcal{A}, \mathcal{B} \rangle = \sum_{p_1=1}^{P_1} \sum_{p_2=1}^{P_2} \cdots \sum_{p_L=1}^{P_L} a_{p_1 \ldots p_L}  b_{p_1 \ldots p_L}  := \sum_{p_1 \cdots p_L}^{P_1 \cdots P_L}  a_{p_1 \ldots p_L}  b_{p_1 \ldots p_L}\, ,$$ and the Frobenius norm of a tensor is  $||\mathcal{A}||_F = \sqrt{\langle \mathcal{A}, 
\mathcal{A} \rangle}$.

\sloppy 
The contracted product of the tensors $\mathcal{A}=\left[ a_{p_1\ldots p_L j_1\dots j_M}\right] \in \mathbb{R}^{P_1  \times \cdots \times  P_L \times J_1  \times \cdots \times  J_M}$ and $\mathcal{B}=\left[ b_{p_1\ldots p_L k_1,\dots,k_N}\right] \in \mathbb{R}^{P_1  \times \cdots \times  P_L\times K_1  \times \cdots \times  K_N}$ along the first $L$ modes is a tensor
$\mathcal{D}=\langle \mathcal{A},\mathcal{B} \rangle_{\lbrace J_1,\dots,J_M;K_1,\dots,K_N \rbrace}=\left[ d_{j_1\ldots j_M k_1\dots k_N}\right] \in \mathbb{R}^{J_1  \times \cdots \times  J_M\times K_1  \times \cdots \times  K_N}$ with 
$d_{j_1\ldots j_M k_1\dots k_N}=\sum_{p_1 \cdots p_L}^{P_1 \cdots P_L}a_{p_1\ldots p_L j_1\dots j_M}b_{p_1\ldots p_L k_1,\dots,k_N}$.

The identity matrix of dimension $P_\ell$ is denoted by $\mathbf{I}_{P_\ell}$. A zero tensor and zero matrix are denoted by $\mathcal{O}_{P_1 \times \cdots \times P_L}$ and $\mathbf{0}_{P_\ell \times K_\ell}$,  where the subscripts indicate their dimensions.

\subsection{Multilinear Principal Component Analysis}
\label{sec:MPCAdef}
Let $\{\mathcal{X}_n=\left[x_{n, p_1\ldots p_L}\right]\}_{n=1}^N$ be a given set of $N$ independent $L$-order tensors. 
The objective of tensor dimension reduction is to find a collection of orthogonal matrices  $\lbrace\mathbf{V}^{(\ell)}\in \mathbb{R}^{P_\ell\times K_\ell}\rbrace_{\ell=1}^L$ of rank $K_\ell \leqslant P_\ell$,  
referred to as the mode-$\ell$ projection matrices, together with a center $\mathcal{C} \in \mathbb{R}^{P_1\times \cdots \times P_L} $ and core tensors $\lbrace\mathcal{U}_n \in \mathbb{R}^{K_1\times\dots\times K_L}\rbrace_{n = 1}^{N}$ such that the reconstructed tensors
\begin{equation} \label{eq:fitted}
\widehat{\mathcal{X}}_n = \mathcal{C} + \mathcal{U}_n \times\lbrace\mathbf{V}\rbrace
\end{equation}
are a good approximation of the original tensors $\mathcal{X}_n$. 

In standard MPCA \citep{Lu:MPCA} the center is not estimated but is fixed to be the mean tensor $\mathcal{C} = \widebar{\mathcal{X}} = \frac{1}{N} \sum_{n = 1}^{N} \mathcal{X}_n$
whereas for fixed projection matrices $\mathbf{V}^{(\ell)}$ the core tensors are given by 
$\widetilde{\mathcal{U}}_n = ({\mathcal{X}}_n - \widebar{\mathcal{X}}) \times \lbrace\mathbf{V}^T\rbrace := \widetilde{\mathcal{X}}_n \times \lbrace\mathbf{V}^T\rbrace$. 
The projection matrices 
are determined such that the resulting core tensors capture most of the variation observed in the original tensors, i.e.\ the total variation $\sum_{n=1}^N\|\widetilde{\mathcal{U}}_n\|_F^2$ is maximized over all projection matrices $\mathbf{V}^{(\ell)}$. 
As explained in Section~\ref{app:MPCAform} of the Supplementary Material, this solution also minimizes the sum of the squared reconstruction errors
\begin{equation}
\label{eq:MPCA}
\sum_{n=1}^N \left\|\mathcal{X}_n - \widehat{\mathcal{X}}_n \right\|_F^2 = 
\sum_{n=1}^N\Big\|\mathcal{X}_n -\mathcal{C} -\mathcal{U}_n \times\lbrace\mathbf{V}\rbrace\Big\|_F^2\, .
\end{equation}
The projection matrices $\mathbf{V}^{(\ell)}$ are computed one by one with all the others kept fixed and consist of the $K_\ell$ eigenvectors corresponding to the largest $K_\ell$ eigenvalues of the matrix 
$\mathbf{\Phi}^{(\ell)} = \Sumn \left( \mathbf{X}_{n(\ell)} - \widebar{\mathbf{X}}_{(\ell)} \right) \mathbf{V}_{\Phi^{(l)}} \mathbf{V}_{\Phi^{(l)}}^T \left( \mathbf{X}_{n(\ell)} - \widebar{\mathbf{X}}_{(\ell)} \right)^T$ with
$ \mathbf{V}_{\Phi^{(\ell)}} = \mathbf{V}^{(\ell+1)} \otimes \mathbf{V}^{(\ell+2)} \otimes \dots \otimes \mathbf{V}^{(L)} \otimes \mathbf{V}^{(1)} \otimes \mathbf{V}^{(2)} \otimes \dots \otimes \mathbf{V}^{(\ell-1)}.$ 
Note that the purpose of centering in this framework is to prevent the mean from dominating the leading eigenvalues of $\mathbf{\Phi}^{(\ell)}$, thereby allowing the tensor decomposition to focus on the genuine low-rank structure.

\begin{figure}[!ht]
    \centering
    \includegraphics[width=0.8\textwidth]{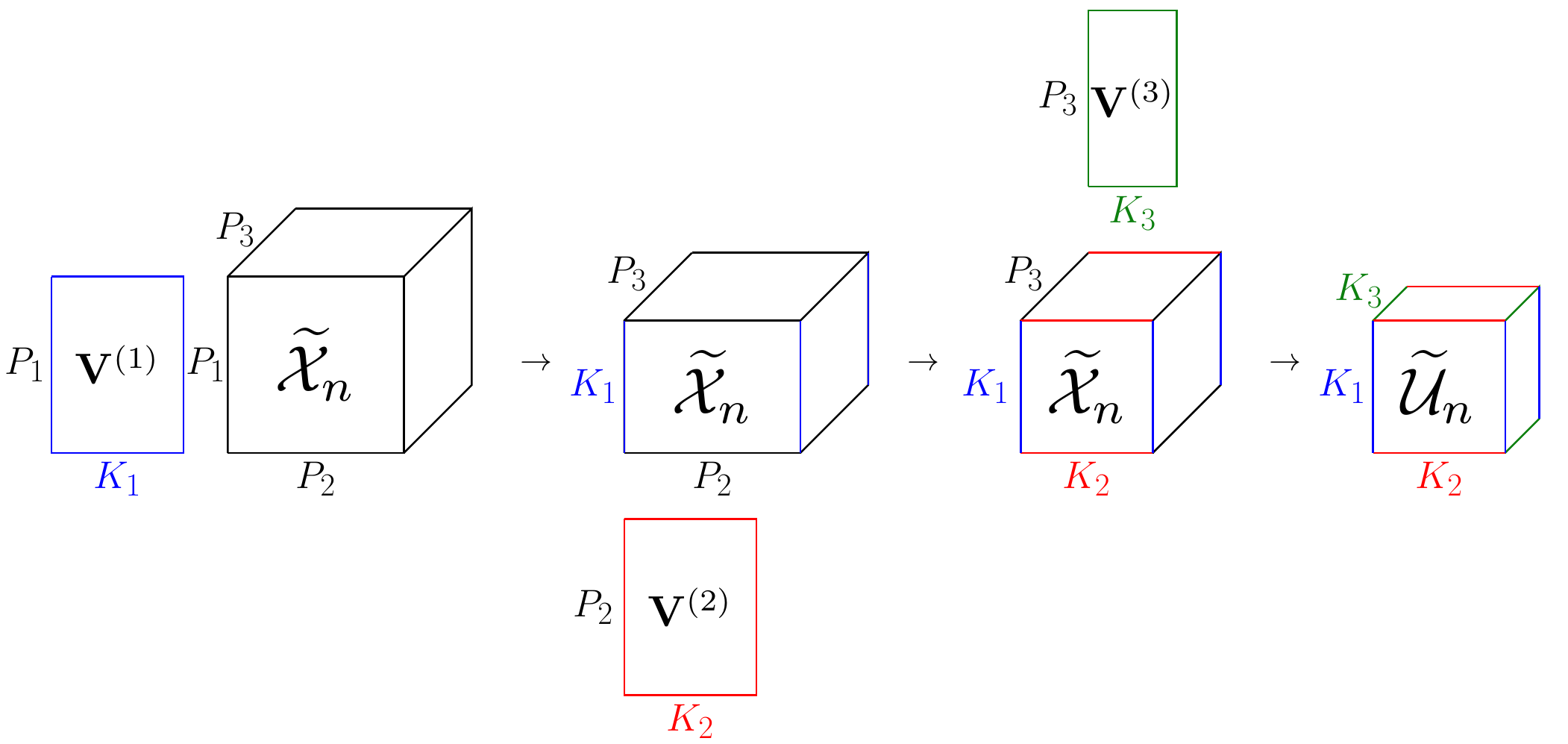} %
    \caption{Illustration of the MPCA decomposition of a mean-centered tensor $\widetilde{\mathcal{X}}_n$.}%
    \label{fig:illustrationMPCA}%
\end{figure}

The MPCA decomposition is illustrated in Figure~\ref{fig:illustrationMPCA}. The projection matrices are common to all samples $\widetilde{\mathcal{X}}_n$ while each of them reduces the dimension of one mode of the $\widetilde{\mathcal{X}}_n$. 
Note that the low-rank constraint is automatically satisfied as $\mathbf{V}^{(\ell)} \in \mathbb{R}^{P_\ell \times K_\ell}$ with $K_\ell \leqslant P_\ell$.

\subsection{Casewise and Cellwise Outliers in Tensor Data}
\label{sec:DefOutl}

\begin{figure}[!ht]
\centering
\includegraphics[width=0.32\textwidth]{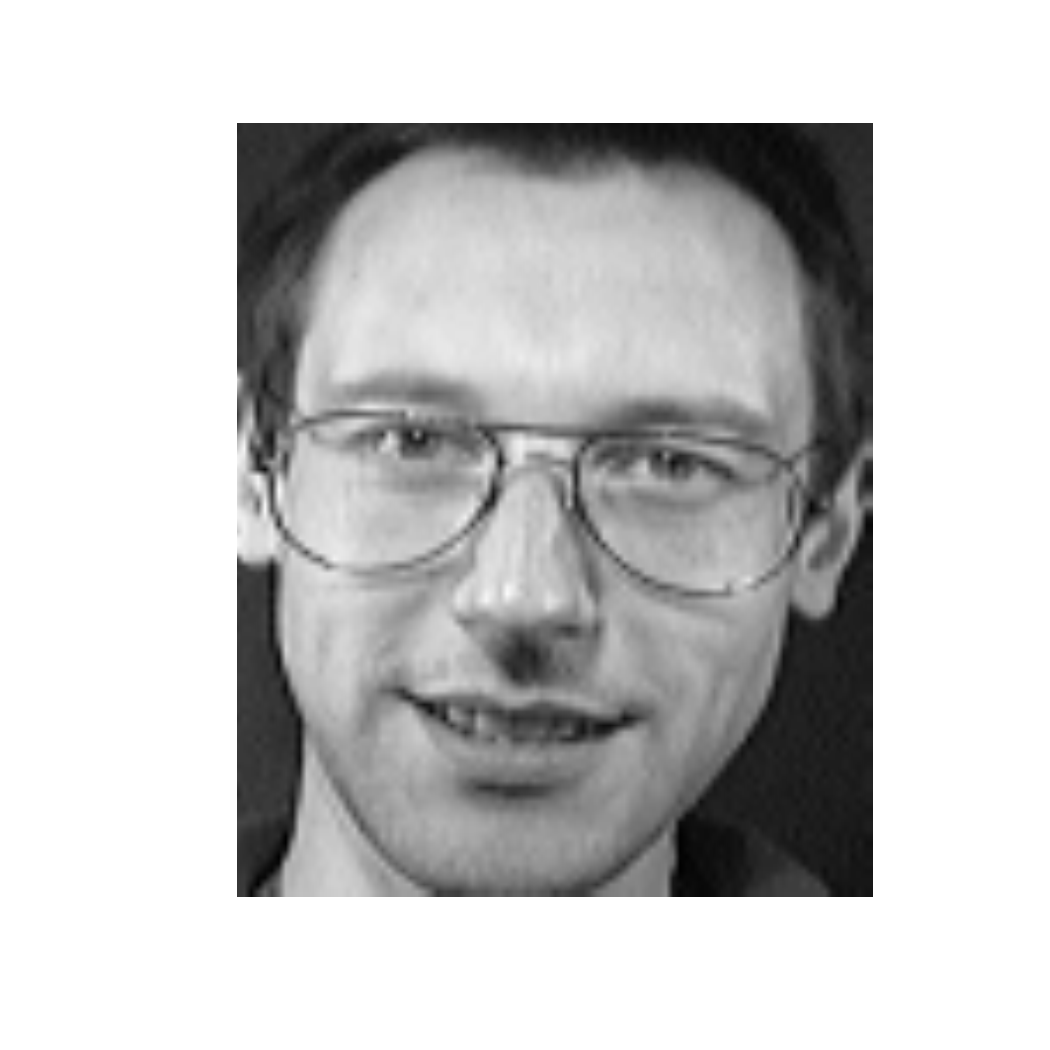}
\includegraphics[width=0.32\textwidth]{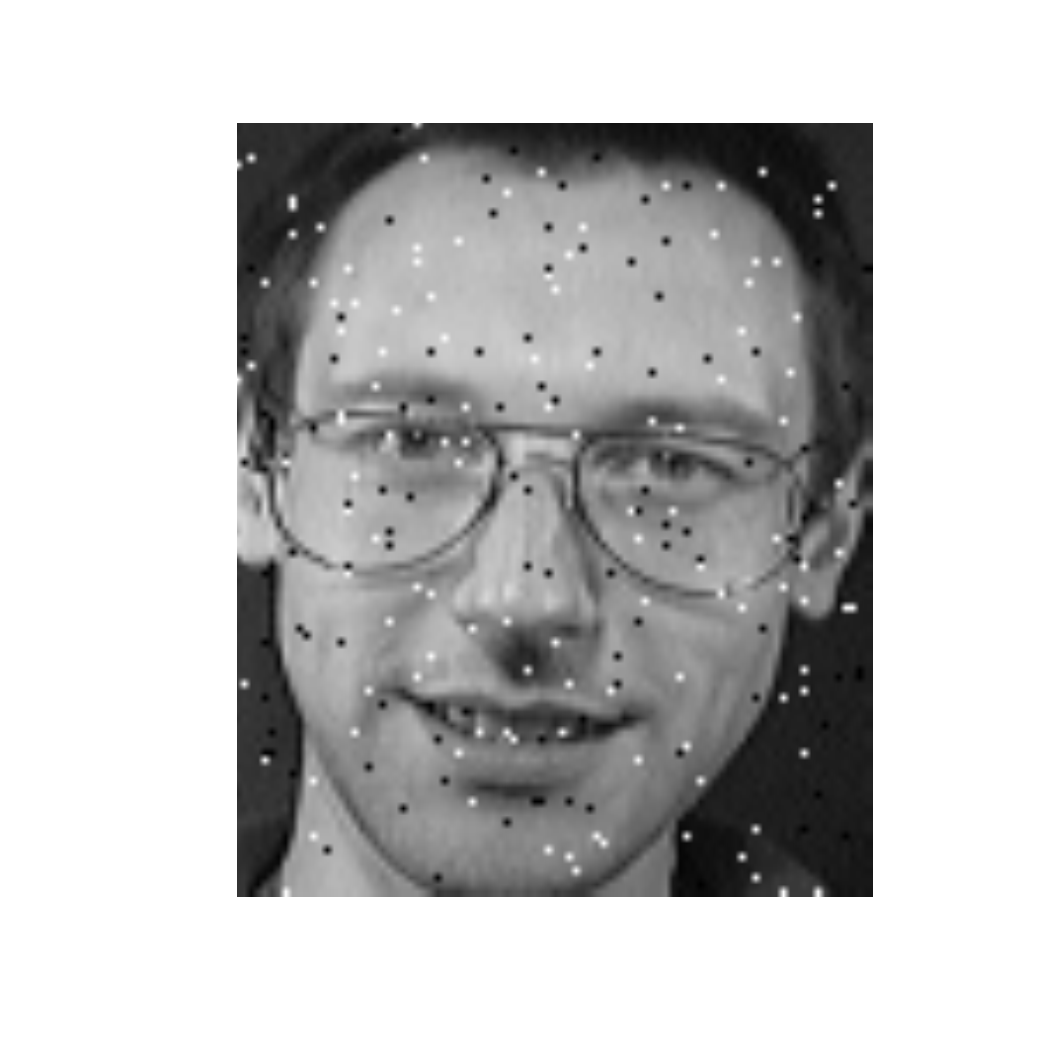}
\includegraphics[width=0.32\textwidth]{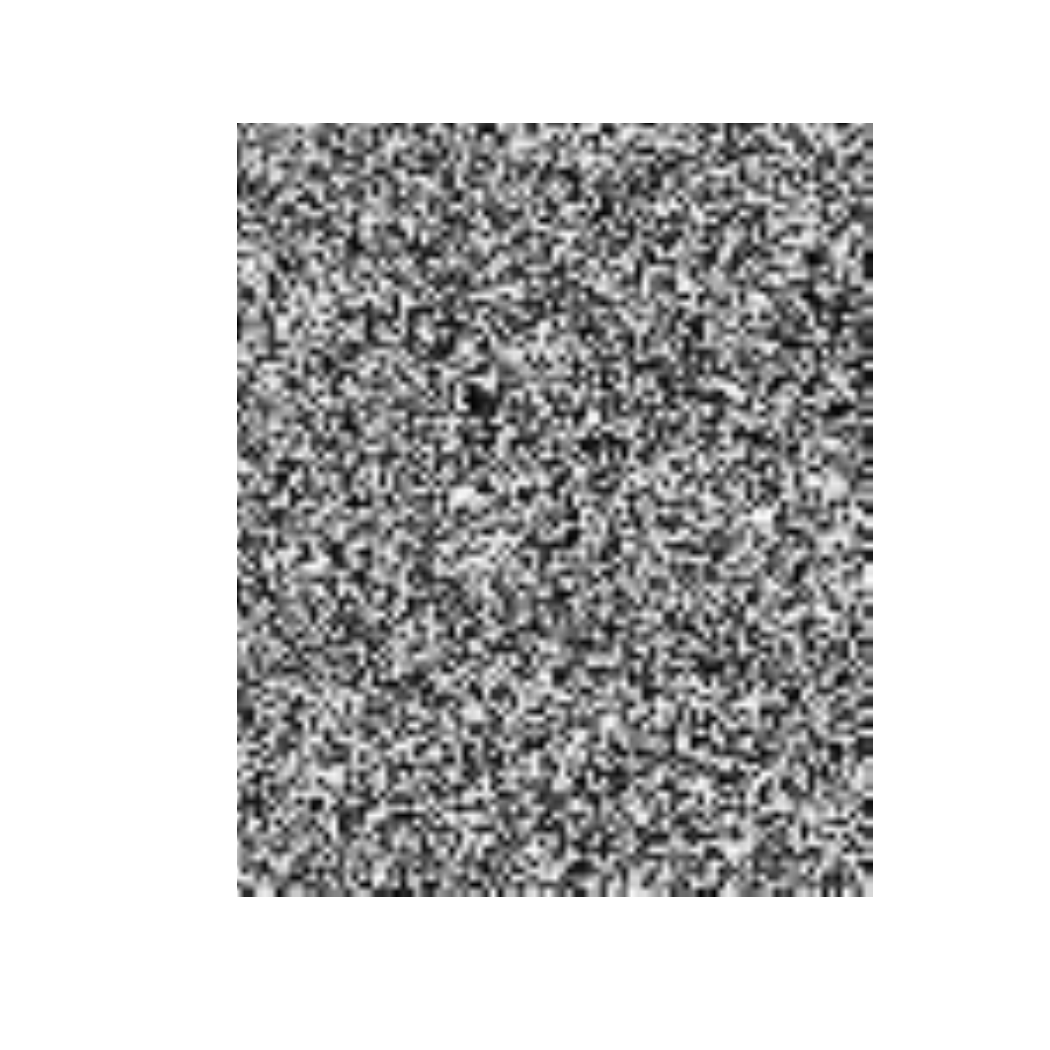}
\vspace{-5mm}
\caption{Face images from the ORL dataset: (left) an individual with a given facial expression,
(middle) the same individual with corrupted pixels, and (right) a noise image.}
\label{fig:ORLdata}
\end{figure}

Traditionally, the term \textit{outlier} refers to an entire outlying case. Such \textit{casewise outliers} are observations that are not generated by the same mechanism as the majority of the data. We retake the example given by \cite{Inoue:RMPCA}. Consider the Olivetti Research Ltd.\ (ORL) face image dataset~\citep{Samaria:ORL}, which contains face images of 40 individuals, each with 10 different facial expressions. Figure~\ref{fig:ORLdata} shows an individual facial expression (left) alongside a casewise outlier (right), which in this example corresponds to a noise image. 
Casewise robust methods assume that fewer than half of the cases contain corrupted values. This assumption is often unrealistic for tensor data. Moreover, casewise robust methods downweight or remove all measurements of the outlying cases, even though the outlying behavior may occur in only a few values. 
This has motivated the study of \textit{cellwise outliers} in recent years. 
These are deviating measurements (cells) that can occur anywhere within the tensor. In the ORL dataset, cellwise outliers are illustrated in the middle panel of Figure~\ref{fig:ORLdata}, where several pixels of the image are corrupted.

In real data, both types of outliers often occur simultaneously, and some measurements may also be missing. Following~\cite{Centofanti:cellpca}, we define the \textit{mixed contaminated and partially observed tensor contamination model}. It assumes that the observed tensor is sampled from a $(P_1 \times \cdots \times P_L)$-variate random variable $\mathcal{X}_{\varepsilon}$, expressed as
\begin{equation} \label{eq:cont_both}
\mathcal{X}_{\varepsilon} = \mathcal{A} \odot \mathcal{X} + (\mathbf{1}_{P_1 \times \cdots \times P_L}-\mathcal{A}) \odot \mathcal{Z}, 
\end{equation}
where $\odot$ denotes the Hadamard (elementwise) product. Here, $\mathcal{X} \sim H_0$ and $\mathcal{Z} \sim H_\mathcal{Z}$, where $H_0$ is the generating distribution of the regular observations and $H_Z$ is an unspecified outlier-generating distribution. The tensor $\mathbf{1}_{P_1 \times \cdots \times P_L}$ contains only ones, whereas
the $(P_1 \times \cdots \times P_L)$-variate variable  $\mathcal{A}$ is defined as $\mathcal{A} = \mathcal{A}^{\text{case}} \odot \mathcal{A}^{\text{cell}} \odot \mathcal{A}^{\text{obs}}$. The variable $\mathcal{A}^{\text{case}}$ has Bernoulli-distributed marginals $A_{p_1 \ldots p_L}^{\text{case}}$ with success parameter $1-\varepsilon^{\text{case}}$, where $0 \leqslant \varepsilon^{\text{case}} \leqslant 0.5$. Its components are fully dependent in the sense that all elements of $\mathcal{A}^{\text{case}}$ take the same value (either all equal to 1 or all equal to 0).
The $(P_1 \times \cdots \times P_L)$-variate variable $\mathcal{A}^{\text{cell}}$ has Bernoulli components $A_{p_1 \ldots p_L}^{\text{cell}}$ with success probabilities $1-\varepsilon^{\text{cell}}_{p_1 \ldots p_L}$. Finally, the components $A_{p_1 \ldots p_L}^{\text{obs}}$ of $\mathcal{A}^{\text{obs}}$ are binary variables taking values in $\lbrace 1, \text{NA}\rbrace$, with $P(A_{p_1 \ldots p_L}^{\text{obs}} = 1) = 1-\varepsilon^{\text{obs}}_{p_1 \ldots p_L}$.
Different assumptions on the dependence structure of $\mathcal{X}$, $\mathcal{Z}$, $\mathcal{A}^{\text{case}}$, $\mathcal{A}^{\text{cell}}$ and $\mathcal{A}^{\text{obs}}$ lead to different contamination models.
For example, when $\mathcal{A}^{\text{obs}}$ is independent of $\mathcal{X}, \mathcal{Z}$ and of the other variables, the values in the dataset are missing completely at random.

\subsection{Robust MPCA}
\label{sec:obj}
In this section, we propose a new method for robust MPCA  which can deal 
with a sample of tensors following model~\eqref{eq:cont_both}, i.e.\ the dataset might contain casewise outlying tensors, cellwise outliers, as well as missing entries. For each given $\mathcal{X}_n$, the tensor $\mathcal{M}_n=[m_{n,p_1 \ldots p_L}]$ has value $m_{n,p_1 \ldots p_L}=1$ if $x_{n,p_1 \ldots p_L}$ is observed, and 0 otherwise. The scalar $m_{n} = \Sumps m_{\eidx}$ denotes the number of observed cells in $\mathcal{X}_n$ and $m = \Sumn m_n$ the total number of observed cells in the set $\{\mathcal{X}_n\}$.
 
The proposed method ROMPCA  estimates projection matrices $\mathbf{V}^{(\ell)}=[v_{p_\ell k_\ell}^{(\ell)}] $, core tensors $\mathcal{U}_n =[u_{n,k_1\ldots k_L}] $ and a robust center $\mathcal{C} = [c_{p_1 \ldots p_L}]$ by minimizing  
\begin{equation}
\label{eq:obj}
\mathcal{L}\left(\lbrace{\mathcal{X}_n \rbrace},\lbrace\mathbf{V}^{(\ell)}\rbrace, \lbrace\mathcal{U}_n\rbrace, \mathcal{C} \right) := \frac{\hat{\sigma}_2^2}{m} \Sumn \ m_n \rho_2\left(\frac{1}{\hat{\sigma}_2}\sqrt{\frac{1}{m_n} \Sumps  m_{\eidx} \scell^2\, \rho_1\left(\frac{r_{\eidx}}{\scell}\right)}\ \right),\\
\end{equation}
where the cellwise residuals are given by
\begin{equation} \label{eq:cellres}
    r_{\eidx}:= \left({x}_{\eidx}- c_{\idx} -  \Sumks 
    {u}_{\Jidx} v_{p_1 k_1}^{(1)} \cdots v_{p_L k_L}^{(L)}\right).
\end{equation}
Each of the scales $\scell$  standardizes the corresponding cellwise residuals $r_{\eidx}$ while  the scale $\hat{\sigma}_2$  standardizes the \textit{casewise deviations}
\begin{equation} \label{eq:caseres}
t_{n} := \sqrt{\frac{1}{m_n} \Sumps m_{\eidx} \scell^2 \rho_1\left(\frac{r_{\eidx}}{\scell}\right)}.
\end{equation}

If $\rho_1(z) = \rho_2(z) = z^2$ and there are no missing entries, the objective function \eqref{eq:obj} reduces to \eqref{eq:MPCA}, the objective function minimized by MPCA. However, to handle both cellwise and casewise outliers simultaneously, we use bounded, even functions $\rho_1(z)$ and $\rho_2(z)$ in ROMPCA with $\rho_j(0)=0$ and $\rho_j(z)>0$ and nondecreasing for $z>0$ ($j=1,2$). 
Specifically, $\rho_1$ is designed to mitigate cellwise outliers. The presence of cellwise outliers in the $n$th tensor would result in large absolute residuals $r_{\eidx}$, but their impact is tempered due to the boundedness of $\rho_1$. Similarly, $\rho_2$ addresses casewise outliers.  A casewise outlier leads to a large deviation $t_{n}$ whose effect is then reduced by $\rho_2$. Note that the presence of $\rho_1$ in  $t_{n}$ reduces the influence of outlying cells in the $n$th tensor and avoids that a single cellwise outlier would always result in a large $t_{n}$\,. 

\begin{figure}[!ht]
    \centering
    \includegraphics[width=0.49\textwidth]{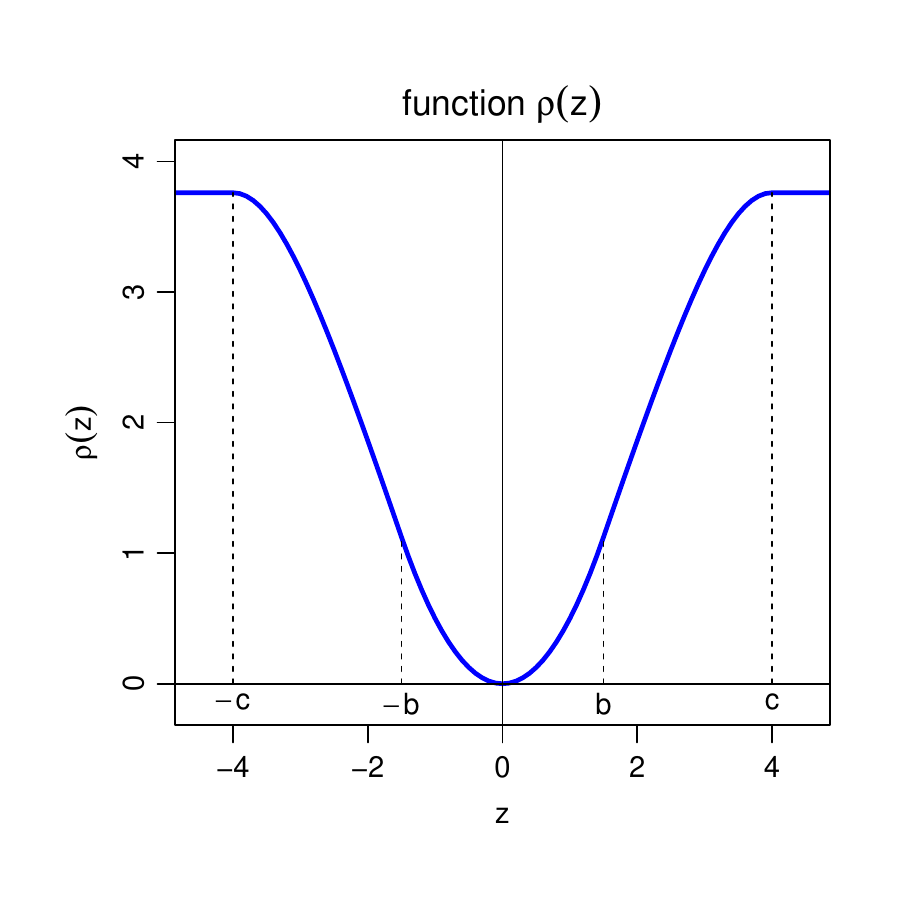} %
    \includegraphics[width=0.49\textwidth]{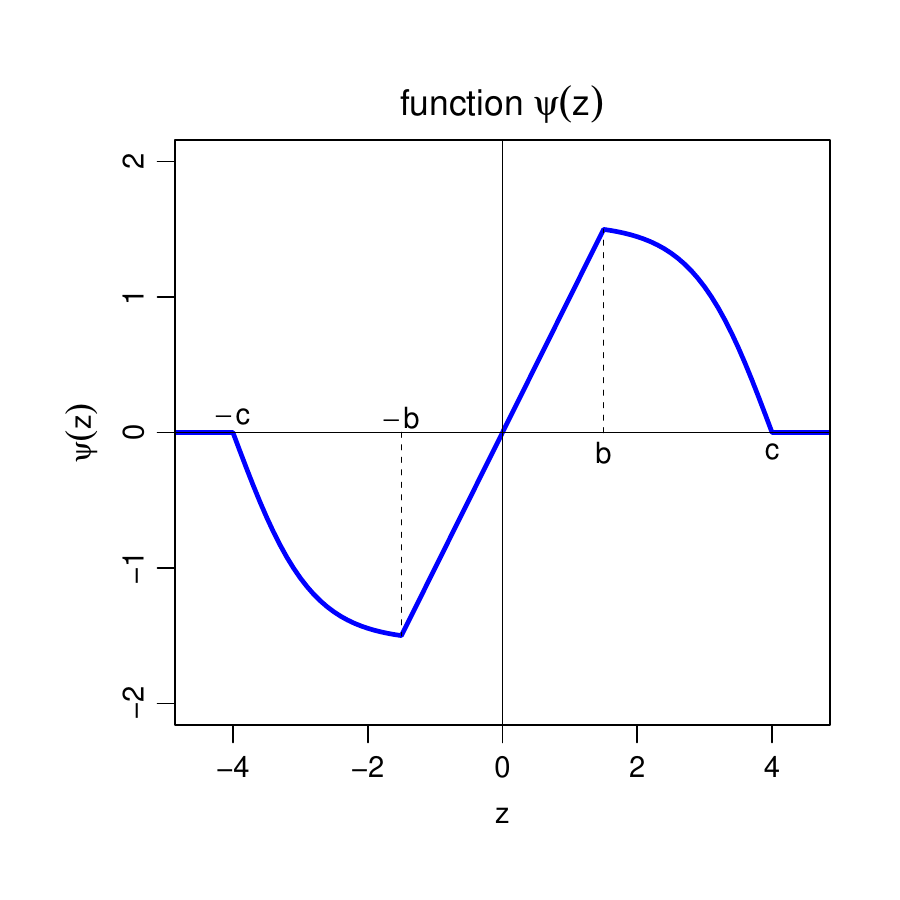} %
    \vspace{-0.5cm}
    \caption{The function $\rho_{b,c}$ with $b = 1.5$ and $c = 4$ (left) and its derivative $\psi_{b,c}$ (right).}%
    \label{fig:HypTan}%
\end{figure}

More precisely, in ROMPCA both $\rho$-functions in~\eqref{eq:obj} are chosen to be the hyperbolic tangent $(tanh)$ \citep{Hampel:CVC}, defined by
\begin{equation*}
\rho_{b,c}(z) = 
\begin{cases} 
    z^2/2 & \text{if } 0 \leqslant |z| \leqslant b, \\
    d - \left(q_1/q_2\right) \ln\left(\cosh\left(q_2\left(c - |z|\right)\right)\right) & \text{if } b \leqslant |z| \leqslant c, \\
    d & \text{if } c \leqslant |z|,
\end{cases}
\end{equation*}
where $d = \left( {b^2}/{2} \right) + \left( q_1/q_2 \right) \ln\left(\cosh\left(q_2 (c - b)\right)\right)$. Its derivative, denoted by $\psi_{b,c} = \rho'_{b,c}$, is
\begin{equation*}
\psi_{b,c}(z) = 
\begin{cases} 
    z & \text{if } 0 \leqslant |z| \leqslant b, \\
    q_1 \tanh\left(q_2\left(c - |z|\right)\right) \, \text{sign}(x) & \text{if } b \leqslant |z| \leqslant c, \\
    0 & \text{if } c \leqslant |z|.
\end{cases}
\end{equation*}
The hyperbolic tangent $\rho$-function and its derivative are shown in Figure~\ref{fig:HypTan} for $b=1.5$, $c=4$, $q_1=1.54$, and $q_2=0.86$. These choices offer a good balance between efficiency and robustness, as discussed in \cite{Hampel:bookIF} and~\cite{Raymaekers:FastCorr}.
At the normal location model, the \textit{tanh} M-estimator attains approximately an efficiency of  95\% and a breakdown value of 50\%. Moreover it has a low gross-error sensitivity and minimal sensitivity in terms of asymptotic variance to infinitesimal contamination.
These choices are further supported by the simulation results in Section~\ref{app:addsim} of the Supplementary Material.
Since $\psi_{1.5,4}(z) = 0$ for $|z| \geqslant 4$, large outlying cells and cases will be completely downweighted in the estimation procedure. See Section~\ref{sec:algo} for more details. 

To correctly reduce the effect of both cellwise and casewise outliers, it is important that the scale estimates $\scell$ and $\hat{\sigma}_{2}$ in~\eqref{eq:obj} are robust as well. Therefore, we will use M-scale estimators based on an initial fit as described in  Section~\ref{sec:init}. 

\subsection{The ROMPCA Algorithm}
\label{sec:algo}
To develop an algorithm to calculate the solution of \eqref{eq:obj}, we first show that it is equivalent to a weighted least squares problem in the sense that they both yield the same first-order conditions.  
Since the loss function in \eqref{eq:obj} is continuously differentiable, we show in Section~\ref{app:firstorder} of the Supplementary Material that its solution needs to satisfy the following first-order conditions
    \begin{align}
    \label{eq:condV}
    \Sumn \Big\langle \left({\mathcal{X}}_n - \mathcal{C} - \mathcal{U}_n \times \lbrace\mathbf{V}\rbrace \right) \odot \mathcal{W}_n \, , \,\mathcal{U}_n^{(-\ell)} \Big\rangle_{\lbrace L^{*};L^{*} \rbrace} & = \mathbf{0}_{P_\ell\times K_\ell}\ , \qquad \quad \ell=1,\ldots,L\\ 
    \label{eq:condU}
    \big( ( {\mathcal{X}}_n - \mathcal{C} - \mathcal{U}_n \times \lbrace \mathbf{V}\rbrace) \odot \mathcal{W}_n \big) \times \lbrace\mathbf{V}^{T}\rbrace & = \mathcal{O}_{K_1 \times \cdots \times K_L}\ , \quad n=1,\ldots,N\\  
    \label{eq:condC}
    \Sumn \left(\mathcal{X}_n - \mathcal{C} - \mathcal{U}_n \times \lbrace \mathbf{V}\rbrace \right) \odot \mathcal{W}_n  &= \mathcal{O}_{P_1\times\cdots \times P_L}\ ,
    \end{align}
where $\mathcal{U}_n^{(-\ell)} = \mathcal{U}_n \times_{-\ell} \lbrace \mathbf{V} \rbrace$ and $L^{*} = \lbrace 1, \ldots, \ell - 1, \ell + 1, \ldots,  L \rbrace$.
For each $n=1\ldots N$, the weight tensor $\mathcal{W}_n= \left[ w_{\eidx} \right]$ is defined as
\begin{equation} \label{eq:Wn}
    \mathcal{W}_n = \Wcell \odot \Wcase \odot\mathcal{M}_{n},
\end{equation}
where $\Wcell$ and  $\Wcase$ contain the cellwise and casewise weights, respectively. Note that, weights corresponding to missing cells are set to zero through $\mathcal{M}_n$, ensuring that these cells are completely downweighted. In particular, using the notation $\psi_i=\rho_i'$ for $i=1,2$, the tensor $\Wcell$ has entries
\begin{equation} \label{eq:weightc}
    \wcell =\psi_1\left(\frac{r_{\eidx}}{\scell}\right)\Big/\frac{r_{\eidx}}{\scell},
\end{equation}
whereas $\Wcase$ is a tensor with the same value  
\begin{equation} \label{eq:weightr}
    w^{\text{case}}_n =\psi_2\left( \frac{{t_n}}{\hat{\sigma}_2} \right)\Big/{\frac{{t_n}}{\hat{\sigma}_2}},
\end{equation}
in each of its cells. We use the standard convention that the ratio $\psi(z)/z$ becomes 1 if $z=0$ such that cells or cases with residual equal to 0 receive full weight 1. 
The conditions on the $\rho$ functions guarantee that the weights $w_{\eidx} \geqslant 0$. We also assume that $\Sumn w_{\eidx}^q > 0$ for each entry, i.e.\ there is no entry whose value is contaminated or missing in all of the $n$ samples.
Note that for the hyperbolic tangent $\rho$-function regular cells and cases (i.e.\ $|z|\leqslant b$) receive full weight 1 while strongly outlying cells and cases ($|z|\geqslant c$) receive weight 0.

For fixed weight tensors $\lbrace\mathcal{W}_n\rbrace$, 
equations \eqref{eq:condV}-\eqref{eq:condC} coincide with the first-order conditions of the weighted least squares problem
\begin{equation}
    \Sumn \Sumps w_{\eidx} \left({x}_{\eidx} - c_{\idx} - \Sumks u_{\Jidx} v_{p_1 k_1}^{(1)} \cdots v_{p_L k_L}^{(L)}\right)^2\,,
    \label{eq:WMPCA}
\end{equation}
as shown in Section~\ref{app:firstorder} of the Supplementary Material. 
While there is no closed form solution of~\eqref{eq:WMPCA}, the solution can be computed efficiently by alternating least squares \citep{Gabriel:LSapprox}. 
Since the weight tensors in \eqref{eq:condV}-\eqref{eq:condC} are not fixed but depend on the estimates, these equations can be solved by an Iteratively Reweighted Least Squares (IRLS) algorithm. 
The IRLS algorithm starts with an initial estimate $\left(\lbrace\mathbf{V}^{(\ell)}_{0}\rbrace, \lbrace\mathcal{U}_{n,0}\rbrace, \mathcal{C}_{0}\right)$ as explained in Section~\ref{sec:init}, and associated weight tensors $\lbrace\mathcal{W}_{n,0}\rbrace$ obtained from 
\eqref{eq:Wn} by using the corresponding residuals \eqref{eq:cellres} and \eqref{eq:caseres} in the weights \eqref{eq:weightc} and \eqref{eq:weightr}. Then, at each iteration step $q = 0,1,2, \ldots$, updated estimates $\lbrace\mathbf{V}^{(\ell)}_{q+1}\rbrace$, $\lbrace\mathcal{U}_{n,q+1}\rbrace$, $\mathcal{C}_{q+1}$ and corresponding weight tensors $\lbrace\mathcal{W}_{n,q+1}\rbrace$ are obtained from the current estimates $\lbrace\mathbf{V}^{(\ell)}_{q}\rbrace$, $\lbrace\mathcal{U}_{n,q}\rbrace$, $\mathcal{C}_{q}$, and $\lbrace\mathcal{W}_{n,q}\rbrace$  by the following procedure, which is derived in Section \ref{app:algo} of the Supplementary Material.

\begin{itemize}
\item[\textbf{(a)}] First, \eqref{eq:WMPCA} is minimized with respect to $\mathbf{V}^{(\ell)}$ by solving \eqref{eq:condV} for  $\ell=1,\ldots,L$ which yields the updated estimates
  \begin{multline}
      \text{vec}\left(\mathbf{V}^{(\ell)}_{q+1}\right) = \left(\Sumn \left(\mathbf{U}_{n(\ell), q}^{(-\ell)} \otimes \mathbf{I}_{P_\ell}\right)\mathbf{W}_{n,q}\left(\mathbf{U}_{n(\ell), q}^{(-\ell)} \otimes \mathbf{I}_{P_\ell}\right)^{T}\right)^{\dagger}\\\left(\Sumn \left(\mathbf{U}_{n(\ell), q}^{(-\ell)} \otimes \mathbf{I}_{P_\ell}\right)\mathbf{W}_{n, q} \text{vec}\left(\mathbf{X}_{n(\ell)} - \mathbf{C}_{(\ell),q}\right)\right),
      \label{eq:solV}
  \end{multline}
  where $^{\dagger}$ denotes the Moore-Penrose generalized inverse, $\mathbf{U}_{n(\ell), q}^{(-\ell)}$ is the mode-$\ell$ unfolding of $\mathcal{U}_{n,q}^{(-\ell)}$ and $\mathbf{C}_{(\ell), q}$ is the mode-$\ell$ unfolding of $\mathcal{C}_{q}$.
  The diagonal weight matrix $\mathbf{W}_{n,q}$ is defined such that its diagonal entries correspond to the vectorization of the mode-$\ell$ unfolding of $\mathcal{W}_{n,q}$.
  The matrix $\widetilde{\mathbf{W}}_{n,q}$ used below is similarly built from $\widetilde{\mathcal{W}}_{n,q} = \mathcal{W}_{n,q}^{\text{cell}} \odot \mathcal{M}_n $. To enforce the orthogonality constraint, the projection matrices $\lbrace \mathbf{V}^{(\ell)}_{q+1} \rbrace$ are orthonormalized at every step.
 
 \item[\textbf{(b)}] 
  Using the updated estimates $\mathbf{V}^{(\ell)}_{q+1}$, \eqref{eq:WMPCA} is then minimized with respect to $\mathcal{U}_n$ by solving \eqref{eq:condU} for $n=1,\ldots,N$ which yields the updated estimates
  \begin{equation}
      \text{vec}\left(\mathcal{U}_{n, q+1}\right) = \left(\left(\sbigotimes_{\ell=1}^{L} {\mathbf{V}_{q+1}^{(\ell)T}}\right)\widetilde{\mathbf{W}}_{n, q}\left(\sbigotimes_{\ell=1}^{L} {\mathbf{V}^{(\ell)}_{q+1}}\right)\right)^{\dagger}\left(\left(\sbigotimes_{\ell=1}^{L} {\mathbf{V}_{q+1}^{(\ell)T}}\right)\widetilde{\mathbf{W}}_{n, q} \text{vec}\left(\mathcal{X}_n - \mathcal{C}_{q}\right)\right)\, .
      \label{eq:solU}
  \end{equation}
 
  \item[\textbf{(c)}] 
   Using the updated estimates $\mathbf{V}^{(\ell)}_{q+1}$ and $\mathcal{U}_{n, q+1}$, the center $\mathcal{C}$ is then updated by solving \eqref{eq:condC} which yields
  \begin{equation}
      \mathcal{C}_{q+1} = \left(\Sumn \left({\mathcal{X}}_n  -\mathcal{U}_{n,q+1} \times \lbrace \mathbf{V}_{q+1}\rbrace\right) \odot \mathcal{W}_{n,q} \right) \odot \mathcal{H}_q,
      \label{eq:solC}
  \end{equation}
  where $\mathcal{H}_q = \left[1 / \Sumn w_{\eidx}^q\right] \in \mathbb{R}^{P_1 \times \ldots \times P_L}$ with $w_{\eidx}^q$ the weights in $\mathcal{W}_{n,q}$.
  
  \item[\textbf{(d)}] Finally, the updated estimates $\lbrace\mathbf{V}^{(\ell)}_{q+1}\rbrace$, $\lbrace\mathcal{U}_{n,q+1}\rbrace$, and $\mathcal{C}_{q+1}$ are used to update the residuals in \eqref{eq:cellres} and \eqref{eq:caseres}, the corresponding weights \eqref{eq:weightc} and \eqref{eq:weightr} and   
  weight tensor $\lbrace\mathcal{W}_{n,q+1}\rbrace$ in  \eqref{eq:Wn}.
\end{itemize}
The algorithm iterates these steps until 
\begin{equation}
      \frac{\mathcal{L}\left(\lbrace \mathcal{X}_{n} \rbrace,\lbrace\mathbf{V}_{q+1}^{(\ell)}\rbrace, \lbrace \mathcal{U}_{n, q+1} \rbrace, \mathcal{C}_{q+1} \right) - \mathcal{L}\left(\lbrace \mathcal{X}_{n} \rbrace,\lbrace\mathbf{V}_{q}^{(\ell)}\rbrace, \lbrace \mathcal{U}_{n, q} \rbrace, \mathcal{C}_{q}\right)}{\mathcal{L}\left(\lbrace \mathcal{X}_{n} \rbrace,\lbrace\mathbf{V}_{q}^{(\ell)}\rbrace, \lbrace \mathcal{U}_{n, q} \rbrace, \mathcal{C}_{q}\right)} \leqslant 10^{-5}.
      \label{eq:ConvCrit}
\end{equation}
The pseudocode of the algorithm is provided in Section~\ref{app:pseudocode} of the Supplementary Material. The following proposition implies that the IRLS algorithm to compute the ROMPCA estimates converges.
\begin{proposition} \label{the_1P}
Each iteration step of the algorithm decreases 
the objective function \eqref{eq:obj}, 
that is, 
 $\mathcal{L}\left(\lbrace \mathcal{X}_{n} \rbrace,\lbrace\mathbf{V}_{q+1}^{(\ell)}\rbrace, \lbrace \mathcal{U}_{n, q+1} \rbrace, \mathcal{C}_{q+1} \right) \leqslant 
 \mathcal{L}\left(\lbrace \mathcal{X}_{n} \rbrace,\lbrace\mathbf{V}_{q}^{(\ell)}\rbrace, \lbrace \mathcal{U}_{n, q} \rbrace, \mathcal{C}_{q}\right)$.
\end{proposition}

The proof is given in Section~\ref{app:proofCOnv} 
of the Supplementary Material, and uses the 
fact that for $z \geqslant 0$ the functions 
$z \rightarrow \rho_1(\sqrt{z})$ 
and $z \rightarrow \rho_2(\sqrt{z})$ are
differentiable and concave.
Since the objective function is decreasing and it has a lower bound of zero, the algorithm must converge.

\subsection{Initialization}
\label{sec:init}

The optimization problem in \eqref{eq:obj} defining our estimator is nonconvex, and, thus, the IRLS algorithm explained in the previous section may converge to a local minimum. To avoid that the algorithm converges to an undesirable
non-robust local minimum, it is important that the initialization of the algorithm is chosen appropriately.

To obtain an initial fit for our IRLS algorithm, we construct two initialization candidates as follows. First we vectorize each tensor $\mathcal{X}_n$ and stack the vectors rowwise into a matrix $\mathbf{X} = [\text{vec}(\mathcal{X}_1) \cdots \text{vec}(\mathcal{X}_N) ]^T$. We apply the Detecting Deviating Cells (DDC) algorithm \citep{Rousseeuw:DDC} to $\mathbf{X}$ using the FastDDC algorithm developed in \cite{Raymaekers:FastCorr}. DDC flags cellwise outliers (represented by an index set $I_{c}$) and provides imputed values for them. It also yields an index set $I_r$ of potential casewise outliers. Note that these index sets are computed from the unfolded data and thus do not fully exploit the multiway structure of the data. Hence, they serve well for the construction of candidates to initialize the IRLS algorithm, but detection of cellwise and casewise outliers in the tensors will be based on the final ROMPCA solution (see Section~\ref{sec:outdetect} for details). We then select the $H=\lceil 0.75N\rceil$ tensors not in $I_{r}$ with the fewest flagged cells in $I_{c}$ and replace their outlying cells by the DDC imputed values. 
Note that the value 0.75 is chosen because it offers a good trade-off between efficiency and robustness, and it is often used as a reference value in the literature \citep{Engelen:RPARAFAC, Hubert:MacroPARAFAC}. If $I_{r}$ contains more than $25\%$ of the samples, then we select all the remaining tensors.
Standard MPCA is applied to these $H$ imputed tensors $\{\mathcal{X}^{\text{\tiny DDC}}_h\}$, yielding initial estimates of  
$\lbrace \mathbf{V}^{(\ell)} \rbrace$ and $\mathcal{C}$. 
For all $n=1,\ldots,N$, initial cellwise weight tensors $\lbrace\Wcell\rbrace$ are obtained as well by setting the entries corresponding to elements of $I_{c}$ equal to zero and all other entries equal to 1. Based on these initial estimates of $\lbrace \mathbf{V}^{(\ell)} \rbrace$, $\mathcal{C}$ and  $\lbrace\Wcell\rbrace$, initial estimates of $\lbrace\mathcal{U}_{n}\rbrace$ for all $N$ tensors are then obtained from~\eqref{eq:solU}. This yields the estimates of the first initialization candidate.

To enhance robustness against cellwise outliers, a second initialization candidate is obtained by solving the objective function~\eqref{eq:obj} with $\rho_1(x) = \left|x\right|$ and $\rho_2(x)=x^2$. To obtain this solution, the IRLS algorithm from Section~\ref{sec:algo} is applied, starting from the first initialization candidate explained above. 
The need for using two distinct initial estimators is further supported by the simulation results reported in Section~\ref{app:addsim} of the Supplementary Material.

For both initialization candidates, the cellwise residuals are calculated using~\eqref{eq:cellres}.   
Each of the corresponding cellwise scales $\scell$ is then estimated by calculating a scale M-estimate of these cellwise residuals. 
A scale M-estimator of a univariate sample $(z_1 , \ldots, z_n)$ is the solution $\hat{\sigma}$ of the equation
\begin{equation}
\frac{1}{n} \sum_{i=1}^{n} \rho \left( \frac{z_i}{a\sigma} \right) = \delta,
\label{eq:Mscale}
\end{equation}
for a given $\delta$. In this paper, the $\rho$-function in~\eqref{eq:Mscale} is the hyperbolic tangent as before. The constant $\delta = 1.88$ for maximal robustness and $a=0.3431$ for consistency at the standard Gaussian distribution. 
Based on the standardized cellwise residuals, the casewise deviations are then calculated using~\eqref{eq:caseres}, and their M-scale $\hat{\sigma}_2$.
The initial estimate $\left(\lbrace\mathbf{V}^{(\ell)}_{0}\rbrace, \lbrace\mathcal{U}_{n,0}\rbrace, \mathcal{C}_{0}\right)$ to start the IRLS algorithm of Section~\ref{sec:algo} is then the initialization candidate 
that yields the smallest scale estimate  $\hat{\sigma}_2$. 

\subsection{Re-estimating the Center}
\label{sec:restim}
In MPCA, the center $\mathcal{C}$ is not uniquely determined. This section further elaborates on this identifiability issue and discusses a possible solution to obtain a meaningful estimate of the center, facilitating interpretation.

Let $\lbrace\mathbf{V}^{(\ell)}\rbrace$, $\lbrace\mathcal{U}_n\rbrace$ and $\mathcal{C}$ denote the estimates obtained by the IRLS algorithm of Section~\ref{sec:algo}, yielding the fitted tensors $\widehat{\mathcal{X}}_n = \mathcal{C} + \mathcal{U}_{n}\times \lbrace \mathbf{V}\rbrace$. 
For any $\mathcal{U}_{0} \in \mathbb{R}^{K_1 \times \ldots \times K_L}$, they can equivalently be computed as
 $   \widehat{\mathcal{X}}_n = {\mathcal{C}}^* + \mathcal{U}_{n}^* \times \lbrace \mathbf{V}\rbrace$,
where ${\mathcal{C}}^* = \mathcal{C} + \mathcal{U}_{0} \times \lbrace \mathbf{V}\rbrace$ and $\mathcal{U}_{n}^*=\mathcal{U}_{n} - \mathcal{U}_{0}$. 
This is known as the transitional ambiguity with respect to the center $\mathcal{C}$. If the primary focus of the analysis is on the fits $\widehat{\mathcal{X}}_n$ or the core tensors $\lbrace\mathcal{U}_{n}\rbrace$, then this is not a concern. 
However, it does mean that the estimate $\mathcal{C}$ does not necessarily represent the robust center of the tensors $\{\mathcal{X}_n\}$. 

To obtain a meaningful estimate of the center, and thus facilitate interpretation, we can adjust the center $\mathcal{C}$ based on the weighted average of the original tensors. In particular, we apply the transformation above for $\mathcal{U}_{0}= \left( \widebar{\mathcal{X}}^{\mathcal{W}} - \mathcal{C}\right)\times \lbrace \mathbf{V}^T\rbrace$ with
\begin{equation*}
\widebar{\mathcal{X}}^{\mathcal{W}} = \left(\Sumn \mathcal{X}_n \odot \mathcal{W}_n \right) \odot \mathcal{H} \, ,
\end{equation*}
where $\mathcal{W}_n$ represents the  final weight tensor \eqref{eq:Wn} corresponding to the estimates $ \lbrace\mathbf{V}^{(\ell)}\rbrace$, $\lbrace\mathcal{U}_n\rbrace$, and $\mathcal{C}$, and $\mathcal{H}=[1/\sum_{n=1}^N w_{n,\idx}].$
This yields the new center 
\begin{equation*}
        {\mathcal{C}}^{*} =  \mathcal{C} +
        \left( \widebar{\mathcal{X}}^{\mathcal{W}} - \mathcal{C}\right)\times \lbrace \mathbf{V}^T\rbrace\times \lbrace \mathbf{V}\rbrace,
    \end{equation*}
and the core tensors are modified accordingly, i.e. 
\begin{equation*}
    \mathcal{U}_{n}^{*} = \mathcal{U}_{n} - \left( \widebar{\mathcal{X}}^{\mathcal{W}} - \mathcal{C}\right)\times \lbrace \mathbf{V}^T\rbrace.
\end{equation*}
The estimates $\lbrace\mathbf{V}^{(\ell)}\rbrace$, $\lbrace\mathcal{U}_n^{*}\rbrace$ and $\mathcal{C}^{*}$ are then reported as the final ROMPCA estimates. 

The time and space complexity of the ROMPCA algorithm are derived in Section~\ref{app:compl} of the Supplementary Material and are given by
$O\!\left(NP \max \Big\lbrace K^2, L^2 K_L, \sum_{\ell=1}^L K_\ell^2 P_\ell^2 \Big\rbrace \right)$ 
and
$O\!\left(NP + PK + \max_{1 \leqslant \ell \leqslant L }\!\left\{K_\ell P_\ell P,\ K_\ell^2 P_\ell^2\right\}\right)
$
where $P = \prod_{\ell = 1}^L P_\ell$ and $K = \prod_{\ell = 1}^L K_\ell$.

\subsection{Determining the Subspace Dimensions}
\label{sec:rank}
Until now, the ranks $K_\ell$, $\ell = 1, \ldots, L$ of the projection matrices $\lbrace \mathbf{V}^{(\ell)} \rbrace$
were assumed fixed. However, in real applications, it is seldom the case that these ranks are known. For standard MPCA \cite{Lu:MPCA} proposed the $Q$-based method to determine these ranks, which is a multilinear extension of the dimension selection strategy based on the cumulative percentage of total variation commonly used in PCA  \citep{Jolliffe:PCA}. To determine the rank $K_\ell$ of $\mathbf{V}^{(\ell)}$, the $Q$-based method considers the ratio
\begin{equation*}
Q^{(\ell)} = \frac{\sum_{p_\ell=1}^{K_\ell} \lambda_{p_\ell}^{(\ell)}}{\sum_{p_\ell=1}^{P_\ell} \lambda_{p_\ell}^{(\ell)}},
\end{equation*}
 where $\lambda_{p_\ell}^{(\ell)}$ are the eigenvalues (in decreasing order) corresponding to the mode-$\ell$ unfolded scatter matrix  $\mathbf{\Phi}^{(\ell)*} = \Sumn \left( \mathbf{X}_{n(\ell)} - \widebar{\mathbf{X}}_{(\ell)} \right)\left( \mathbf{X}_{n(\ell)} - \widebar{\mathbf{X}}_{(\ell)} \right)^T$. The $Q^{(\ell)}$ value thus measures the remaining portion of the mode-$\ell$ total scatter when removing the last $P_\ell-K_\ell$ components.

The rank for each mode may be chosen graphically by plotting the cumulative eigenvalues and determining the elbow point in this plot. Alternatively, it can be chosen as the smallest dimension $K_\ell$ for which $Q^{(\ell)}$ achieves a fixed threshold (e.g.\ 0.8), ensuring that the desired minimum explained variance is achieved.
 We cannot rely on the eigenvalues $\lambda_{p_\ell}^{(\ell)}$ of the unfolded scatter matrices estimated from $\{\mathcal{X}_n\}$ as they can be strongly affected by outliers and cannot handle missing data.
Therefore, we compute the eigenvalues from $\{\mathcal{X}^{\text{\tiny DDC}}_h\}$ defined in Section~\ref{sec:init}. Section~\ref{app:addrank} of the Supplementary Material illustrates the effectiveness of the proposed procedure on a simulated dataset.

\subsection{Imputed Tensors}
\label{sec:Impute}
If a tensor $\mathcal{X}_n$ has entries with cellwise weights strictly smaller than 1 and/or missing entries, we provide an imputed version $\mathcal{X}_n^{\text{\tiny imp}}$ of the tensor. 
Clearly, we do not want to modify entries of $\mathcal{X}_n$ with weight $w_{\eidx}^{\text{cell}} = 1$, hence
$x_{\eidx}^{\text{\tiny imp}} = x_{\eidx}$ for these entries.
The other entries of $\mathcal{X}_n$ correspond to cellwise outliers or missing values and we use the ROMPCA solution to replace these entries by more regular values. 

Equation \eqref{eq:condU2} in the Supplementary Material can equivalently be written as 
\begin{equation*}
\big(\widetilde{\mathcal{W}}_n  \odot (\mathcal{X}_n - \widehat{\mathcal{X}}_n ) \big) \times \lbrace \mathbf{V}^T \rbrace 
= \mathcal{O}_{K_1 \times \dots \times K_L}\,. 
\end{equation*}
Therefore, the imputed tensor is constructed as 
\begin{equation*} 
\mathcal{X}_n^{\text{\tiny imp}} := \widehat{\mathcal{X}}_n + \widetilde{\mathcal{W}}_n \odot (\mathcal{X}_n - \widehat{\mathcal{X}}_n ). 
\end{equation*} 
It then immediately follows from the previous equation that the multilinear projection of $\mathcal{X}_n^{\text{\tiny imp}}$ equals $\widehat{\mathcal{X}}_n$. Hence, $\mathcal{X}_n$ and $\mathcal{X}_n^{\text{\tiny imp}}$ result in the same reconstructed tensor $\widehat{\mathcal{X}}_n$, but the imputed tensor has no missing or outlying entries and thus can be safely used in further analysis such as tensor regression or classification. 

\section{Simulation Study}
\label{sec:simu}
In this section, we evaluate the empirical performance of ROMPCA  in the presence of cellwise outliers and/or casewise outliers, both with and without missing values. The data generation process for the simulation study is inspired by \cite{Lu:MPCA}. 
Samples of 3-order tensors $\mathcal{X}_n \in \mathbb{R}^{P_1 \times P_2 \times P_3}$, $n=1,\dots,N$, are generated according to the  model  $\mathcal{X}_n = \mathcal{U}_n \times_1 \mathbf{V}^{(1)} \times_2 \mathbf{V}^{(2)} \times_3 \mathbf{V}^{(3)} + \mathcal{E}_n$, 
where $\mathcal{U}_n \in \mathbb{R}^{K_1 \times K_2 \times K_3}$ are the core tensors, $\mathbf{V}^{(\ell)}\in \mathbb{R}^{P_\ell \times K_\ell}, \ell=1,2,3,$ are the mode-$\ell$ projection matrices, and $\mathcal{E}_n \in \mathbb{R}^{P_1 \times P_2 \times P_3}$ are noise tensors. 
The elements $u_{\eidx}$ of $\mathcal{U}_n$ are drawn from the standard Gaussian distribution. These elements are then multiplied by $[(P_1, P_2, P_3)/(p_1 \cdot p_2 \cdot p_3)]^{0.9}$ to control the size of the eigenvalues.
 The columns of the matrices $\mathbf{V}^{(\ell)}$ are the eigenvectors corresponding to the  $K_\ell$ largest eigenvalues of the $P_\ell \times P_\ell$ matrix $\mathbf{\Sigma}^{(\ell)}$ with elements $\sigma_{i, j}^{(\ell)} = (-0.9)^{|i-j|}$. The entries of the noise tensor $\mathcal{E}_n$ are sampled independently from a Gaussian distribution with zero mean and variance $0.1$. We consider two settings: (i)  $(P_1, P_2, P_3) = (30, 20, 5)$ and $(K_1, K_2, K_3) = (8, 6, 2)$, and (ii) $(P_1, P_2, P_3) = (15, 10, 5)$ and $(K_1, K_2, K_3) = (4, 3, 2)$. Because results for both scenarios are analogous, here we report only the results for setting (i) with higher dimensions. Section~\ref{app:addsim} of the Supplementary Material 
provides the results for setting (ii).

Three contamination scenarios are considered. 
In the cellwise outliers scenario, $\varepsilon^{\text{cell}}=20\%$ of all $NP_1P_2P_3$ cells are randomly selected and each of these entries $x_{n,\idx}$ is replaced by $\gamma_{\text{cell}} s_{\idx}$, where $s_{\idx}$ is the standard deviation of $\{x_{n,\idx}\}_{n=1}^N$ and $\gamma_{\text{cell}}$ ranges from $0$ to $7$.
In the casewise outliers scenario, $\varepsilon^{\text{case}}=20\%$ of the $N$ tensors are replaced by tensors generated from the model $\mathcal{X}_n^{\star} = \mathcal{U}_n^{\star} \times_1 \mathbf{V}^{*(1)} \times_2 \mathbf{V}^{*(2)} \times_3 \mathbf{V}^{*(3)} + \mathcal{E}_n$, where $\mathcal{U}_n^* \in \mathbb{R}^{\left(K_1+1\right) \times \left(K_2+1\right) \times \left(K_3+1\right)}$ is a tensor whose entries are 1 if all indices are odd and all other entries are set to zero.
The columns of the matrices $\mathbf{V}^{(\ell)}$ for each mode are the first $K_\ell+1$ odd-indexed eigenvectors of the matrix $\mathbf{\Sigma}^{(\ell)}$ when ordered according to decreasing eigenvalues. 
This ensures that the resulting tensors have large variability along directions that differ from the bulk of the data. The resulting tensor $\mathcal{X}_n^{\star}$ is then multiplied by $\gamma_{\text{case}}=3\gamma_{\text{cell}}$ where $\gamma_{\text{cell}}$ ranges from $0$ to $7$ as before.
Hence, the parameters $\gamma_{\text{cell}}$ and $\gamma_{\text{case}}$ control the magnitude of contamination. Note that the data are clean when $\gamma_{\text{cell}} = \gamma_{\text{case}} = 0$.
In the combined scenario, the first $10\%$ of tensors are replaced by casewise outliers as in the previous scenario. The remaining tensors are then contaminated by $10\%$ of cellwise outliers as in the first scenario.  
Here, we set $\gamma_{\text{case}}=6\gamma_{\text{cell}}$, where  $\gamma_{\text{cell}}$ again varies from 0 to 7. When we consider settings with missing values, additionally, $\varepsilon^{\text{obs}}=10\%$ of the entries are set completely at random to a missing value.

The performance is evaluated by the mean squared error (MSE) computed on the regular cells, given by
\begin{equation*}
    \text{MSE} = \Sumn \Sumps \delta_{\eidx} \left(x_{\eidx} - \hat{x}_{\eidx}\right)^2 \Big/ \Sumn \Sumps \delta_{\eidx}\, ,
\end{equation*}
where $\delta_{\eidx}$ is $ 0$ if $\mathcal{X}_n$ is generated as a casewise outlier or if entry $x_{\eidx}$ is a generated cellwise outlier or missing entry, and $\delta_{\eidx}$ is $1$ otherwise. 
We report for each scenario the 
mean MSE over 50 samples of $N=100$ observations for each of the methods as a function of the outlier size $\gamma_{\text{cell}}$.

First, we consider settings without missing values. 
ROMPCA is compared to standard MPCA and to two alternative methods that are related to the methods proposed by \cite{Inoue:RMPCA} that are designed to address either casewise or cellwise outliers, but not both at the same time. Both methods can be seen as special cases of ROMPCA.  The first one, OnlyCase-MPCA,  uses $\rho_1(z)=z^2$ and the second method, OnlyCell-MPCA, has $\rho_2(z)=z^2$. In our simulations it turned out that they yielded more robust results than the original proposals of \cite{Inoue:RMPCA}. 

\begin{figure}[!ht]
    \centering
    \includegraphics[width=.32\textwidth]{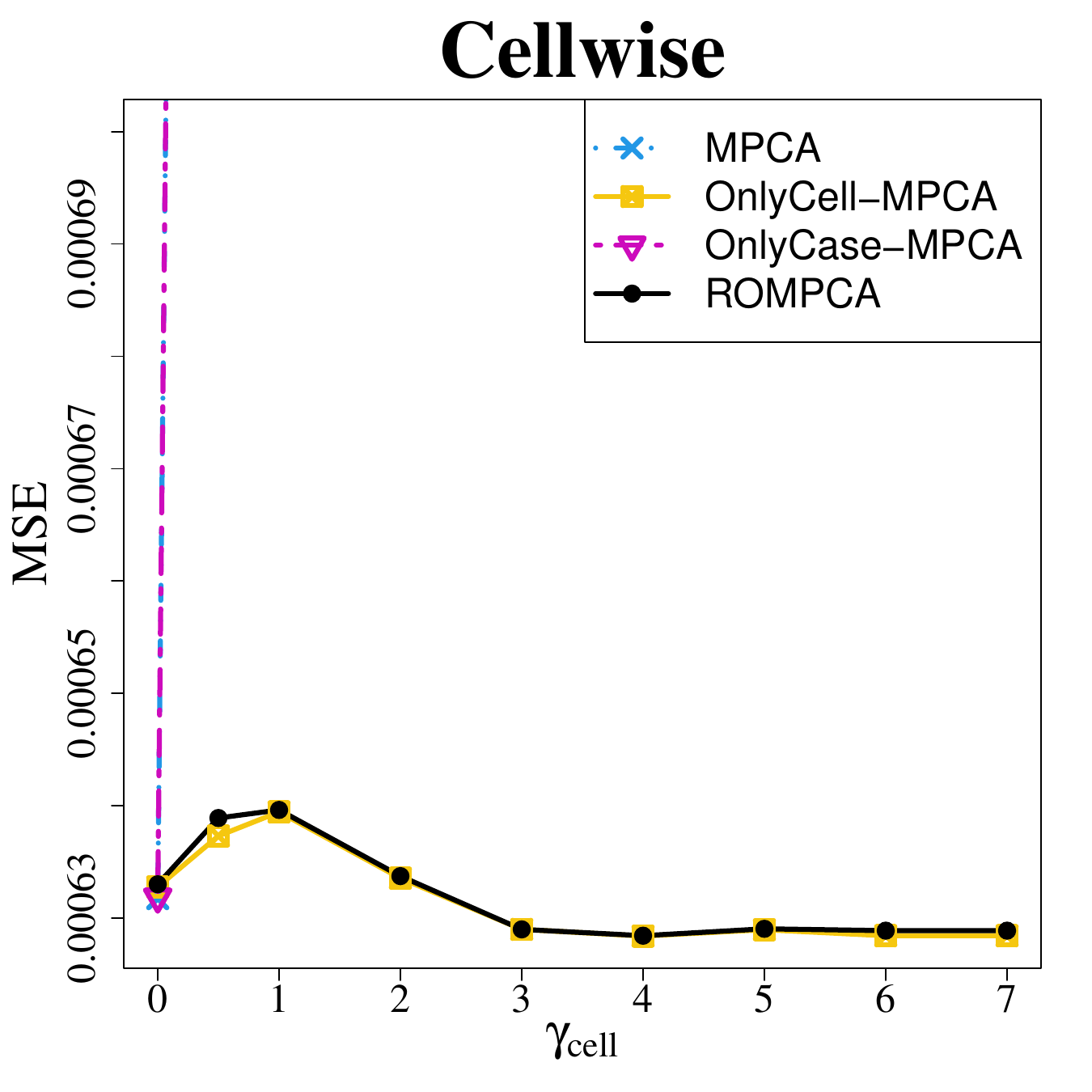} %
    \includegraphics[width=.32\textwidth]{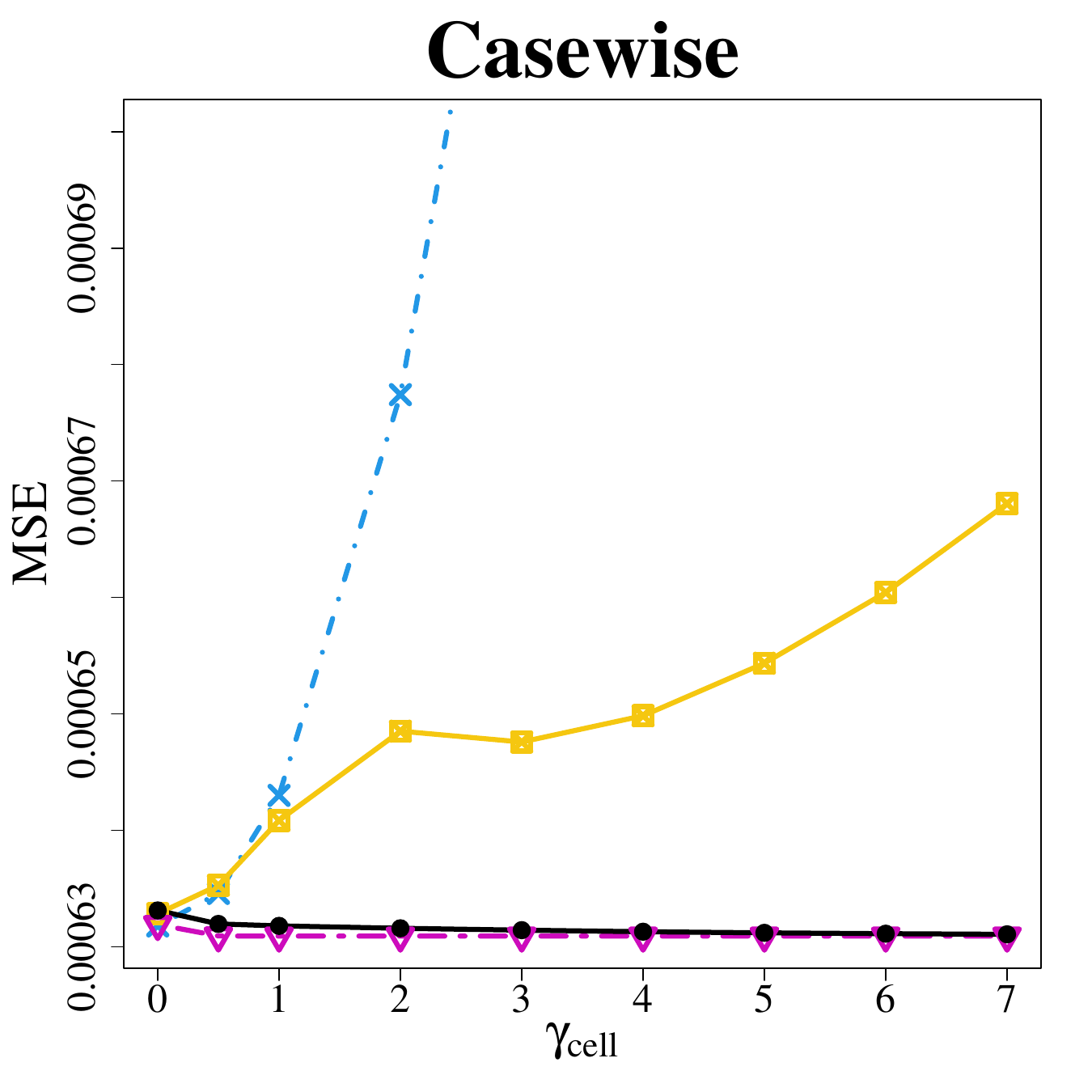} %
    \includegraphics[width=.32\textwidth]{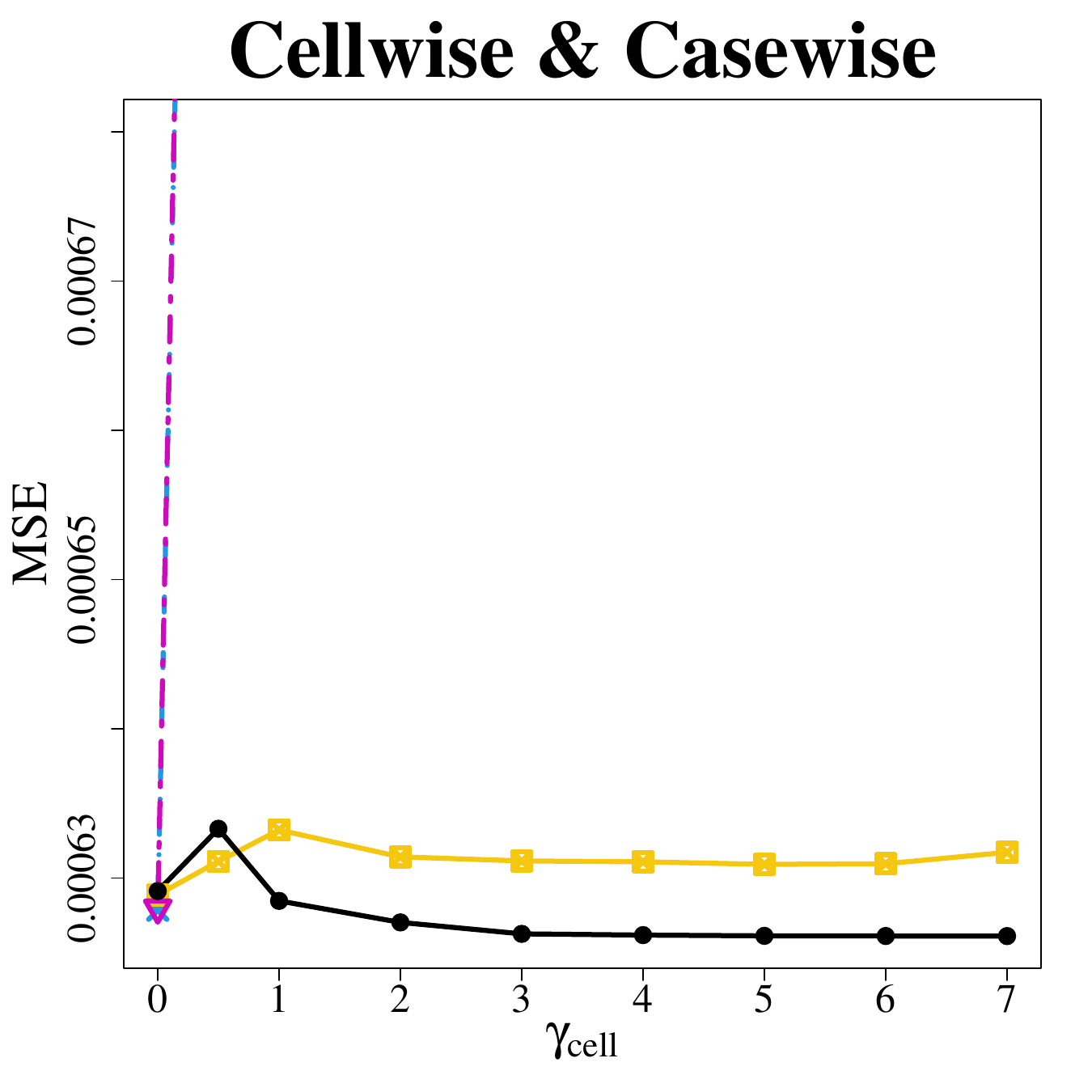} %
 \caption{Mean MSE attained by MPCA, OnlyCase-MPCA, OnlyCell-MPCA, and ROMPCA  under cellwise contamination, casewise contamination or both in function of $\gamma_{\text{cell}}$ for data without missing values.}%
\label{fig:simulMSE_A09}%
\end{figure}

Figure \ref{fig:simulMSE_A09} shows the mean MSE of the four methods for all three contamination scenarios.
Without contamination ($\gamma_{\text{cell}}=0$ in the plots of Figure~\ref{fig:simulMSE_A09}),  standard MPCA slightly outperforms all other methods, as expected. 
From the left plot of Figure~\ref{fig:simulMSE_A09}, it can be seen that MPCA and OnlyCase-MPCA are highly affected by the presence of cellwise contamination, while ROMPCA and OnlyCell-MPCA show a much more robust behavior. Cellwise outliers have a slight effect on these methods when they are close by, but this effect quickly disappears when the cellwise outliers are more distinct from the regular measurements. 

The middle plot of Figure~\ref{fig:simulMSE_A09}) shows that standard 
MPCA also cannot resist casewise outliers. OnlyCell-MPCA also struggles to resist the effect of this type of contamination for which it is not designed. In contrast, both ROMPCA and
OnlyCase-MPCA are hardly affected by the presence of casewise contamination with OnlyCase-MPCA performing slightly better than ROMPCA in this case. 

When both cellwise outliers and casewise outliers are present, ROMPCA generally outperforms all other methods as soon as the outliers are distinguishable from the regular data. As expected, the other methods do not perform well in this setting, either because they are not robust to cellwise outliers (OnlyCase-MPCA), casewise outliers (OnlyCell-MPCA), or both (MPCA).

\begin{figure}[!ht]
    \centering
    \includegraphics[width=.25\paperwidth]{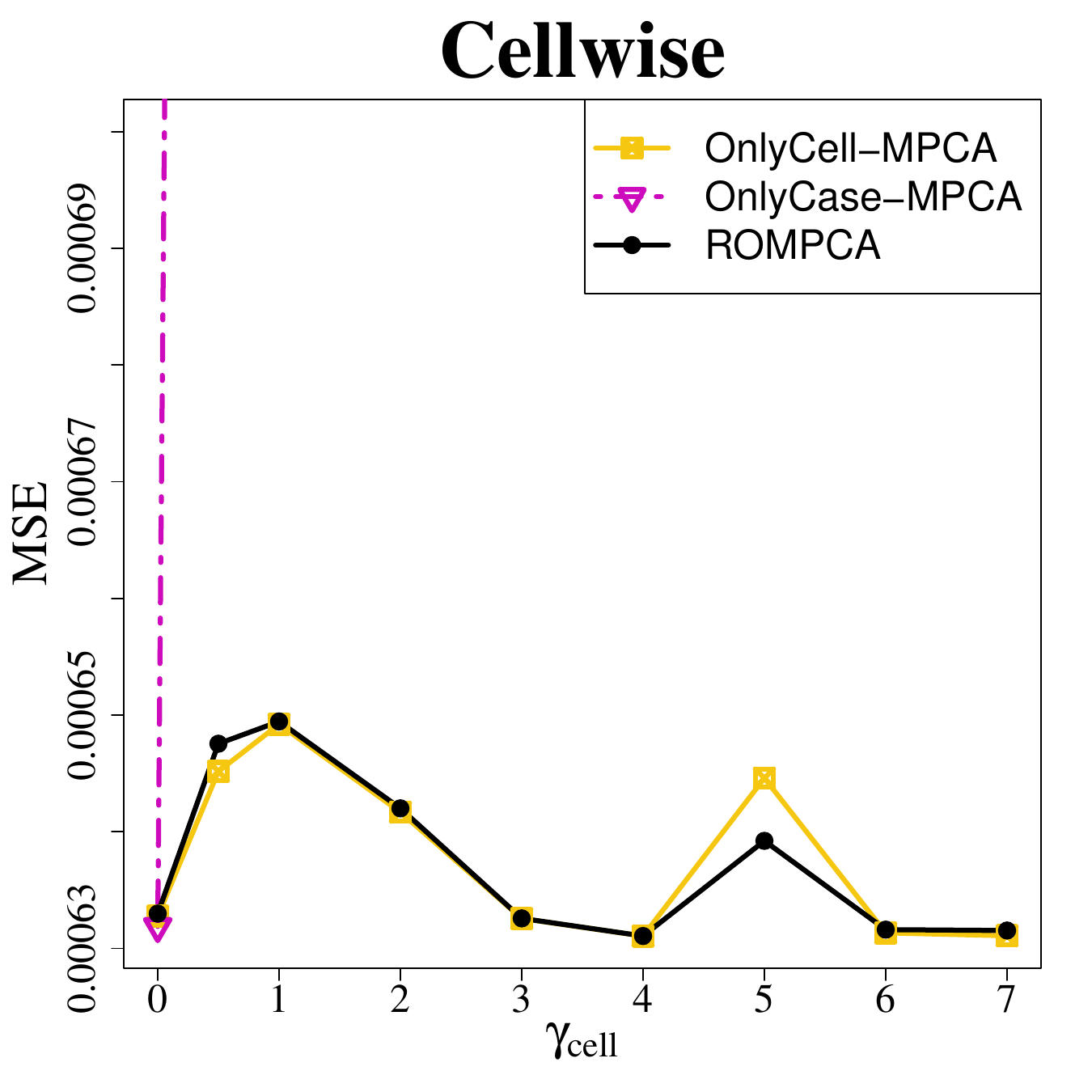} %
    \includegraphics[clip,width=.25\paperwidth]{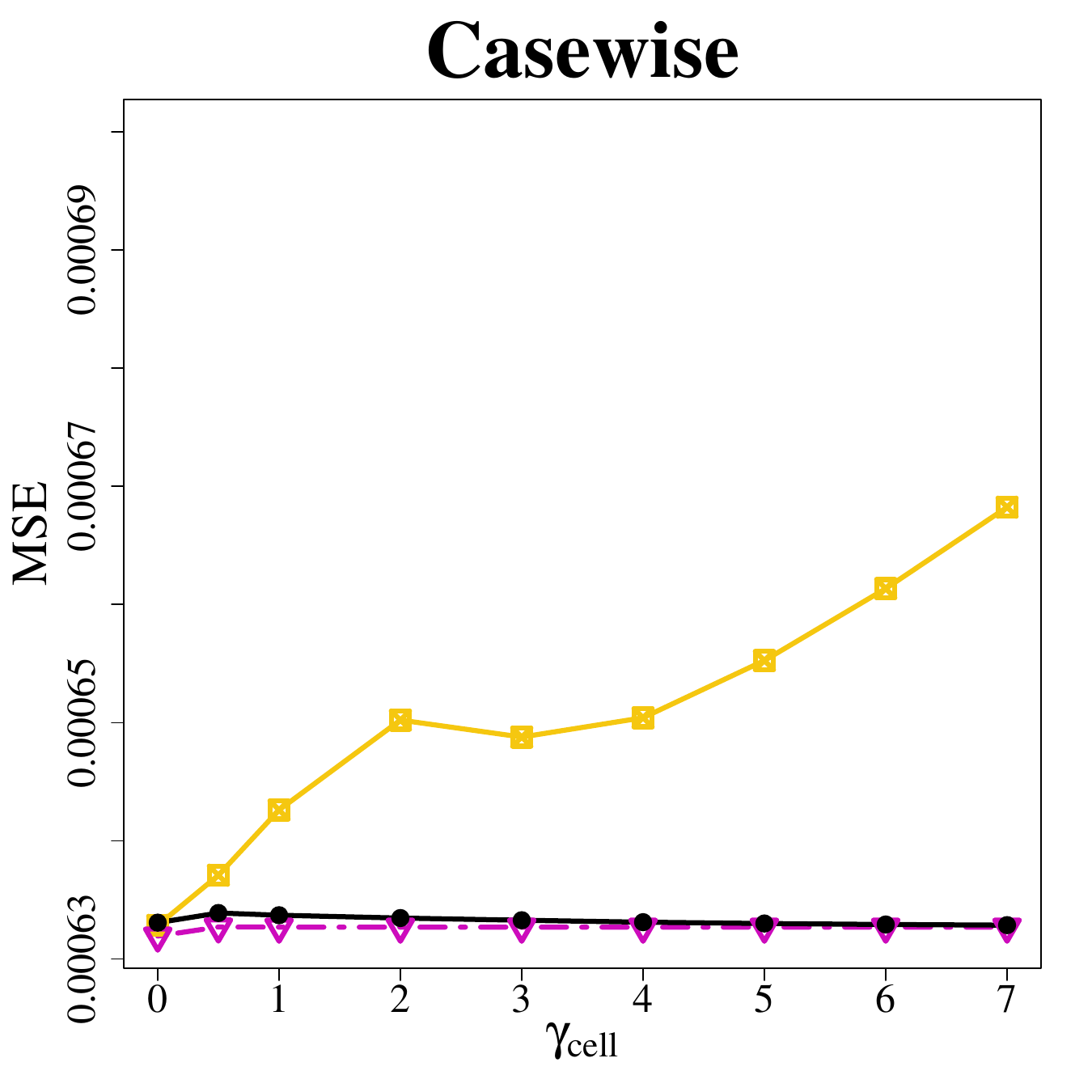} %
    \includegraphics[clip,width=.25\paperwidth]{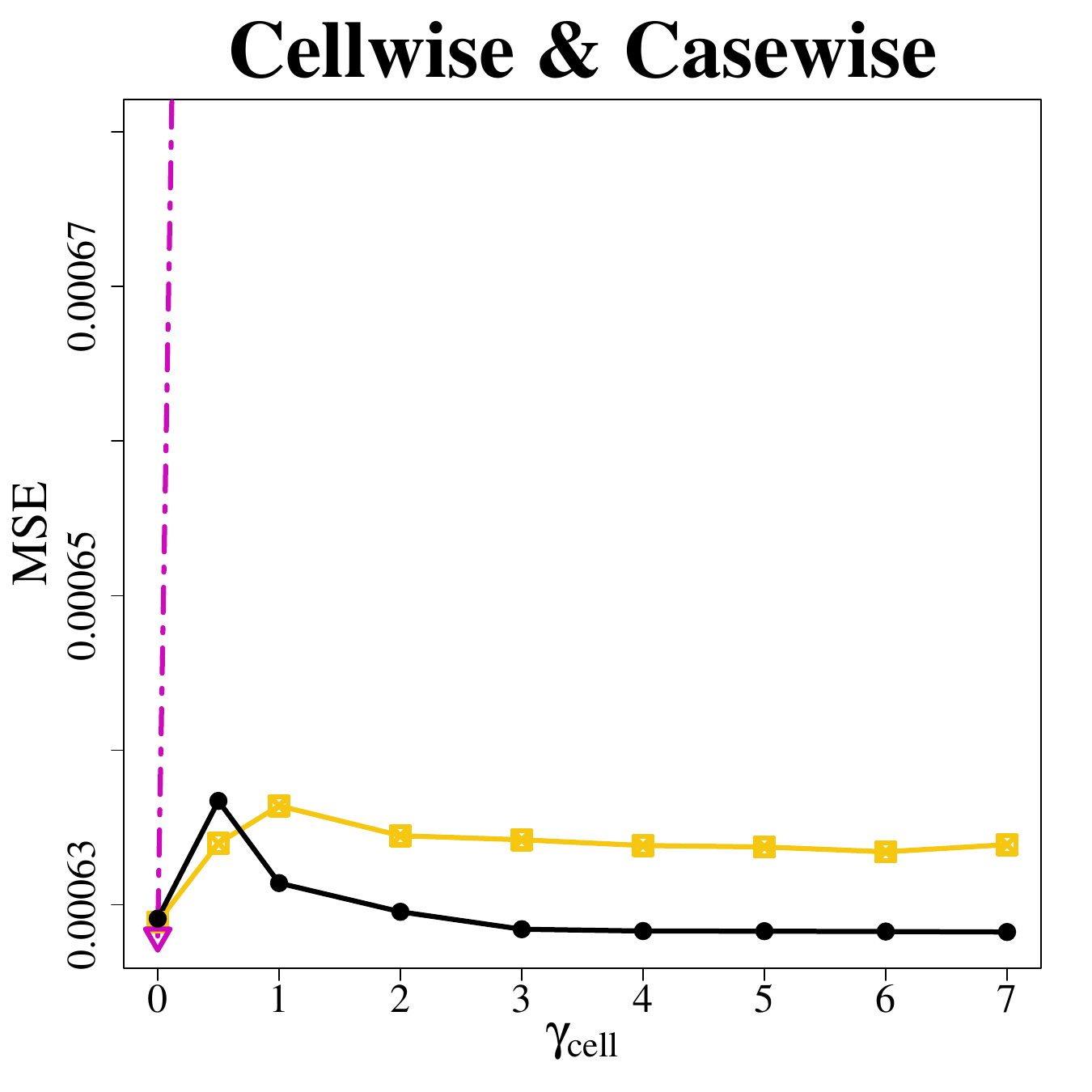} %
 \caption{Mean MSE attained by OnlyCase-MPCA, OnlyCell-MPCA, and ROMPCA  under cellwise contamination, casewise contamination or both in function of $\gamma_{\text{cell}}$ for data with $10\%$ of missing values.}%
\label{fig:simulMSE_A09_NA}%
\end{figure}

Figure~\ref{fig:simulMSE_A09_NA} shows the MSE results for the same contamination scenarios when also missing values are added. Standard MPCA is not included in this comparison as this method cannot cope with missing values. 
Comparing these results with Figure~\ref{fig:simulMSE_A09}, it can be seen that the presence of missing values generally has little effect on the performance of the methods. 
Overall, ROMPCA shows the best behavior across all scenarios.

Section~\ref{app:addsim} of the Supplementary Material presents additional simulation results concerning the choice of tuning constants, the $\rho$-function, and the use of two distinct initial estimators.

In the cellwise outliers scenario, each variable retains only $(1 - \varepsilon^{\text{cell}}) \times 100\% $ clean values, where $\varepsilon^{\text{cell}}=0.2$.
As a result, the probability that a case contains at least one outlying cell equals $1-(1-\varepsilon^{\text{cell}})^{\prod^L_1 P_\ell}$, which in this simulation setting is almost one. Hence, this corresponds to a very severe contamination level, since none of the cases is entirely uncontaminated. 
The simulation results show that ROMPCA maintains good performance even under such conditions, indicating that the algorithm is robust and numerically stable in this challenging scenario.

The average running times over 50 repetitions of ROMPCA in these settings are 21.23 and 139.56 seconds for the settings $(P_1, P_2, P_3) = (15, 10, 5)$ and $(K_1, K_2, K_3) = (4, 3, 2)$, and $(P_1, P_2, P_3) = (30, 20, 5)$ and $(K_1, K_2, K_3) = (8, 6, 2)$, respectively, with $N = 100$.

\section{Outlier Detection}
\label{sec:outdetect}
Based on the ROMPCA estimates we can construct numerical and graphical diagnostics to gain more insight into the outlying cells and anomalous cases in the data.
First the scale estimates $\hat{\sigma}_{1,\idx}$  are recomputed as M-scales $\tilde{\sigma}_{\idx}$ of the ROMPCA cellwise residuals~\eqref{eq:cellres}. This yields for each observation its final standardized residual tensor $\mathcal{\widetilde{R}}_n = [\tilde{r}_{\eidx}] = [ r_{\eidx}/\tilde{\sigma}_{\idx}]$.

The residual cellmap is a heatmap which visualizes the cellwise outlyingness of entries in the data \citep{Rousseeuw:DDC, Hubert:MacroPARAFAC}. 
Standardized cellwise residuals whose absolute value is smaller than the threshold $c_{\text{cell}} = \sqrt{\chi^2_{1,0.998}} = 3.09$ are considered to be regular and are colored yellow. The other cells are flagged as cellwise outliers. To indicate their direction and degree of outlyingness, cells with positive standardized cellwise residual exceeding $c_{\text{cell}}$ are colored from light orange to red, while cells with negative standardized cellwise residual below $-c_{\text{cell}}$ are colored from purple to dark blue. Cells whose value is missing are colored white. 

We can display a residual cellmap of all individual data cells by vectorizing each observed tensor and stacking these vectors rowwise into a matrix (as in Section~\ref{sec:init}). 
As this may result in a very wide matrix (if $\prod_{\ell=1}^L P_\ell$ is large), it is possible to aggregate adjacent cells and display the average color. 
As an illustration, let us consider a dataset with $N=100$ simulated tensors according to the setting of Section~\ref{sec:simu} with $(P_1, P_2, P_3) = (15, 10, 5)$. We leave 10 out of the 100 tensors completely uncontaminated, change 10 tensors into casewise outliers and then contaminate the remaining cells with 10\% of cellwise outliers and 1\% of missing values. We use $\gamma_{\text{cell}}=5$ and $\gamma_{\text{case}}=6\gamma_{\text{cell}}$.

\begin{figure}[!ht]
\begin{center}
\includegraphics[width=1\textwidth]{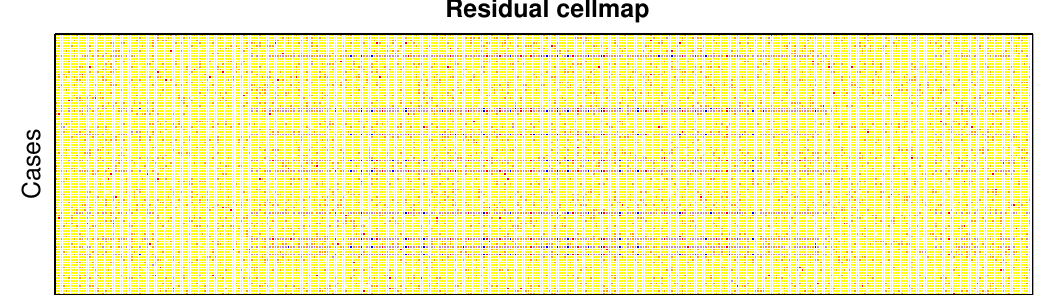}
\end{center}
\vspace{-0.5cm}
\caption{Residual cellmap of a simulated dataset with aggregated columns.}
\label{fig:Simulation_UnfResMap}
\end{figure}

Figure~\ref{fig:Simulation_UnfResMap} shows the residual cellmap for this dataset, where each cell has the average color of two columns of the vectorized tensors. The $10$ outlying cases are clearly visible by the horizontal lines with mostly outlying cells. 

\begin{figure}[!ht]
\begin{center}
\includegraphics[width=1\textwidth]{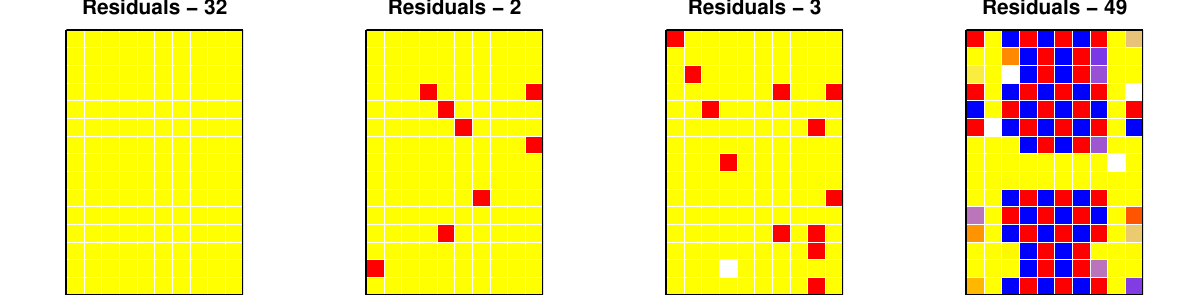}
\end{center}
\vspace{-0.5cm}
\caption{Residual cellmaps of slices from the 3th mode of four observations from the simulated dataset.}
\label{fig:Simulation_ResMaps}
\end{figure}

While this residual cellmap makes it possible to easily identify cases with many cellwise outliers, the high dimensionality of this matrix and/or the merging of columns makes it difficult to identify the positions of cellwise outliers in the data tensors. A more detailed insight is provided by making residual cellmaps of slices of the standardized residual tensors. This is illustrated in Figure~\ref{fig:Simulation_ResMaps}, which shows the residual cellmaps of slices from the 3rd mode of the standardized residual tensors corresponding to four different observations. The first observation is an uncontaminated tensor, the second one has cellwise outliers, the third one additionally has a missing value while the fourth tensor is a casewise outlier. 

\begin{figure}[!ht]
\begin{center}
    \includegraphics[width=0.75\textwidth]{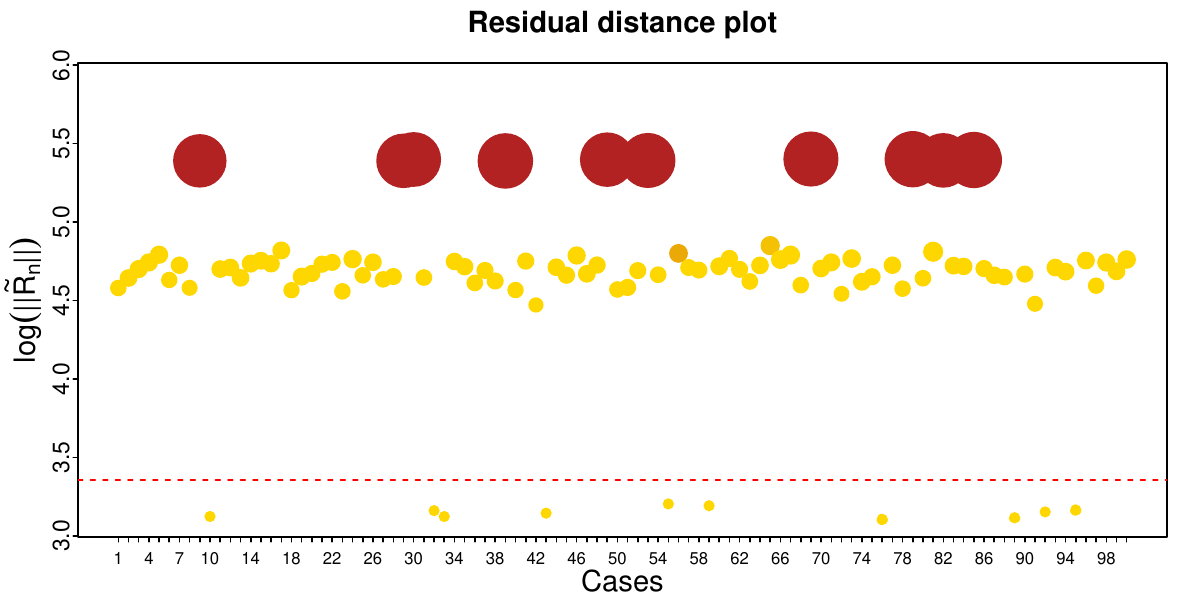}
\end{center}
\vspace{-0.5cm}
\caption{Plot of the ROMPCA residual distances for a simulated dataset.}
\label{fig:Simulation_ResDistPlot}
\end{figure}

To focus more on the casewise outlyingness
of the observed tensors, a new residual plot is proposed based on the ROMPCA fit. It displays (on the log scale) the observations' residual distance, defined as $\|\widetilde{\mathcal{R}}_n\|_F$, versus their index. 
The size of each point in this plot is proportional to the Percentage of Outlying Cells (POC) in the observation, which for each observation is defined as 
\begin{equation*}
    \text{POC}_n = \frac{1}{\prod_\ell P_\ell} \Sumps I \left(|\tilde{r}_{\eidx}| > c_{\text{cell}} \right),
\end{equation*}
where $I$ is the indicator function \citep{Hubert:MacroPARAFAC}. Larger points thus correspond to observations with many outlying cells. Finally, points are colored based on their casewise weight. To be consistent with the residual cellmap, the color is yellow when $\wcase = 0$, red when $\wcase = 1$, and orange for intermediate weights. The red horizontal dotted line is the cutoff $c_{\text{case}}$, i.e., the 0.99-quantile of the distribution of $\|\mathcal{\widetilde{R}}_n\|_F$ for uncontaminated errors.
In particular, a large set of tensors of the same size as $\widetilde{\mathcal{R}}_n$ with standard normal entries is generated. The 0.99 quantile of their Frobenius norms is then calculated to obtain $c_{\text{case}}$. The 0.99 quantile is selected as it is considered small enough to identify tensors with unusually large residual distances, yet high enough to avoid incorrectly flagging regular tensors. 
This choice is further motivated by related work that has adopted the same threshold to identify extreme values of the residual distance \citep{Centofanti:cellpca}.
Cases with a residual distance smaller than $c_{\text{case}}$ are considered to be regular observations. Figure~\ref{fig:Simulation_ResDistPlot} shows the residual distance plot for the simulated dataset of Figure~\ref{fig:Simulation_UnfResMap}. We can clearly distinguish cases with contamination (above the cutoff line) from the regular cases. The difference between tensors with cellwise outliers on the one hand and casewise outliers on the other hand is also clearly seen from the color and size of the points.

\section{Real Data Examples}
\label{sec:realdata}
\subsection{Dog Walker Data} \label{Ex:Dog walker}
\begin{figure}[!ht]
\begin{center}
       \includegraphics[width=1\textwidth]{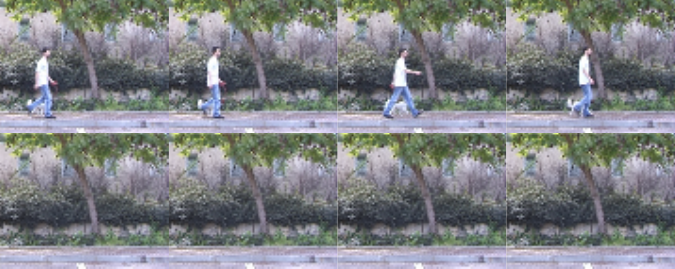}
\end{center}
\vspace{-0.5cm}
\caption{Frames from the Dog Walker data (upper row) and reconstructed frames using ROMPCA (lower row).}
\label{fig:DogWalker_Reconst_DataFrames}
\end{figure}

The Dog Walker data (\url{https://www.wisdom.weizmann.ac.il/~vision/SpaceTimeActions.html}) consists of surveillance video data of a man walking his dog.  The video was filmed using a static camera and consists of $54$ colored frames. Each frame has $(144 \times 180)$ pixels and is stored using the RGB (Red, Green, Blue) color model. For the sake of computation time, every frame is down-sampled by image decimation such that it has $(72 \times 90)$ pixels.  The upper row of Figure~\ref{fig:DogWalker_Reconst_DataFrames} shows four frames from the video. 

This dataset can be represented as a collection of $3$-order tensors, where the first two modes correspond to the pixel space, while the third mode corresponds to the color space. We approximate the data tensors according to the following decomposition
\begin{equation*}
    \widehat{\mathcal{X}}_n = \mathcal{C} + \mathcal{U}_n \times_1 \mathbf{V}^{\text{\tiny v}} \times_2 \mathbf{V}^{\text{\tiny h}} \times_3 \mathbf{V}^{\text{\tiny col}}.
\end{equation*}
Here, $\mathbf{V}^{\text{\tiny col}}$ denotes the projection matrix for the color space, whereas $\mathbf{V}^{\text{\tiny v}}$ and $\mathbf{V}^{\text{\tiny h}}$ denote the projection matrices for the vertical axis and horizontal axis of the pixel space, respectively. 
The ranks are chosen equal to $(1,1,1)$ based on a plot of the cumulative eigenvalues as explained in Section~\ref{sec:rank}. See  Section~\ref{app:addDogRes} of the Supplementary Material for more details.

In Figure~\ref{fig:DogWalker_Reconst_DataFrames}, the observed frames are compared to the fitted frames using ROMPCA. It can immediately be seen that the man walking the dog does not appear anymore on the fitted frames. In this setting, the man walking his dog is successfully identified as outlying as it is the only part that significantly changes in the successive frames. 

\begin{figure}[!ht]
\begin{center}
      \includegraphics[width=.6\textwidth]{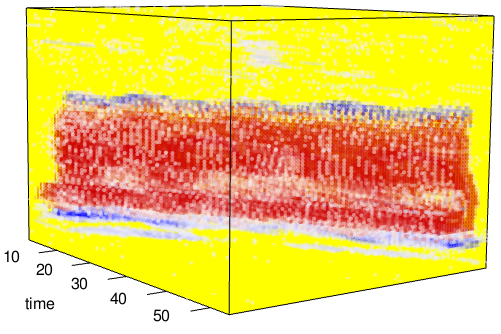}
\end{center}
\vspace{-0.5cm}
\caption{ROMPCA 3D residual cellmap of the Dog Walker dataset.}
\label{fig:DogWalker_Cellmap}
\end{figure}

The 2D residual cellmap (as in Figure~\ref{fig:Simulation_UnfResMap}) is shown in Figure~\ref{fig:DogWalker_cellmap_comp} of the Supplementary Material~\ref{app:addDogRes}. Here, we present in Figure~\ref{fig:DogWalker_Cellmap} a 3D residual cellmap of the aggregated slices corresponding to the blue channel. 
 A clear diagonal pattern of cellwise outliers can be observed, corresponding to the man walking his dog, whose position changes across the frames.

\begin{figure}[!ht]
\begin{center}
    \includegraphics[width=1\textwidth]{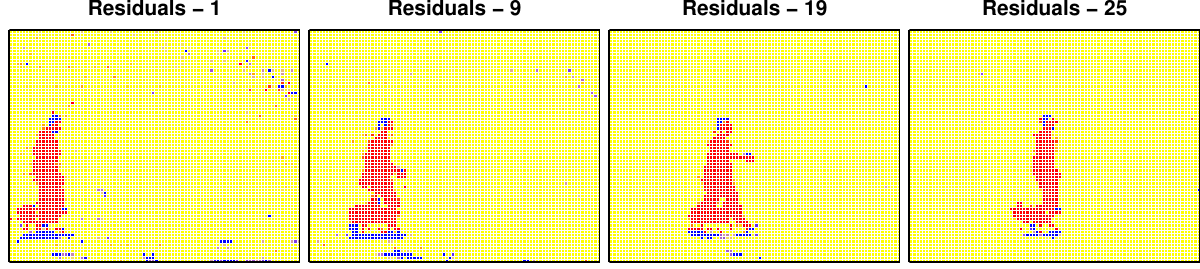}
\end{center}
\vspace{-0.5cm}
\caption{ROMPCA residual cellmaps of several slices (corresponding to the blue color) of the Dog Walker data.}
\label{fig:DogWalker_Cellmap_30_blue}
\end{figure}

For a more detailed view, we look at the residual cellmap of some slices. Figure~\ref{fig:DogWalker_Cellmap_30_blue} shows the residual cellmap for the blue color slice of four frames. In these residual cellmaps, we can clearly see that most of the cellwise outliers correspond to the man walking his dog. The first frames also show numerous deviating pixels in the background. They mostly result from the wind moving the tree leaves.

\begin{figure}[!ht]
\begin{center}
    \includegraphics[width=0.5\textwidth]{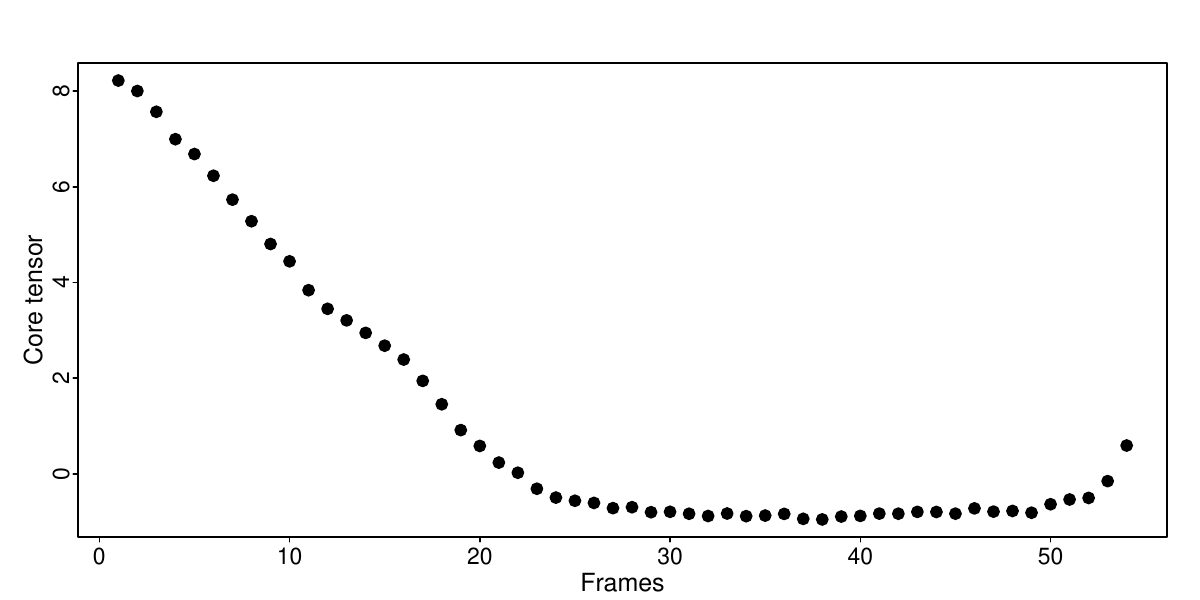}
\end{center}
\vspace{-0.5cm}
\caption{ROMPCA core tensors of the Dog Walker data.}
\label{fig:DogWalker_Scores_ROMPCA}
\end{figure}

Finally we take a closer look at the core tensors, that here, consist of  scalars. They are displayed in Figure~\ref{fig:DogWalker_Scores_ROMPCA} and clearly show a non-constant pattern, mostly in the first half. Thus, the background is not constant, as one might initially suspect from watching the video.  ROMPCA has detected that there is smooth change in the illumination.   

Section~\ref{app:addDogRes} of the Supplementary Material contains more results for this example. There, we show that MPCA is not able to clearly distinguish the man from the background. Furthermore, we also show that DDC applied to the vectorized tensors  does not capture the change in the illumination. A multilinear approach is thus better in modeling this variation.  

\subsection{Dorrit Data}
As a second example, we consider the Dorrit dataset  \citep{Baunsgaard:dorrit}. Data are laboratory-generated and consist of $27$ different mixtures of fluorophores (hydroquinone, tryptophan, phenylalanine and dopa). For each mixture, the Excitation Emission Matrix (EEM) is measured with emission wavelengths from $250$ nm to $482$ nm at $2$ nm intervals, and excitation wavelengths from $230$ nm to $315$ nm at $5$ nm intervals. This yields a collection of $27$ matrices of size $116 \times 18$. 
Samples 2, 3, and 5 are known to be highly anomalous observations. They have very high concentrations for some fluorophores, resulting in large intensities and missing entries.
Data are preprocessed to remove the Rayleigh and Raman scattering that affects all samples \citep[see e.g.][]{Engelen:Scatter}.

The plot of the cumulative eigenvalues (see Section~\ref{app:addDorRes} of the Supplementary Material) suggests to choose the ranks $(K_1, K_2) = (4,4)$. This is in line with the ranks used by \cite{Heng:RobustTensor} in their analysis. 

\begin{figure}[!ht]
\centering
    \includegraphics[width=0.9\textwidth]{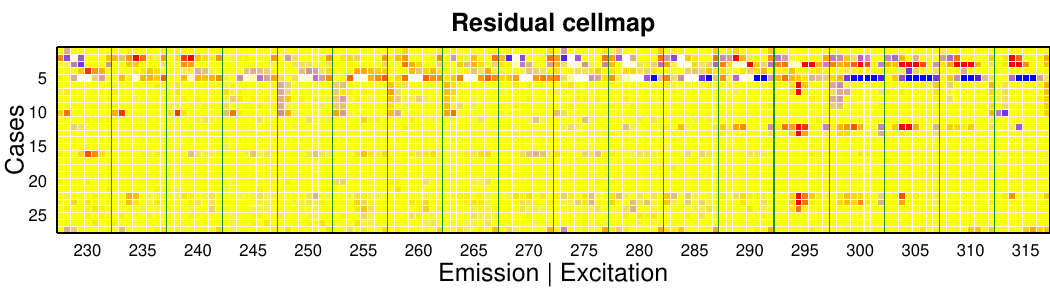}
\caption{ROMPCA residual cellmap of the Dorrit data.}
\label{fig:Dorrit_EnhancedCellmap}
\end{figure}

Figure~\ref{fig:Dorrit_EnhancedCellmap} shows the ROMPCA residual cellmap where cells in adjacent columns have been aggregated such that the first 8 columns in the residual map correspond to the emission wavelengths at an excitation wavelength of 230 nm, the next 8 columns correspond to an excitation wavelength of 235 nm, and so on. This results in 18 blocks, which are separated by thin green lines. We see that ROMPCA has identified cellwise outliers in many observations, but mostly in the cases 2, 3 and 5. 

\begin{figure}
\begin{center}
    \includegraphics[width=0.75\textwidth]{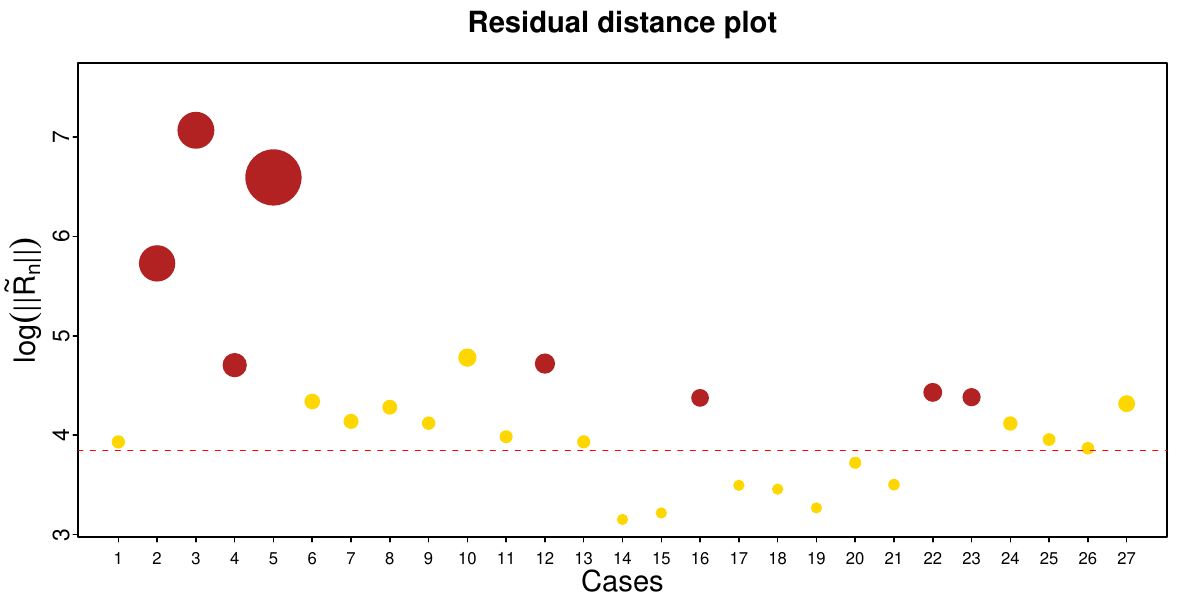}
\end{center}
\vspace{-0.5cm}
\caption{Plot of the ROMPCA residual distances of the Dorrit data.}
\label{fig:Dorrit_RMCPA_ResDist}
\end{figure}

These three observations also most pronounced stand out in the residual distance plot in Figure~\ref{fig:Dorrit_RMCPA_ResDist}. Five more mixtures are colored red and thus flagged as casewise outliers. They contain a substantial fraction of cells that are mildly outlying. 
Only seven mixtures are identified as having a low residual distance. The remaining 12 mixtures exceed the  

\begin{figure}[!ht]
 \begin{center}
 \begin{tabular}{cc}
 \includegraphics[width=0.35\textwidth]{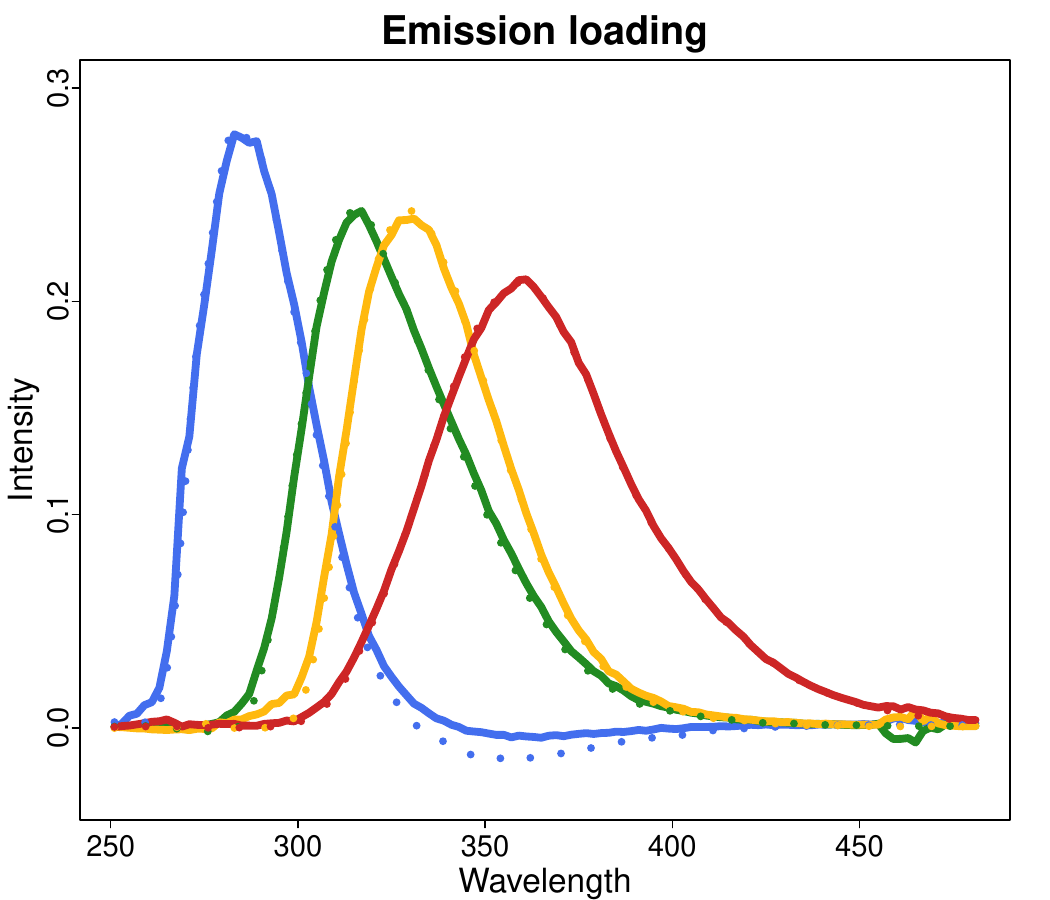} &
 \includegraphics[width=0.35\textwidth]{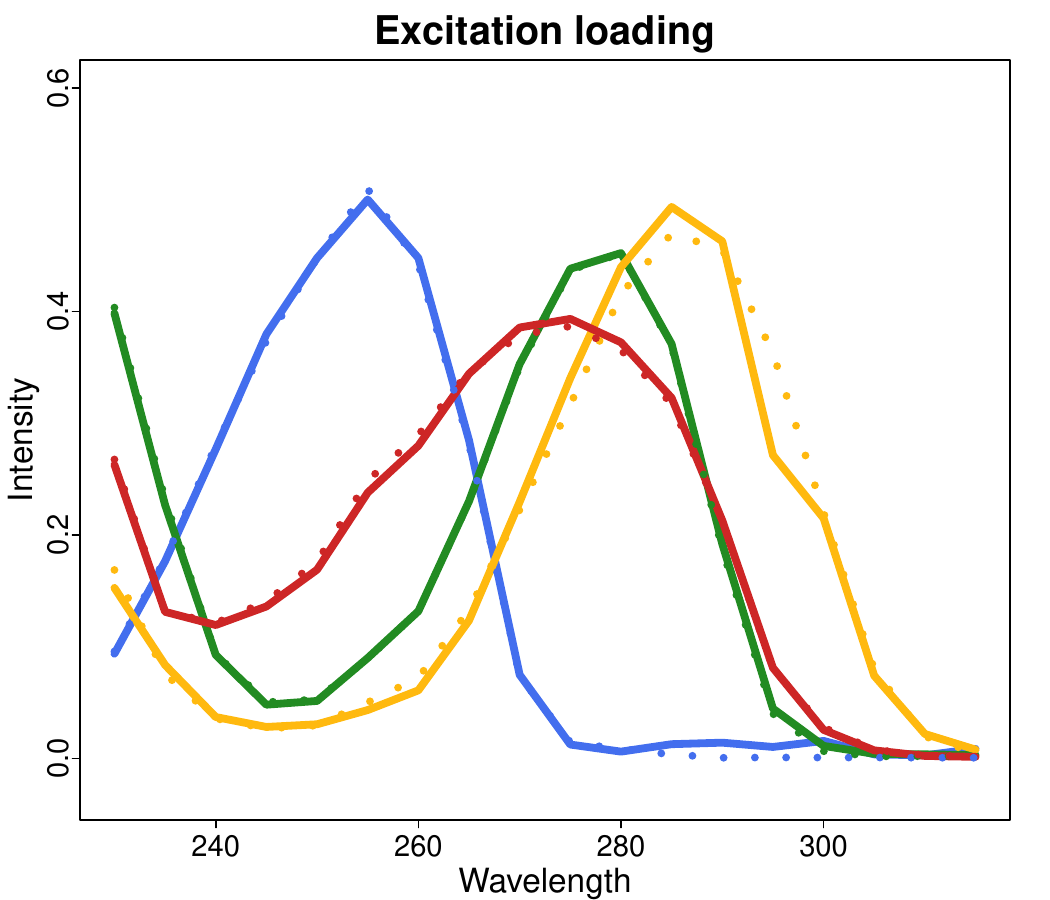}
 \end{tabular}
 \end{center}
 \vspace{-0.5cm}
 \caption{Emission (left) and excitation (right) loadings for the Dorrit dataset. The solid lines represent the estimated loadings based on ROMPCA, while the dotted lines indicate the pure loadings.}
 \label{fig:Dorrit_ROMPCA_loadings}
 \end{figure}

Following \cite{Heng:RobustTensor} we next apply a PARAFAC decomposition to the ROMPCA fitted matrices $\widehat{\mathcal{X}}_n$ that correspond to the $\widetilde{N}=19$ cases with a casewise weight of 1. 
We construct the 3-order tensor $\widetilde{\mathcal{X}}$ where the first mode represents the observations, so $\widetilde{\mathcal{X}}_{\tilde{n},:,:}=\widehat{\mathcal{X}}_{\tilde{n}}$ for $\tilde{n}=1,\ldots,\widetilde{N}$.
The PARAFAC model decomposes this tensor into the sum of $F$ rank-1 tensors: 
$$\widetilde{\mathcal{X}} \approx \sum_{j=1}^F \mathbf{a}_f \otimes \mathbf{b}_f \otimes \mathbf{c}_f$$
where $\otimes$ denotes the outer product between vectors. Here we take $F=4$ since it is known that the mixture is composed of four fluorophores. The emission $\mathbf{b}_f$ and excitation profiles $\mathbf{c}_f$ of pure samples were obtained earlier and serve as a reference \citep{Baunsgaard:dorrit}. 
Figure~\ref{fig:Dorrit_ROMPCA_loadings} compares the estimated emission and excitation loadings (solid lines) with the benchmark loadings (dotted lines). It can be seen that all loading vectors are well aligned with the pure loadings. ROMPCA has thus been been able to extract well the main structure from the data, and to provide an outlier-free reconstructed EEM dataset to which PARAFAC could be applied. 

\begin{figure}[!ht]
\begin{center}
\begin{tabular}{cc}
\includegraphics[width=0.35\textwidth]{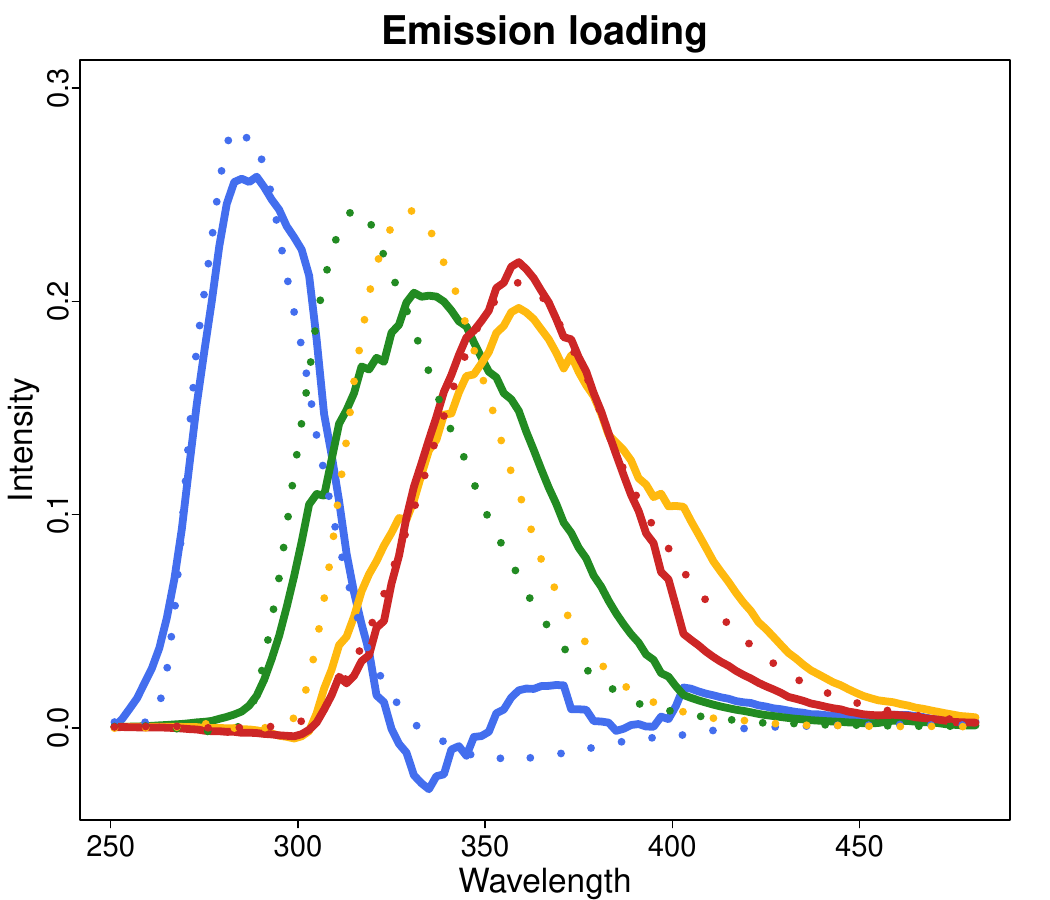} &
\includegraphics[width=0.35\textwidth]{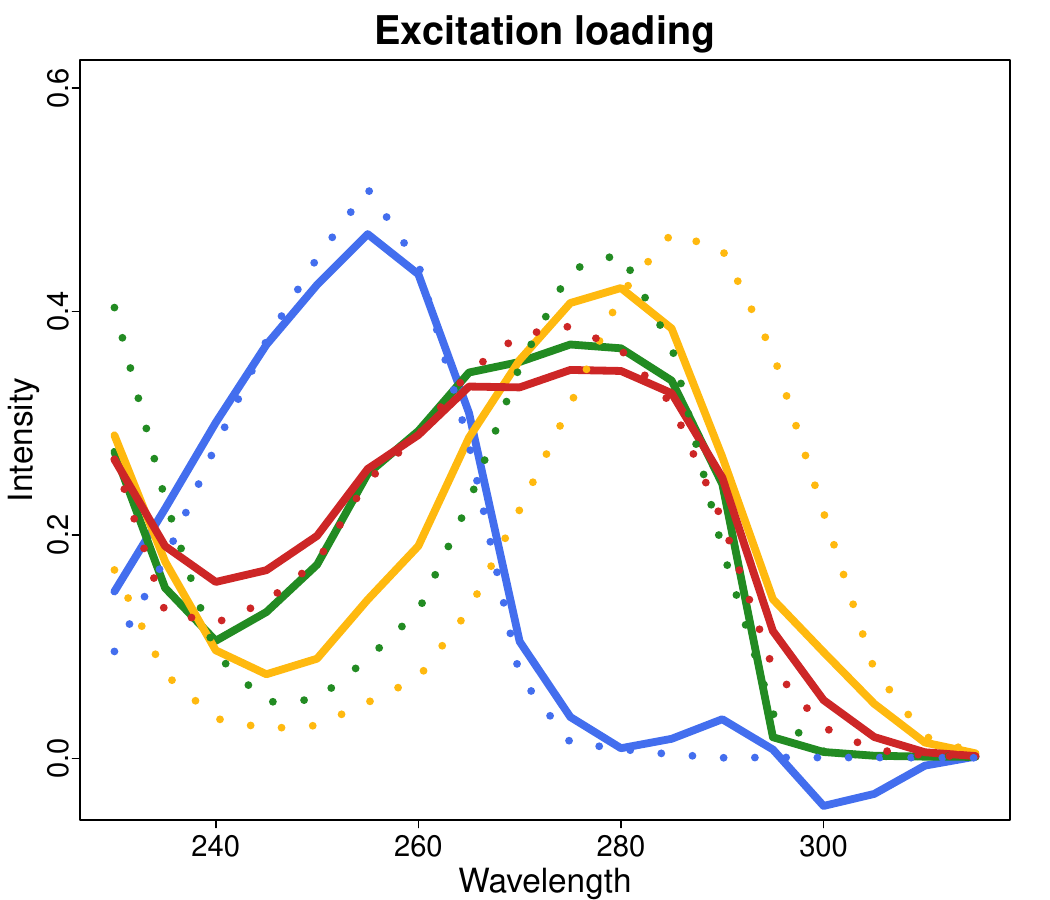} %
\end{tabular}
\end{center}
\vspace{-0.5cm}
\caption{Emission (left) and excitation (right) loadings for the Dorrit dataset. The solid lines represent the estimated loadings based on MPCA, while the dotted lines indicate the pure loadings.}
\label{fig:Dorrit_CMPCA_loadings}
\end{figure}

In contrast, Figure~\ref{fig:Dorrit_CMPCA_loadings} shows the estimated emission and excitation loadings (solid lines) obtained from the PARAFAC decomposition of the standard MPCA fitted tensors. 
They are clearly highly distorted by the outlying measurements. 

\section{Conclusion}
\label{sec:disc}
We proposed a novel robust MPCA method for tensor data, the first to simultaneously handle casewise and cellwise outliers, along with missing values. Its objective function combines two robust loss functions to reduce the effect of cellwise and casewise outliers, respectively. 

The ROMPCA fit can be computed via an iteratively reweighted least squares algorithm which uses alternating weighted least squares as its building block. 
The algorithm yields cellwise and casewise weights, which, together with the standardized cellwise residuals, can be used to construct 
graphical outlier detection tools. We developed residual cellmaps and a residual distance plot to visualize both types of outliers. The good performance of ROMPCA is demonstrated via simulations and applications to real datasets. 

Another advantage of ROMPCA is that it also returns imputed tensors in which the missing and outlying entries have been replaced by imputed values. If the original tensors are replaced by these imputed tensors in further analysis such as tensor regression or classification, then standard methods, which cannot handle missing data and are not robust, can be applied directly. The availability of imputed tensors further enables the extension of advanced graphical displays, originally developed for cellPCA in the multivariate setting, to tensor-valued data \citep{Hirari:cellGraphics}.

As stated in Section~\ref{sec:restim}, the center must be corrected to address translational ambiguity. We handle this by proposing a specific recentering procedure. In particular, to obtain a meaningful estimate of the center and facilitate interpretation, we adjust $\mathcal{C}$ based on the weighted average of the observed tensors. However, other methods to estimate the center can be considered, such as jointly estimating the center and the low-rank component while imposing structural assumptions on the center, such as sparsity or boundedness. Exploring such approaches in our setting could lead to a more interpretable estimate of the center and represents an interesting direction for future work.

In some applications, tensor data exhibit a temporal structure. Although in MPCA this is handled by adapting the tensor representation so that the samples remain independent, developing an extension of ROMPCA that explicitly accounts for temporal dependence is nontrivial as discussed in Section~\ref{app:disc} of the Supplementary Material. Although our proposed method already shows good computational performance in both simulations and real-data case studies, computational scalability remains an important consideration for tensor-based methods. In Section~\ref{app:disc} of the Supplementary Material, we outline possible strategies to address scalability in high-dimensional tensors. However, there is still room for improvement,  and developing new strategies to effectively handle temporally dependent tensors while enhancing computational scalability represents a promising direction for future research.\\

\noindent{\bf Software availability.} The \textsf{R} code and the datasets are publicly available from the webpage\linebreak \url{
https://wis.kuleuven.be/statdatascience/robust/software}.


\vspace{3mm}
\noindent{\bf Acknowledgement.} 
Part of Fabio Centofanti’s contribution to this work was developed while he was employed at the University of Naples Federico II. \\

%

\begin{thebibliography}{}

\bibitem[Alqallaf et~al., 2009]{Alqallaf:cell}
Alqallaf, F., Van~Aelst, S., Yohai, V.~J., and Zamar, R.~H. (2009).
\newblock Propagation of outliers in multivariate data.
\newblock {\em The Annals of Statistics}, 37:311--331.

\bibitem[Ballard and Kolda, 2025]{BallardKolda:bookTensor}
Ballard, G. and Kolda, T.~G. (2025).
\newblock {\em Tensor Decompositions for Data Science}.
\newblock Cambridge University Press.

\bibitem[Baunsgaard, 1999]{Baunsgaard:dorrit}
Baunsgaard, D. (1999).
\newblock {\em Factors affecting 3-way modelling ({PARAFAC}) of fluorescence
  landscapes}.
\newblock PhD thesis, Royal Veterinary and Agricultural University, Department
  of Dairy and Food technology, Frederiksberg, Denmark.

\bibitem[Carroll and Chang, 1970]{Carroll:PARanalysis}
Carroll, J.~D. and Chang, J.-J. (1970).
\newblock Analysis of individual differences in multidimensional scaling via an
  n-way generalization of “{E}ckart-{Y}oung” decomposition.
\newblock {\em Psychometrika}, 35(3):283--319.

\bibitem[Centofanti et~al., 2025a]{Centofanti:RODESSA}
Centofanti, F., Hubert, M., Palumbo, B., and Rousseeuw, P.~J. (2025a).
\newblock Multivariate singular spectrum analysis by robust diagonalwise
  low-rank approximation.
\newblock {\em Journal of Computational and Graphical Statistics}, 34:360--373.

\bibitem[Centofanti et~al., 2024]{Centofanti:cellpca}
Centofanti, F., Hubert, M., and Rousseeuw, P.~J. (2024).
\newblock Robust principal components by casewise and cellwise weighting.
\newblock {\em arXiv preprint arXiv:2408.13596}.

\bibitem[Centofanti et~al., 2025b]{Centofanti:cellRCOV}
Centofanti, F., Hubert, M., and Rousseeuw, P.~J. (2025b).
\newblock Cellwise and casewise robust covariance in high dimensions.
\newblock {\em arXiv preprint arXiv:2505.19925}.

\bibitem[Croux and Ruiz-Gazen, 2005]{Croux:Proj}
Croux, C. and Ruiz-Gazen, A. (2005).
\newblock High breakdown estimators for principal components: the
  projection-pursuit approach revisited.
\newblock {\em Journal of Multivariate Analysis}, 95:206--226.

\bibitem[De~La~Torre and Black, 2003]{Torre:Robframework}
De~La~Torre, F. and Black, M.~J. (2003).
\newblock A framework for robust subspace learning.
\newblock {\em International Journal of Computer Vision}, 54:117--142.

\bibitem[De~Lathauwer et~al., 2000]{Lathauwer:MSVD}
De~Lathauwer, L., De~Moor, B., and Vandewalle, J. (2000).
\newblock A multilinear singular value decomposition.
\newblock {\em SIAM journal on Matrix Analysis and Applications},
  21(4):1253--1278.

\bibitem[Engelen et~al., 2007]{Engelen:Scatter}
Engelen, S., Frosch~M{\o}ller, S., and Hubert, M. (2007).
\newblock Automatically identifying scatter in fluorescence data using robust
  techniques.
\newblock {\em Chemometrics and Intelligent Laboratory Systems}, 86:35--51.

\bibitem[Engelen and Hubert, 2011]{Engelen:RPARAFAC}
Engelen, S. and Hubert, M. (2011).
\newblock Detecting outlying samples in a parallel factor analysis model.
\newblock {\em Analytica Chimica Acta}, 705(1-2):155--165.


\bibitem[Gabriel, 1978]{Gabriel:LSapprox}
Gabriel, K.~R. (1978).
\newblock Least squares approximation of matrices by additive and
  multiplicative models.
\newblock {\em Journal of the Royal Statistical Society Series B},
  40(2):186--196.

\bibitem[Goldfarb and Qin, 2014]{Goldfarb:L1}
Goldfarb, D. and Qin, Z. (2014).
\newblock Robust low-rank tensor recovery: Models and algorithms.
\newblock {\em SIAM Journal on Matrix Analysis and Applications},
  35(1):225--253.

\bibitem[Hampel et~al., 1986]{Hampel:bookIF}
Hampel, F.~R., Ronchetti, E.~M., Rousseeuw, P.~J., and Stahel, W.~A. (1986).
\newblock {\em {Robust Statistics: the Approach based on Influence Functions}}.
\newblock Wiley.

\bibitem[Hampel et~al., 1981]{Hampel:CVC}
Hampel, F.~R., Rousseeuw, P.~J., and Ronchetti, E. (1981).
\newblock The change-of-variance curve and optimal redescending {M}-estimators.
\newblock {\em Journal of the American Statistical Association},
  76(375):643--648.

\bibitem[Han and Bhanu, 2005]{Han:IndividualGait}
Han, J. and Bhanu, B. (2005).
\newblock Individual recognition using gait energy image.
\newblock {\em IEEE Transactions on Pattern Analysis and Machine Intelligence},
  28(2):316--322.

\bibitem[Harshman, 1970]{Harshman:PARAFAC}
Harshman, R.~A. (1970).
\newblock Foundations of the {PARAFAC} procedure: Models and conditions for an
  “explanatory” multi-modal factor analysis.
\newblock {\em UCLA working papers in phonetics}, 16(1):84.

\bibitem[Heng et~al., 2023]{Heng:RobustTensor}
Heng, Q., Chi, E.~C., and Liu, Y. (2023).
\newblock Robust low-rank tensor decomposition with the {L}2 criterion.
\newblock {\em Technometrics}, 65(4):537--552.

\bibitem[Hirari et~al., 2025]{Hirari:cellGraphics}
Hirari, M., Hubert, M., and Rousseeuw, P.~J. (2025).
\newblock Graphical tools for visualizing cellwise and casewise outliers.
\newblock {\em Journal of Data Science, Statistics, and Visualisation}, 5(10).

\bibitem[Hubert and Hirari, 2024]{Hubert:MacroPARAFAC}
Hubert, M. and Hirari, M. (2024).
\newblock {MacroPARAFAC} for handling rowwise and cellwise outliers in
  incomplete multiway data.
\newblock {\em Chemometrics and Intelligent Laboratory Systems}, 251:105170.

\bibitem[Hubert et~al., 2019]{Hubert:MacroPCA}
Hubert, M., Rousseeuw, P.~J., and {Van den Bossche}, W. (2019).
\newblock {MacroPCA}: An all-in-one {PCA} method allowing for missing values as
  well as cellwise and rowwise outliers.
\newblock {\em Technometrics}, 61:459--473.

\bibitem[Hubert et~al., 2005]{Hubert:ROBPCA}
Hubert, M., Rousseeuw, P.~J., and Vanden~Branden, K. (2005).
\newblock {ROBPCA}: a new approach to robust principal component analysis.
\newblock {\em Technometrics}, 47:64--79.

\bibitem[Hubert et~al., 2012]{Hubert:RPARAFAC-SI}
Hubert, M., Van~Kerckhoven, J., and Verdonck, T. (2012).
\newblock Robust {PARAFAC} for incomplete data.
\newblock {\em Journal of Chemometrics}, 26:290--298.

\bibitem[Inoue et~al., 2009]{Inoue:RMPCA}
Inoue, K., Hara, K., and Urahama, K. (2009).
\newblock Robust multilinear principal component analysis.
\newblock In {\em 2009 IEEE 12th International Conference on Computer Vision},
  pages 591--597.

\bibitem[Jolliffe, 2011]{Jolliffe:PCA}
Jolliffe, I. (2011).
\newblock {\em {Principal Component Analysis}}.
\newblock Springer.

\bibitem[Kroonenberg, 2008]{Kroonenberg:BookMulti}
Kroonenberg, P.~M. (2008).
\newblock {\em Applied Multiway Data Analysis}.
\newblock Wiley Series in Probability and Statistics. Wiley-Interscience,
  Hoboken, NJ.

\bibitem[Liu et~al., 2015]{Liu:GenHOOI}
Liu, Y., Shang, F., Fan, W., Cheng, J., and Cheng, H. (2015).
\newblock Generalized higher order orthogonal iteration for tensor learning and
  decomposition.
\newblock {\em IEEE Transactions on Neural Networks and Learning Systems},
  27(12):2551--2563.

\bibitem[Locantore et~al., 1999]{Locantore:RobPCA}
Locantore, N., Marron, J., Simpson, D., Tripoli, N., Zhang, J., Cohen, K.,
  Boente, G., Fraiman, R., Brumback, B., Croux, C., et~al. (1999).
\newblock Robust principal component analysis for functional data.
\newblock {\em Test}, 8:1--73.

\bibitem[Lu et~al., 2008]{Lu:MPCA}
Lu, H., Plataniotis, K.~N., and Venetsanopoulos, A.~N. (2008).
\newblock {MPCA}: Multilinear principal component analysis of tensor objects.
\newblock {\em IEEE Transactions on Neural Networks}, 19(1):18--39.

\bibitem[Maronna and Yohai, 2008]{Maronna:RobElement}
Maronna, R.~A. and Yohai, V.~J. (2008).
\newblock Robust low-rank approximation of data matrices with elementwise
  contamination.
\newblock {\em Technometrics}, 50(3):295--304.

\bibitem[Raymaekers and Rousseeuw, 2021]{Raymaekers:FastCorr}
Raymaekers, J. and Rousseeuw, P.~J. (2021).
\newblock Fast robust correlation for high-dimensional data.
\newblock {\em Technometrics}, 63(2):184--198.

\bibitem[Raymaekers and Rousseeuw, 2025]{Raymaekers:Challenges}
Raymaekers, J. and Rousseeuw, P.~J. (2025).
\newblock Challenges of cellwise outliers.
\newblock {\em Econometrics and Statistics}.
\newblock To appear, https://doi.org/10.1016/j.ecosta.2024.02.002\,.

\bibitem[{Rousseeuw} and {Van den Bossche}, 2018]{Rousseeuw:DDC}
{Rousseeuw}, P.~J. and {Van den Bossche}, W. (2018).
\newblock Detecting deviating data cells.
\newblock {\em Technometrics}, 60:135--145.

\bibitem[Samaria and Harter, 1994]{Samaria:ORL}
Samaria, F.~S. and Harter, A.~C. (1994).
\newblock Parameterisation of a stochastic model for human face identification.
\newblock In {\em Proceedings of 1994 IEEE Workshop on Applications of Computer
  Vision}, pages 138--142. IEEE.

\bibitem[Sarkar et~al., 2005]{Sarkar:Humanid}
Sarkar, S., Phillips, P.~J., Liu, Z., Vega, I.~R., Grother, P., and Bowyer,
  K.~W. (2005).
\newblock The humanid gait challenge problem: Data sets, performance, and
  analysis.
\newblock {\em IEEE Transactions on Pattern Analysis and Machine Intelligence},
  27(2):162--177.

\bibitem[Serneels and Verdonck, 2008]{Serneels:RobPCA}
Serneels, S. and Verdonck, T. (2008).
\newblock Principal component analysis for data containing outliers and missing
  elements.
\newblock {\em Computational Statistics \& Data Analysis}, 52(3):1712--1727.

\bibitem[Shi et~al., 2018]{Shi:FeatureExt}
Shi, Q., Cheung, Y.-M., Zhao, Q., and Lu, H. (2018).
\newblock Feature extraction for incomplete data via low-rank tensor
  decomposition with feature regularization.
\newblock {\em IEEE Transactions on Neural Networks and Learning Systems},
  30(6):1803--1817.

\bibitem[Smilde et~al., 2004]{Smilde:Bookmultiway}
Smilde, A., Bro, R., and Geladi, P. (2004).
\newblock {\em Multi-way Analysis with Applications in the Chemical Sciences}.
\newblock Wiley, England.

\bibitem[Todorov et~al., 2023]{Todorov:PARAFAC}
Todorov, V., Simonacci, V., Gallo, M., and Trendafilov, N. (2023).
\newblock A novel estimation procedure for robust {CANDECOMP/PARAFAC} model
  fitting.
\newblock {\em Econometrics and Statistics}.
\newblock To appear, https://doi.org/10.1016/j.ecosta.2023.07.001.

\bibitem[Tucker, 1966]{Tucker:Tucker}
Tucker, L.~R. (1966).
\newblock Some mathematical notes on three-mode factor analysis.
\newblock {\em Psychometrika}, 31(3):279--311.

\bibitem[Vasilescu and Terzopoulos, 2007]{Vasilescu:MultiliProj}
Vasilescu, M. A.~O. and Terzopoulos, D. (2007).
\newblock Multilinear projection for appearance-based recognition in the tensor
  framework.
\newblock In {\em 2007 IEEE 11th International Conference on Computer Vision},
  pages 1--8.

\bibitem[Ye et~al., 2004]{Ye:GPCA}
Ye, J., Janardan, R., and Li, Q. (2004).
\newblock {GPCA}: An efficient dimension reduction scheme for image compression
  and retrieval.
\newblock In {\em Proceedings of the tenth ACM SIGKDD International Conference
  on Knowledge Discovery and Data Mining}, pages 354--363.

\bibitem[Zhao et~al., 2015]{Zhao:BayesianCP}
Zhao, Q., Zhang, L., and Cichocki, A. (2015).
\newblock Bayesian {CP} factorization of incomplete tensors with automatic rank
  determination.
\newblock {\em IEEE Transactions on Pattern Analysis and Machine Intelligence},
  37(9):1751--1763.

\end{thebibliography}

\begin{thebibliography}{}
\providecommand{\natexlab}[1]{#1}
\providecommand{\url}[1]{\texttt{#1}}
\expandafter\ifx\csname urlstyle\endcsname\relax
  \providecommand{\doi}[1]{doi: #1}\else
  \providecommand{\doi}{doi: \begingroup \urlstyle{rm}\Url}\fi
	
\bibitem[Acar et~al., 2007]{Acar:CPDEEG}
Acar, E., Aykut-Bingol, C., Bingol, H., Bro, R., and Yener, B. (2007).
\newblock Multiway analysis of epilepsy tensors.
\newblock {\em Bioinformatics}, 23(13):i10--i18.

\bibitem[Cichocki et~al., 2016]{Cichocki:TensNet}
Cichocki, A., Lee, N., Oseledets, I., Phan, A.-H., Zhao, Q., Mandic, D.~P.,
  et~al. (2016).
\newblock Tensor networks for dimensionality reduction and large-scale
  optimization: Part 1 low-rank tensor decompositions.
\newblock {\em Foundations and Trends{\textregistered} in Machine Learning},
  9(4-5):249--429.

\bibitem[Cong et~al., 2015]{Cong:TensEEG}
Cong, F., Lin, Q.-H., Kuang, L.-D., Gong, X.-F., Astikainen, P., and
  Ristaniemi, T. (2015).
\newblock Tensor decomposition of {EEG} signals: a brief review.
\newblock {\em Journal of Neuroscience Methods}, 248:59--69.

\bibitem[Fackler, 2019]{Fackler:KronAlgo}
Fackler, P.~L. (2019).
\newblock Algorithm 993: Efficient computation with {K}ronecker products.
\newblock {\em ACM Transactions on Mathematical Software (TOMS)}, 45(2):1--9.

\bibitem[Fartash et~al., 2015]{Fartash:SpeechMPCA}
Fartash, M., Setayeshi, S., and Razzazi, F. (2015).
\newblock A noise robust speech features extraction approach in
  multidimensional cortical representation using multilinear principal
  component analysis.
\newblock {\em International Journal of Speech Technology}, 18(3):351--365.

\bibitem[Huber, 1981]{Huber:RobStat}
Huber, P.~J. (1981).
\newblock {\em {Robust Statistics}}.
\newblock John Wiley \& Sons.

\bibitem[Latchoumane et~al., 2012]{Latchoumane:TuckerEEG}
Latchoumane, C.-F.~V., Vialatte, F.-B., Sol{\'e}-Casals, J., Maurice, M.,
  Wimalaratna, S.~R., Hudson, N., Jeong, J., and Cichocki, A. (2012).
\newblock Multiway array decomposition analysis of {EEG}s in {A}lzheimer's
  disease.
\newblock {\em Journal of Neuroscience Methods}, 207(1):41--50.

\bibitem[Maronna et~al., 2019]{maronna2019robust}
Maronna, R.~A., Martin, R.~D., Yohai, V.~J., and Salibi{\'a}n-Barrera, M.
  (2019).
\newblock {\em Robust Statistics: Theory and Methods (with R)}.
\newblock John Wiley \& Sons.


\bibitem[Rolinger et~al., 2019]{Rolinger:PerfTensDecomp}
Rolinger, T.~B., Simon, T.~A., and Krieger, C.~D. (2019).
\newblock Performance considerations for scalable parallel tensor
  decomposition.
\newblock {\em Journal of Parallel and Distributed Computing}, 129:83--98.


\bibitem[Wu et~al., 2017]{Wu:MPCAnet}
Wu, J., Qiu, S., Zeng, R., Kong, Y., Senhadji, L., and Shu, H. (2017).
\newblock Multilinear principal component analysis network for tensor object
  classification.
\newblock {\em IEEE Access}, 5:3322--3331.

\end{thebibliography}

\spacingset{1}

\clearpage

\pagenumbering{arabic}
\appendix
\begin{center}
\large{Supplementary Materials to: \\ Casewise and Cellwise Robust Multilinear Principal Component Analysis}\\
\vspace{7mm}
\normalsize{Mehdi Hirari, Fabio Centofanti, Mia Hubert, Stefan Van Aelst} 
\end{center}
\vspace{3mm}

\renewcommand{\thesection}{\Alph{section}}
\renewcommand\thefigure{S.\arabic{figure}}   
\renewcommand{\theequation}{S.\arabic{equation}} 
 
\setcounter{figure}{0}    
\setcounter{equation}{0}    

\section{MPCA formulation}
\label{app:MPCAform}

The multilinear principal component analysis (MPCA) problem of \cite{Lu:MPCA} can be reformulated as a minimization problem as shown in Section 2 of~\cite{Inoue:RMPCA}. 
In particular, following their derivation yields the following result.

Let $\{\mathcal{X}_n\}_{n=1}^N$ be a set of tensors, where
$\mathcal{X}_n \in \mathbb{R}^{P_1\times\cdots\times P_L}$ is an $L$th-order tensor. The total scatter of these tensors is defined as $\Psi_{\mathcal{X}} \;=\; \sum_{n=1}^N \|\mathcal{X}_n - \widebar{\mathcal{X}}\|_F^2$, where $\widebar{\mathcal{X}}= \frac{1}{N}\sum_{n=1}^N \mathcal{X}_n$ is the sample mean. Let $\{\mathbf{V}^{(\ell)}\}_{\ell=1}^L$ be a collection of orthogonal matrices, where $\mathbf{V}^{(\ell)}  \in \mathbb{R}^{P_\ell\times K_\ell}$ with $K_\ell \leqslant P_\ell$ for $\ell=1,\ldots,L$. Define $\mathcal{U}_n = \mathcal{X}_n \times \{\mathbf{V}^T\}$ and $\widebar{\mathcal{U}} = \tfrac{1}{N}\sum_{n=1}^N \mathcal{U}_n$.

The MPCA objective is to determine the projection matrices $\{\mathbf{V^{(\ell)}}\}$ that maximize the total scatter of the $\{\mathcal{U}_n\}_{n=1}^N$:
\begin{align}
  \max_{\{\mathbf{V^{(\ell)}}\}} \quad & \Psi_{\mathcal{U}} = \sum_{n=1}^N \|\mathcal{U}_n - \widebar{\mathcal{U}}\|_F^2 \label{eq:max-var}\\
  \text{s.t.}\quad & \mathbf{V}^{(\ell)T}\mathbf{V}^{(\ell)} = \mathbf{I}_{K_\ell}, \qquad \ell=1,\ldots,L,
  \nonumber 
\end{align}
where $\mathbf{I}_{K_\ell}$ denotes the $K_\ell \times K_\ell$ identity matrix.
Since $\Psi_{\mathcal{X}}$ is a constant, \eqref{eq:max-var} is equivalent to 
\begin{equation*}
  \min_{\{\mathbf{V^{(\ell)}}\}} \;\; \Psi_{\mathcal{X}} - \Psi_{\mathcal{U}}\,.
\end{equation*}
If we denote $\widetilde{\mathcal{X}}_n = \mathcal{X}_n - \widebar{\mathcal{X}}$, 
$\widetilde{\mathcal{U}}_n = \mathcal{U}_n - \widebar{\mathcal{U}}$, and $\widetilde{\mathcal{U}}_n = \widetilde{\mathcal{X}}_n \times \{\mathbf{V}^T\}$
 we can rewrite the MPCA objective: 
\begin{align*}
  \Psi_{\mathcal{X}} - \Psi_{\mathcal{U}}
  &= \sum_{n=1}^N \|\widetilde{\mathcal{X}}_n\|_F^2 \;-\; \sum_{n=1}^N \|\widetilde{\mathcal{U}}_n\|_F^2 
  \nonumber \\
  &= \sum_{n=1}^N \|\widetilde{\mathcal{X}}_n\|_F^2 \;-\; \sum_{n=1}^N \|\widetilde{\mathcal{X}}_n \times \{\mathbf{V}^T\}\|_F^2 
  \nonumber \\
  &= \sum_{n=1}^N \Big(
       \|\widetilde{\mathcal{X}}_n\|_F^2
       - 2\,\|\widetilde{\mathcal{X}}_n \times \{\mathbf{V}^T\}\|_F^2
       + \|\widetilde{\mathcal{X}}_n \times \{\mathbf{V}^T\}\|_F^2
     \Big) 
     \nonumber \\
   & =  \sum_{n=1}^N \Big(
   \|\widetilde{\mathcal{X}}_n\|_F^2
       - 2\, < \widetilde{\mathcal{X}}_n \times \{\mathbf{V}^T\}, \widetilde{\mathcal{X}}_n \times \{\mathbf{V}^T\} > 
       + \|\widetilde{\mathcal{X}}_n \times \{\mathbf{V}^T\}\|_F^2
     \Big) 
     \nonumber \\
      & \overset{(*)}{=}  \sum_{n=1}^N \Big(
   \|\widetilde{\mathcal{X}}_n\|_F^2
       - 2\, < \widetilde{\mathcal{X}}_n, \big(\widetilde{\mathcal{X}}_n \times \{\mathbf{V}^T\}\big) \times \{\mathbf{V}\} > 
       + \|\big(\widetilde{\mathcal{X}}_n \times \{\mathbf{V}^T\}\big) \times \{\mathbf{V}\}\|_F^2
     \Big) 
     \nonumber \\
  &= \sum_{n=1}^N \Big\|
       \widetilde{\mathcal{X}}_n
       - \big(\widetilde{\mathcal{X}}_n \times \{\mathbf{V}^T\}\big) \times \{\mathbf{V}\}
     \Big\|_F^2 \nonumber \\
  &= \sum_{n=1}^N \big\|\widetilde{\mathcal{X}}_n - \widetilde{\mathcal{U}}_n \times \{\mathbf{V}\}\big\|_F^2  
  \nonumber \\
  &= \sum_{n=1}^N \big\|\mathcal{X}_n - \widebar{\mathcal{X}} - \widetilde{\mathcal{U}}_n \times \{\mathbf{V}\}\big\|_F^2
\end{align*}
which corresponds to \eqref{eq:MPCA}. Note that $(*)$ holds because the projection matrices $\mathbf{V}^{(\ell)}$ are orthogonal.

\section{First-Order Conditions}
\label{app:firstorder}

In the following, the derivation of the first-order
necessary conditions \eqref{eq:condV}-\eqref{eq:condC} is presented.
The objective function is given by
\begin{equation*}
	\mathcal{L}\left(\lbrace \mathcal{X}_{n} \rbrace,\lbrace\mathbf{V}^{(\ell)}\rbrace, \lbrace\mathcal{U}_n\rbrace, \mathcal{C} \right) = \frac{\hat{\sigma}_2^2}{m} \Sumn \ m_n \rho_2\left(\frac{1}{\hat{\sigma}_2}\sqrt{\frac{1}{m_n} \Sumps  m_{\eidx} \scell^2\, \rho_1\left(\frac{r_{\eidx}}{\scell}\right)}\ \right),
\end{equation*}

The derivative of $\mathcal{L}$ with respect to $v_{p_l k_l}^{(\ell)}$ is
 \begin{align*}
           \frac{\partial \mathcal{L}}{\partial v_{p_\ell k_\ell}^{(\ell)}}
            &=\frac{\hat{\sigma}_2^2}{m} \Sumn m_n \rho'_2\left( t_n/\hat{\sigma}_2\right) \frac{1}{\hat{\sigma}_2}\frac{\partial}{\partial v_{p_\ell k_\ell}^{(\ell)}} \sqrt{\frac{1}{m_n} \Sumps m_{\eidx} \scell^2\, \rho_1\left(\frac{r_{\eidx}}{\scell}\right)} \\ 
            &=\frac{1}{2m} \Sumn \frac{\psi_2\left( t_n/\hat{\sigma}_2\right)}{t_n/\hat{\sigma}_2} \frac{\partial}{\partial v_{p_\ell k_\ell}^{(\ell)}} \left(\Sumps m_{\eidx} \scell^2 \rho_{1}\left(\frac{r_{\eidx}}{\scell} \right)\right) \\ 
             &= \frac{1}{2m} \Sumn \wcase\, 
    \sum_{p_\ell,\ell \in L^{*}} \scell^2\, \psi_1(r_{\eidx}/\scell)
     \frac{m_{\eidx}}{\scell}
     \frac{\partial r_{\eidx}}{\partial v_{p_\ell k_\ell}^{(\ell)}} \nonumber\\
     &= -\frac{1}{2m} \Sumn   
     \sum_{p_\ell,\ell \in L^{*}} \Big[\wcase  \wcell m_{\eidx} r_{\eidx}\\&\hspace{4cm}\sum_{k_\ell,\ell \in L^{*}}\left({u}_{\Jidx} v_{p_1 k_1}^{(1)} \cdots v_{p_{\ell-1} k_{\ell-1}}^{(\ell-1)} v_{p_{\ell+1} k_{\ell+1}}^{(\ell+1)}\cdots v_{p_L k_L}^{(L)}\right)\Big], \nonumber
    \end{align*}
     where $L^{*} = \lbrace1, \ldots, \ell-1, \ell+1, \ldots, L \rbrace$.
This leads to the first first-order condition
    \begin{align*}
           \frac{\partial \mathcal{L}}{\partial {\mathbf{V}}^{(\ell)}} &= \Sumn \Big\langle \left({\mathcal{X}}_n - \mathcal{C} - {\mathcal{U}}_n \times \lbrace\mathbf{V}\rbrace \right) \odot \mathcal{W}_n , {\mathcal{U}}_{n}^{(-\ell)} \Big\rangle_{\lbrace L^{*};L^{*} \rbrace}=\mathbf{0}_{P_\ell\times K_\ell}.
    \end{align*}

The derivative of $\mathcal{L}$ with respect to ${u}_{\Jidx}$ is
 \begin{align*}
           \frac{\partial \mathcal{L}}{\partial {u}_{\Jidx}}
            &=\frac{\hat{\sigma}_2^2}{m} m_n \rho'_2\left( t_n/\hat{\sigma}_2\right) \frac{1}{\hat{\sigma}_2}\frac{\partial}{\partial {u}_{\Jidx}} \sqrt{\frac{1}{m_n} \Sumps m_{\eidx} \scell^2\, \rho_1\left(\frac{r_{\eidx}}{\scell}\right)} \\ 
            &=\frac{1}{2m} \frac{\psi_2\left( t_n/\hat{\sigma}_2\right)}{t_n/\hat{\sigma}_2} \frac{\partial}{\partial {u}_{\Jidx}} \left(\Sumps m_{\eidx} \scell^2 \rho_{1}\left(\frac{r_{\eidx}}{\scell} \right)\right) \\ 
             &= \frac{1}{2m} \Sumps \wcase\, 
     \scell^2\, \psi_1\left(r_{\eidx}/\scell\right)
     \frac{m_{\eidx}}{\scell}
     \frac{\partial r_{\eidx}}{\partial {u}_{\Jidx}} \nonumber\\
  &= -\frac{1}{2m}\Sumps   \wcase\,
     \scell\, \frac{\psi_1\left(r_{\eidx}/\scell\right)}{r_{\eidx}/\scell}
     \frac{m_{\eidx}}{\scell}r_{\eidx}v_{p_1 k_1}^{(1)} \cdots v_{p_L k_L}^{(L)} \nonumber\\
  &= -\frac{1}{2m}  
     \Sumps  \wcase  \wcell m_{\eidx} r_{\eidx}
     v_{p_1 k_1}^{(1)} \cdots v_{p_L k_L}^{(L)} \,,
    \end{align*}
which results in  the second first-order condition
 \begin{align*}
            \frac{\partial \mathcal{L}}{\partial {\mathcal{U}}_{n}} &= \big( \left({\mathcal{X}}_n - \mathcal{C} -  {\mathcal{U}}_n \times \lbrace\mathbf{V}\rbrace \right) \odot \mathcal{W}_n \big) \times \lbrace \mathbf{V}^T\rbrace=\mathcal{O}_{K_1\times \ldots \times K_L}\, .
    \end{align*}

     The derivative of $\mathcal{L}$ with respect to $c_{\idx}$
is
    \begin{align*}
          \frac{\partial \mathcal{L}}{\partial c_{\idx}}
            &=\frac{\hat{\sigma}_2^2}{m}\Sumn m_n \rho'_2\left( t_n/\hat{\sigma}_2\right) \frac{1}{\hat{\sigma}_2}\frac{\partial}{\partial c_{\idx}} \sqrt{\frac{1}{m_n} \Sumps  m_{\eidx} \scell^2\, \rho_1\left(\frac{r_{\eidx}}{\scell}\right)} \\ 
            &=\frac{1}{2m} \Sumn  \frac{\psi_2\left( t_n/\hat{\sigma}_2\right)}{t_n/\hat{\sigma}_2} \frac{\partial}{\partial c_{\idx}} \left(\Sumps m_{\eidx} \scell^2 \rho_{1}\left(\frac{r_{\eidx}}{\scell} \right)\right) \\ 
             &= \frac{1}{2m}\Sumn \wcase \, 
     \scell^2\, \psi_1\left(r_{\eidx}/\scell\right)
     \frac{m_{\eidx}}{\scell}
     \frac{\partial}{\partial c_{\idx}} r_{\eidx}\nonumber\\
  &= -\frac{1}{2m}\Sumn \wcase \, 
     \scell \, \frac{\psi_1\left(r_{\eidx}/\scell\right)}{r_{\eidx}/\scell}
     \frac{m_{\eidx}}{\scell}r_{\eidx}\nonumber\\
  &= -\frac{1}{2m}\Sumn 
     \wcase \wcell m_{\eidx}
     r_{\eidx}\,,
    \end{align*}
    so the gradient with respect to the tensor $\mathcal{C}$ becomes
\begin{align*} 
 \frac{\partial \mathcal{L}}{\partial \mathcal{C}}
  &= \Sumn \left({\mathcal{X}}_n - \mathcal{C} - {\mathcal{U}}_n \times \lbrace \mathbf{V}\rbrace \right) \odot \mathcal{W}_n=\mathcal{O}_{P_1 \times \cdots \times P_L}.
\end{align*}

In what follows, it is shown that the first order conditions of the ROMPCA objective function~\eqref{eq:obj} are identical to those of the  weighted MPCA objective function~\eqref{eq:WMPCA}  given by
\begin{align}
	\widetilde{\mathcal{L}}\left(\lbrace \mathcal{X}_{n} \rbrace,\lbrace\mathbf{V}^{(\ell)}\rbrace, \lbrace{\mathcal{U}}_n\rbrace, \mathcal{C}\right) & = \Sumn \Sumps w_{\eidx} \left({x}_{\eidx}-c_{\idx} - \Sumks {u}_{\Jidx} v_{p_1 k_1}^{(1)} \cdots v_{p_L k_L}^{(L)}\right)^2 \nonumber \\
    & = \sum_{n=1}^N \left\| \mathcal{W}_n^{1/2} \odot \left({\mathcal{X}}_n - \mathcal{C} - {\mathcal{U}}_n \times \lbrace \mathbf{V}\rbrace \right) \right\|_F^2 \label{eq:WMPCA_obj}
\end{align}
The derivative of $\widetilde{\mathcal{L}}$ with respect to $v_{p_\ell k_\ell}^{(\ell)}$ is
 \begin{align*}
          \frac{\partial \widetilde{\mathcal{L}}}{\partial v_{p_\ell k_\ell}^{(\ell)}}
            &=\Sumn \frac{\partial}{\partial v_{p_\ell k_\ell}^{(\ell)}} \Sumps w_{\eidx} r_{\eidx}^2 \\ 
            &=2\Sumn \sum_{p_\ell,\ell \in L^{*}} w_{\eidx} r_{\eidx} \frac{\partial r_{\eidx}}{\partial v_{p_\ell k_\ell}^{(\ell)}} \\ 
            &=-2 \Sumn \sum_{p_\ell,\ell\in L^{*}} \Big[w_{\eidx} r_{\eidx}
            \sum_{k_\ell,\ell \in L^{*}}\left({u}_{\Jidx} v_{p_1 k_1}^{(1)} \cdots v_{p_{\ell-1} k_{\ell-1}}^{(\ell-1)} v_{p_{\ell+1} k_{\ell+1}}^{(\ell+1)}\cdots v_{p_L k_L}^{(L)}\right)\Big]
    \end{align*}
which results in
    \begin{align*}
           \frac{\partial \widetilde{\mathcal{L}}}{\partial {\mathbf{V}}^{(\ell)}} &= \Sumn \Big\langle \left({\mathcal{X}}_n - \mathcal{C} - {\mathcal{U}}_n \times \lbrace\mathbf{V}\rbrace\right) \odot \mathcal{W}_n \, , \, {\mathcal{U}}_{n}^{(-\ell)} \Big\rangle_{\lbrace L^{*};L^{*} \rbrace} 
           = \mathbf{0}_{P_\ell\times K_\ell}
    \end{align*}
that is identical to~\eqref{eq:condV}.
The derivative of $\widetilde{\mathcal{L}}$ with respect to ${u}_{\Jidx}$ is
 \begin{align*}
            \frac{\partial \tilde{\mathcal{L}}}{\partial {u}_{\Jidx}}
            &=\frac{\partial}{\partial {u}_{\Jidx}} \Sumps w_{\eidx} \left({x}_{\eidx} -c_{\idx} - \Sumks {u}_{\Jidx} v_{p_1 k_1}^{(1)} \cdots v_{p_L k_L}^{(L)}\right)^2 \\ 
            &=2\Sumps w_{\eidx} r_{\eidx} \frac{\partial r_{\eidx}}{\partial {u}_{\Jidx}} \\ 
            &=-2 \Sumps  w_{\eidx} r_{\eidx} v_{p_1 k_1}^{(1)} \cdots v_{p_L k_L}^{(L)} \,,
    \end{align*}
hence the second first-order condition is equal to~\eqref{eq:condU}:
    \begin{align*}
            \frac{\partial \widetilde{\mathcal{L}}}{\partial {\mathcal{U}}_{n}} = \big( \left({\mathcal{X}}_n - \mathcal{C} -  {\mathcal{U}}_n \times \lbrace\mathbf{V}\rbrace \right) \odot \mathcal{W}_n \big) \times \lbrace {\mathbf{V}}^T\rbrace = \mathcal{O}_{K_1 \times \cdots \times K_L}\, .
\end{align*} 
    The derivative of $\widetilde{\mathcal{L}}$ with respect to $c_{\idx}$ is
    \begin{align*}
           \frac{\partial \tilde{\mathcal{L}}}{\partial c_{\idx}}
            &=\Sumn w_{\eidx} \frac{\partial}{\partial c_{\idx}} \left({x}_{\eidx} - c_{\idx} - \Sumks {u}_{\Jidx} v_{p_1 k_1}^{(1)} \cdots v_{p_L k_L}^{(L)}\right)^2 \\ 
            &= -2 \Sumn w_{\eidx} r_{\eidx}\,,
     \end{align*}
    which results in condition~\eqref{eq:condC}:
\begin{align*} 
\frac{\partial \widetilde{\mathcal{L}}}{\partial \mathcal{C}}
&= \Sumn \left({\mathcal{X}}_n - {\mathcal{C}} - {\mathcal{U}}_n \times \lbrace \mathbf{V}\rbrace \right) \odot \mathcal{W}_n 
= \mathcal{O}_{P_1 \times \cdots \times P_L}.
\end{align*}

\section{Description of the Algorithm}
\label{app:algo}
The IRLS  starts with the initial estimates 
$\lbrace\mathbf{V}^{(\ell)}_{0}\rbrace, \lbrace{\mathcal{U}}_{n,0}\rbrace$ and $\mathcal{C}_{0}$ defined in Section~\ref{sec:init}, and the corresponding $\lbrace\mathcal{W}_{n,0}\rbrace$ obtained from 
\eqref{eq:Wn}, \eqref{eq:weightc}, \eqref{eq:weightr}, \eqref{eq:cellres} and \eqref{eq:caseres}. Then, for each $q = 0, 1,2, \ldots$, new estimates $\lbrace\mathbf{V}^{(\ell)}_{q+1}\rbrace$, $\lbrace{\mathcal{U}}_{n,q+1}\rbrace$, $\mathcal{C}_{q+1}$ and $\lbrace\mathcal{W}_{n,q+1}\rbrace$ are obtained from $\lbrace\mathbf{V}^{(\ell)}_{q}\rbrace$, $\lbrace{\mathcal{U}}_{n,q}\rbrace$, $\mathcal{C}_{q}$ and 
$\lbrace\mathcal{W}_{n,q}\rbrace$
by the following procedure:
\begin{itemize}
\item[\textbf{(a)}]  Minimize \eqref{eq:WMPCA} with respect to $\mathbf{V}^{(\ell)}$ by substituting $\mathcal{C}_{q}$, $\lbrace{\mathcal{U}}_{n,q}\rbrace$, and $\lbrace\mathcal{W}_{n,q}\rbrace$ in \eqref{eq:condV}.
That is, for $\ell=1,\dots,L$,
    \begin{align*}
            \Sumn \Big\langle \left(\mathcal{X}_n - \mathcal{C}_{q} -{\mathcal{U}}_{n,q} \times \lbrace{\mathbf{V}}\rbrace\right) \odot \mathcal{W}_{n,q}\, , \,{\mathcal{U}}_{n,q}^{(-\ell)}  \Big\rangle_{\lbrace L^{*};L^{*} \rbrace} & = \mathbf{0}_{P_\ell \times K_\ell}\, ; \\  
            \Sumn \Big\langle \left({\mathcal{U}}_{n,q} \times \lbrace{\mathbf{V}}\rbrace\right) \odot \mathcal{W}_{n,q}\, , \,{\mathcal{U}}_{n,q}^{(-\ell)}  \Big\rangle_{\lbrace L^{*};L^{*} \rbrace} &= \Sumn  \Big\langle \left(\mathcal{X}_n - \mathcal{C}_{q}\right) \odot \mathcal{W}_{n,q}\, , \, {\mathcal{U}}_{n,q}^{(-\ell)}  \Big\rangle_{\lbrace L^{*};L^{*} \rbrace}\, ; \\  
            \Sumn \Big\langle \left({\mathcal{U}}_{n,q}^{(-\ell)} \times_\ell {\mathbf{V}^{(\ell)}}\right) \odot \mathcal{W}_{n,q}\, , \, {\mathcal{U}}_{n,q}^{(-\ell)}  \Big\rangle_{\lbrace L^{*};L^{*} \rbrace} & = \Sumn  \Big\langle \left(\mathcal{X}_n - \mathcal{C}_{q}\right) \odot \mathcal{W}_{n,q}\, , \, {\mathcal{U}}_{n,q}^{(-\ell)}  \Big\rangle_{\lbrace L^{*};L^{*} \rbrace}\, ,
    \end{align*}
    with $L^{*} = \lbrace1, \ldots, \ell-1, \ell+1, \ldots, L \rbrace$.
Through unfolding, we have that
    \begin{align*}
             \Sumn \Big\langle \left({\mathbf{V}^{(\ell)}} {\mathbf{U}}_{n(\ell),q}^{(-\ell)}\right) \odot \mathbf{W}_{n(\ell),q} \, , \,{\mathbf{U}}_{n(\ell),q}^{(-\ell)}  \Big\rangle_{\lbrace L^{*};L^{*} \rbrace}  &=\Sumn  \Big\langle \left(\mathbf{X}_{n(\ell)} - \mathbf{C}_{(\ell),q}\right) \odot \mathbf{W}_{n(\ell),q} \, , \, {\mathbf{U}}_{n(\ell),q}^{(-\ell)}  \Big\rangle_{\lbrace L^{*};L^{*} \rbrace}\, ;\\
              \Sumn \bigg(\left({\mathbf{V}^{(\ell)}} {\mathbf{U}}_{n(\ell),q}^{(-\ell)}\right) \odot \mathbf{W}_{n(\ell),q}\bigg)  {\mathbf{U}}_{n(\ell),q}^{(-\ell)T} &= \Sumn \left((\mathbf{X}_{n(\ell)} - \mathbf{C}_{(\ell),q}) \odot \mathbf{W}_{n(\ell),q}\right) {\mathbf{U}}_{n(\ell),q}^{(-\ell)T}\, .
    \end{align*}

By vectorizing this last equation, we have that
    \begin{align*}
             \Sumn  \left({\mathbf{U}}_{n(\ell),q}^{(-\ell)} \otimes \mathbf{I}_{P_\ell}\right)&\left(\text{vec}\left({\mathbf{V}^{(\ell)}} {\mathbf{U}}_{n(\ell),q}^{(-\ell)}\right)\odot \text{vec} \left(\mathbf{W}_{n(\ell),q}\right)\right)\\
              &\qquad \qquad =\Sumn \left({\mathbf{U}}_{n(\ell),q}^{(-\ell)} \otimes \mathbf{I}_{P_\ell}\right)  \left(\text{vec}\left(\mathbf{X}_{n(\ell)} - \mathbf{C}_{(\ell),q}\right) \odot\text{vec}\left(\mathbf{W}_{n(\ell),q}\right)\right);\\
              \Sumn  \left({\mathbf{U}}_{n(\ell),q}^{(-\ell)} \otimes \mathbf{I}_{P_\ell}\right) &\mathbf{W}_{n,q} \text{vec}\left({\mathbf{V}^{(\ell)}} {\mathbf{U}}_{n(\ell),q}^{(-\ell)}\right)  \\&\qquad \qquad =  \Sumn \left({\mathbf{U}}_{n(\ell),q}^{(-\ell)} \otimes \mathbf{I}_{P_\ell}\right) \mathbf{W}_{n,q} \text{vec}\left(\mathbf{X}_{n(\ell)} - \mathbf{C}_{(\ell),q}\right)\, ;  \\
             \Bigg(\Sumn \left({\mathbf{U}}_{n(\ell),q}^{(-\ell)} \otimes \mathbf{I}_{P_\ell}\right) &\mathbf{W}_{n,q} \left({\mathbf{U}}_{n(\ell),q}^{(-\ell)} \otimes \mathbf{I}_{P_\ell}\right)^{T}\Bigg)\text{vec}\left({\mathbf{V}^{(\ell)}}\right) \\&\qquad \qquad =   \Sumn \left({\mathbf{U}}_{n(\ell),q}^{(-\ell)} \otimes \mathbf{I}_{P_\ell}\right) \mathbf{W}_{n,q} \text{vec}\left(\mathbf{X}_{n(\ell)} - \mathbf{C}_{(\ell),q}\right)\, .
             \end{align*}
         
Thus,
\begin{multline*}
     \text{vec}\left(\mathbf{V}^{(\ell)}_{q+1}\right) = \left(\Sumn \left({\mathbf{U}}_{n(\ell), q}^{(-\ell)} \otimes \mathbf{I}_{P_\ell}\right)\mathbf{W}_{n,q}\left({\mathbf{U}}_{n(\ell), q}^{(-\ell)} \otimes \mathbf{I}_{P_\ell}\right)^{T}\right)^{\dagger}\\\left(\Sumn \left({\mathbf{U}}_{n(\ell), q}^{(-\ell)} \otimes \mathbf{I}_{P_\ell}\right)\mathbf{W}_{n, q} \text{vec}\left(\mathbf{X}_{n(\ell)} - \mathbf{C}_{(\ell),q}\right)\right).
    \end{multline*}
 
where $^{\dagger}$ denotes the Moore-Penrose generalized inverse. Here the use of the generalized inverse is justified by the fact that  $\Sumn ({\mathbf{U}}_{n(\ell),q}^{(-\ell)} \otimes \mathbf{I}_{P_\ell})\mathbf{W}_{n,q}({\mathbf{U}}_{n(\ell),q}^{(-\ell)} \otimes \mathbf{I}_{P_\ell})^{T}$ can be singular, e.g, when the weight matrices $\mathbf{W}_{n,q}$ have too many zeros.

\item[\textbf{(b)}] Minimize \eqref{eq:WMPCA} with respect to ${\mathcal{U}}_n$ by substituting $\lbrace\mathbf{V}^{(\ell)}_{q+1}\rbrace$, $\mathcal{C}_{q}$, and $\lbrace{\mathcal{W}}_{n,q}\rbrace$ in \eqref{eq:condU}.  That is, for $n=1,\ldots,N$,
\begin{equation}\label{eq:condU2w}
  \big( \left( \mathcal{X}_n - \mathcal{C}_{q} - {\mathcal{U}}_n \times \lbrace \mathbf{V}_{q+1}\rbrace\right) \odot \mathcal{W}_{n,q} \big) \times \lbrace\mathbf{V}_{q+1}^{T}\rbrace = \mathcal{O}_{K_1 \times \ldots \times K_L}\, .        
\end{equation}
             
Instead we solve
\begin{equation} 
 \left( \left({\mathcal{X}}_n - \mathcal{C}_q -  {\mathcal{U}}_n \times \lbrace\mathbf{V}_{q+1}\rbrace \right) \odot \widetilde{\mathcal{W}}_{n,q} \right) \times \lbrace \mathbf{V}_{q+1}^T\rbrace=\mathcal{O}_{K_1\times \ldots \times K_L} \label{eq:condU2}
     \end{equation}
 with $\widetilde{\mathcal{W}}_{n,q} = \mathcal{W}_{n,q}^{\text{cell}} \odot \mathcal{M}_n$. This is justified since \eqref{eq:condU2w} is equivalent to \eqref{eq:condU2} when $w^{\text{case}}_n \ne 0$. For a tensor with $w^{\text{case}}_n=0$, the term $\left\| \mathcal{W}_n^{1/2} \odot \left({\mathcal{X}}_n - \mathcal{C} - {\mathcal{U}}_n \times \lbrace \mathbf{V}\rbrace \right) \right\|_F^2$ in \eqref{eq:WMPCA_obj} is minimal as it is zero. 
             Then \begin{align*}
             \left(\left({\mathcal{U}}_n \times \lbrace \mathbf{V}_{q+1}\rbrace\right) \odot \widetilde{\mathcal{W}}_{n,q}\right) \times \lbrace \mathbf{V}_{q+1}^{T}\rbrace &= \left(\left(\mathcal{X}_n - \mathcal{C}_{q} \right)\odot \widetilde{\mathcal{W}}_{n,q} \right) \times \lbrace\mathbf{V}_{q+1}^{T}\rbrace\, .
    \end{align*}

By vectorizing this last equation, we have that
    \begin{align*}
              \left(\sbigotimes_{\ell=1}^{L} \mathbf{V}^{(\ell)T}_{q+1}\right)\left(\text{vec}\left({\mathcal{U}}_n \times \lbrace {\mathbf{V}_{q+1}}\rbrace\right) \odot \text{vec}\left(\widetilde{\mathcal{W}}_{n,q}\right)\right) & = \left(\sbigotimes_{\ell=1}^{L} \mathbf{V}^{(\ell)T}_{q+1}\right)\left( \text{vec}\left( \mathcal{X}_n - \mathcal{C}_{q}\right) \odot \text{vec}\left(\widetilde{\mathcal{W}}_{n,q}\right)\right)\, ; \\ \left(\sbigotimes_{\ell=1}^{L} \mathbf{V}^{(\ell)T}_{q+1}\right)\widetilde{\mathbf{W}}_{n,q} \left(\sbigotimes_{\ell=1}^{L} \mathbf{V}^{(\ell)}_{q+1}\right)\text{vec}\left({\mathcal{U}}_n\right) & = \left(\sbigotimes_{\ell=1}^{L} \mathbf{V}^{(\ell)T}_{q+1}\right) \widetilde{\mathbf{W}}_{n,q} \text{vec}\left( \mathcal{X}_n - \mathcal{C}_{q}\right)\, .  
     \end{align*}
Thus, 
   \begin{equation*}
   \text{vec}\left({\mathcal{U}}_{n, q+1}\right) = \left[\left(\sbigotimes_{\ell=1}^{L} \mathbf{V}^{(\ell)T}_{q+1}\right)\widetilde{\mathbf{W}}_{n, q}\left(\sbigotimes_{\ell=1}^{L} {\mathbf{V}^{(\ell)}_{q+1}}\right)\right]^{\dagger}\left[\left(\sbigotimes_{\ell=1}^{L} {\mathbf{V}^{(\ell)T}_{q+1}}\right)\widetilde{\mathbf{W}}_{n, q} \text{vec}\left(\mathcal{X}_n - \mathcal{C}_{q}\right)\right]\, .
\end{equation*}

\item[\textbf{(c)}] Minimize \eqref{eq:WMPCA} with respect to $\mathcal{C}$ by substituting $\lbrace\mathbf{V}^{(\ell)}_{q+1}\rbrace$, $\lbrace{\mathcal{U}}_{n,q+1}\rbrace$, and $\lbrace\mathcal{W}_{n,q}\rbrace$ in \eqref{eq:condC}.  That is
\begin{align*}
    &\Sumn \left(\mathcal{X}_n  - {\mathcal{U}}_{n,q+1} \times \lbrace \mathbf{V}_{q+1}\rbrace - \mathcal{C}\right) \odot \mathcal{W}_{n,q} = \mathcal{O}_{P_1 \times \ldots \times P_L}\, ;\\
    & \mathcal{C} \odot \left(\Sumn \mathcal{W}_{n,q}\right)  = \Sumn \left(\mathcal{X}_n  - {\mathcal{U}}_{n,q+1} \times \lbrace \mathbf{V}_{q+1}\rbrace\right) \odot \mathcal{W}_{n,q}\, .
\end{align*}
Thus,
\begin{equation*}
    \mathcal{C}_{q+1} = \left(\Sumn \left(\mathcal{X}_n  - {\mathcal{U}}_{n,q+1} \times \lbrace \mathbf{V}_{q+1}\rbrace \right) \odot \mathcal{W}_{n,q}\right) \odot \mathcal{H}_q\, .
\end{equation*}

\item[\textbf{(d)}] Update $\{\mathcal{W}_{n}\}$ using \eqref{eq:Wn}, \eqref{eq:weightc}, \eqref{eq:weightr}, \eqref{eq:cellres} and \eqref{eq:caseres} with $\lbrace\mathbf{V}^{(\ell)}_{q+1}\rbrace$, $\lbrace{\mathcal{U}}_{n,q+1}\rbrace$, and $\mathcal{C}_{q+1}$.

\end{itemize}

\newpage
\section{Proof of Proposition \ref{the_1P}}
\label{app:proofCOnv}
In this section Proposition \ref{the_1P} is proved, which ensures that each step of the algorithm decreases the objective function~\eqref{eq:obj}. 

Given $\{\mathcal{X}_n\}$ and potential estimates $(\{\mathbf{V}^{(\ell)}\}, \{\mathcal{U}_n\}, \mathcal{C})$, we denote the tensor of fitted values $\Theta_n = [\theta_{\eidx}] = \widehat{\mathcal{X}}_n = \mathcal{C} + {\mathcal{U}}_n \times \lbrace \mathbf{V}\}$ and a potential fit as $\boldsymbol{\theta} = \{ \text{vec}(\Theta_n)  \rbrace_{n = 1}^N$ where all elements are stacked into a vector. Similarly we define the vector $\mathbf{f}(\boldsymbol{\theta})= \lbrace \text{vec}\big((\mathcal{X}_n - \Theta_n) \odot (\mathcal{X}_n - \Theta_n)\big) \rbrace_{n = 1}^N$. It has $D = N\prod_\ell P_\ell$ entries with values $\left(x_{\eidx} - \theta_{\eidx} \right)^2$ which we denote by $f_{d_{\eidx}}$. 

We can then write the ROMPCA objective function \eqref{eq:obj} as $L \left(\mathbf{f} \left(\boldsymbol{\theta} \right) \right) := \mathcal{L} \left(\lbrace\mathcal{X}_n\rbrace, \lbrace\mathbf{V}^{(\ell)}\rbrace, \lbrace{\mathcal{U}}_n\rbrace, \mathcal{C} \right)$.

The proof of Proposition~\ref{the_1P} follows the steps of \cite{Centofanti:cellpca} and is based on the following lemmas.\\

\textbf{Lemma 1.} For given weight tensors $\lbrace\mathcal{W}_{n,q}\rbrace$, each of the update steps (a), (b), and (c) of the algorithm in Section~\ref{app:algo} decrease the weighted MPCA objective function~\eqref{eq:WMPCA}.\\

\textbf{Lemma 2.} The function $\mathbf{f} \rightarrow L\left(\mathbf{f} \right)$ is concave.\\

\textit{Proof.} It is shown in \cite{Centofanti:cellpca} that the univariate function $h:\mathbb{R}_{+} \rightarrow \mathbb{R}_{+} : z \rightarrow \rho_{b,c} \left(\sqrt{z} \right)$, in which $\rho_{b,c}$ is the hyperbolic tangent $\rho$-function, is concave.

By the definition of concavity of a multivariate function, we need to prove that for any column vectors ${\mathbf{f}}^1$, ${\mathbf{f}}^2$ in $\mathbb{R}^{D}_{+}$ and any $\lambda$ in $\left(0,1 \right)$ it holds that $L \left(\lambda \mathbf{f}^1 + \left(1 - \lambda \right){\mathbf{f}}^2 \right) \geqslant  \lambda L\left(\mathbf{f}^1 \right) + \left(1 - \lambda \right) L\left({\mathbf{f}}^2 \right)$.
This follows from
\begin{align*}
    L \left(\lambda \mathbf{f}^{1} + \left(1 - \lambda \right)\mathbf{f}^{2} \right) &= 
    \frac{\hat{\sigma}_{2}^2}{m} \Sumn h_2 \left( \frac{1}{m_n \hat{\sigma}_{2}^2} \Sumps m_{\eidx} \scell^2 h_1 \left( \frac{\lambda f_{d_{\eidx}}^{1} + (1 - \lambda)f_{d_{\eidx}}^{2}}{\scell^2} \right) \right) \\
    &\geqslant 
      \frac{\hat{\sigma}_{2}^2}{m} \Sumn h_2 \left( \frac{1}{m_n \hat{\sigma}_{2}^2} \Sumps m_{\eidx} \scell^2 \bigg[ \lambda h_1 \left( \frac{f_{d_{\eidx}}^{1}}{\scell^2} \right) \right. \\
    &  \hspace{8cm} + \left. (1 - \lambda) h_1 \left( \frac{f_{d_{\eidx}}^{2}}{\scell^2} \right) \bigg] \right) \\
    &= 
    \frac{\hat{\sigma}_{2}^2}{m} \Sumn h_2 \left( \lambda \frac{1}{m_n \hat{\sigma}_{2}^2} \Sumps m_{\eidx} \scell^2 h_1 \left( \frac{f_{d_{\eidx}}^{1}}{\scell^2} \right) \right. \\
    &  \qquad\qquad\qquad + \left. (1 - \lambda) \frac{1}{m_n \hat{\sigma}_{2}^2} \Sumps m_{\eidx} \scell^2 h_1 \left( \frac{f_{d_{\eidx}}^{2}}{\scell^2} \right) \right) \\
    &\geqslant 
    \frac{\hat{\sigma}_{2}^2}{m} \Sumn \left[ \lambda h_2 \left( \frac{1}{m_n \hat{\sigma}_{2}^2} \Sumps m_{\eidx} \scell^2 h_1 \left( \frac{f_{d_{\eidx}}^{1}}{\scell^2} \right) \right) \right. \\
    & \qquad\qquad\qquad + \left. (1 - \lambda) h_2 \left( \frac{1}{m_n \hat{\sigma}_{2}^2} \Sumps m_{\eidx} \scell^2 h_1 \left( \frac{f_{d_{\eidx}}^{2}}{\scell^2} \right) \right) \right] \\
    &= \lambda L\left(\mathbf{f}^{1} \right) + \left(1 - \lambda \right) L\left(\mathbf{f}^{2} \right).
\end{align*}
The first inequality follows from the concavity of $h_1$ and the fact that $h_2$ is nondecreasing. The second inequality comes from the concavity of $h_2$ . Thus, $L$ is a concave function. \hfill $\square$\\

The weighted MPCA objective \eqref{eq:WMPCA} can also be written as a function of $\mathbf{f}$. Denote it as $L_{\mathbf{w}} (\mathbf{f}):=  \mathbf{w}^T \mathbf{f}$, where $\mathbf{w}=\{\text{vec}(\mathcal{W}_n)\}_{n=1}^N$ turns the weight tensors into a vector in the same way as the column vector $\mathbf{f}$. Lemma $3$ links the weighted MPCA objective $L_{\mathbf{w}}$ with the original objective $L$. \\

\textbf{Lemma 3.} If two column vectors $\mathbf{f}^{1}, \mathbf{f}^{2} \in \mathbb{R}_{+}^{D}$ satisfy $L_{\mathbf{w}} \left(\mathbf{f}^{1} \right) \leqslant L_{\mathbf{w}}\left( \mathbf{f}^{2} \right)$, then $L(\mathbf{f}^{1}) \leqslant L\left( \mathbf{f}^{2} \right)$.\\

\textit{Proof.} By using Lemma 2 and \cite{Centofanti:cellpca}.
\\

\textit{Proof of Proposition \ref{the_1P}.} When the fit is updated from $\boldsymbol{\theta}_{q}$ to $\boldsymbol{\theta}_{q+1}$, Lemma $1$ shows that $L_{\mathbf{w}_{q}} \left(\mathbf{f} \left(\boldsymbol{\theta}_{q+1} \right) \right) \leqslant L_{\mathbf{w}_{q}} \left( \mathbf{f} \left(\boldsymbol{\theta}_{q} \right) \right)$, so using Lemma $3$ it follows that $L\left( \mathbf{f} \left(\boldsymbol{\theta}_{q+1} \right) \right) \leqslant L\left( \mathbf{f} \left(\boldsymbol{\theta}_{q} \right) \right)$. \hfill $\square$

\newpage
\section{Pseudocode of the ROMPCA algorithm}
\label{app:pseudocode}

Algorithm~\ref{alg:irls} provides the pseudocode of 
the steps in Sections~\ref{sec:obj}--\ref{sec:algo} of the 
main text.

\begin{algorithm}[H]
\caption{IRLS algorithm for ROMPCA} \label{alg:irls}
\begin{algorithmic}[1]
  \STATE{Retain $\lbrace \mathbf{V}^{(\ell)}_0 \rbrace$, $\lbrace \mathcal{U}_{n,0} \rbrace$, $\mathcal{C}_0$, $\scell$ and 
     $\hat{\sigma}_2$ from the initial fit.}
    \STATE{Compute the initial weight tensors $\lbrace\mathcal{W}_{n,0}\rbrace$ using \eqref{eq:Wn}, \eqref{eq:weightc}, and \eqref{eq:weightr}.}
  \STATE{Set $q = 0$.}
  \REPEAT  
     \STATE{(a) For $\ell=1,\ldots,L$, update $\lbrace \mathbf{V}^{(\ell)} \rbrace$:  
         \begin{multline*}
      \text{vec}\left(\mathbf{V}^{(\ell)}_{q+1}\right) = \left(\Sumn \left(\mathbf{U}_{n(\ell), q}^{(-\ell)} \otimes \mathbf{I}_{P_\ell}\right)\mathbf{W}_{n,q}\left(\mathbf{U}_{n(\ell), q}^{(-\ell)} \otimes \mathbf{I}_{P_\ell}\right)^{T}\right)^{\dagger}\\\left(\Sumn \left(\mathbf{U}_{n(\ell), q}^{(-\ell)} \otimes \mathbf{I}_{P_\ell}\right)\mathbf{W}_{n, q} \text{vec}\left(\mathbf{X}_{n(\ell)} - \mathbf{C}_{(\ell),q}\right)\right).
    \end{multline*}
      Orthonormalize $\mathbf{V}^{(\ell)}_{q+1}$.
      }
     \STATE{(b) For $n=1,\ldots,N$, update $\lbrace \mathcal{U}_{n} \rbrace$:   
        \begin{equation*}
            \text{vec}\left(\mathcal{U}_{n, q+1}\right) = \left(\left(\sbigotimes_{\ell=1}^{L} {\mathbf{V}_{q+1}^{(\ell)T}}\right)\widetilde{\mathbf{W}}_{n, q}\left(\sbigotimes_{\ell=1}^{L} {\mathbf{V}^{(\ell)}_{q+1}}\right)\right)^{\dagger}\left(\left(\sbigotimes_{\ell=1}^{L} {\mathbf{V}_{q+1}^{(\ell)T}}\right)\widetilde{\mathbf{W}}_{n, q} \text{vec}\left(\mathcal{X}_n - \mathcal{C}_{q}\right)\right)\, .
        \end{equation*}
     }
     \STATE{(c) Update $\mathcal{C}$:   
        \begin{equation*}
            \mathcal{C}_{q+1} = \left(\Sumn \left({\mathcal{X}}_n  -\mathcal{U}_{n,q+1} \times \lbrace \mathbf{V}_{q+1}\rbrace\right) \odot \mathcal{W}_{n,q} \right) \odot \mathcal{H}_q.
        \end{equation*}
     }
     \STATE{(d) Update $\mathcal{W}_{n}$: Compute $\lbrace\mathcal{W}_{n,q+1}\rbrace$ using \eqref{eq:Wn}, \eqref{eq:weightc}, and \eqref{eq:weightr}.}
     \STATE{Increment $q$: $q = q + 1$.}
  \UNTIL{
  \begin{equation*}
      \frac{\mathcal{L}\left(\lbrace \mathcal{X}_{n} \rbrace,\lbrace\mathbf{V}_{q+1}^{(\ell)}\rbrace, \lbrace \mathcal{U}_{n, q+1} \rbrace, \mathcal{C}_{q+1} \right) - \mathcal{L}\left(\lbrace \mathcal{X}_{n} \rbrace,\lbrace\mathbf{V}_{q}^{(\ell)}\rbrace, \lbrace \mathcal{U}_{n, q} \rbrace, \mathcal{C}_{q}\right)}{\mathcal{L}\left(\lbrace \mathcal{X}_{n} \rbrace,\lbrace\mathbf{V}_{q}^{(\ell)}\rbrace, \lbrace \mathcal{U}_{n, q} \rbrace, \mathcal{C}_{q}\right)} \leqslant 10^{-5}.
  \end{equation*}
  }
\end{algorithmic} 
\end{algorithm} 

\newpage
\section{Complexity of the ROMPCA algorithm}
\label{app:compl}
We will first study the time complexity of the IRLS algorithm 
described in Section~\ref{sec:algo} and summarized in the pseudocode in the Supplementary Material~\ref{app:pseudocode}. We denote $P = \prod_{\ell=1}^L P_\ell$ and $K = \prod_{\ell=1}^L K_\ell$. 

The complexity for updating $\mathbf{V}^{(\ell)}$ as in \eqref{eq:solV} is obtained as follows:
\begin{enumerate}
\item \textbf{Compute $\mathrm{vec}\left(\mathbf{X}_{n(\ell)} - \mathbf{C}_{(\ell)}\right)$.}  
Unfolding is a simple reshape with $O(1)$ cost, since no data is moved and only indices are reinterpreted.  
Elementwise subtraction of two matrices of size $P_\ell \times \prod_{i \neq \ell} P_i$ requires $O\!\left(P\right)$ operations.
\item \textbf{Compute $\mathbf{W}_{n}\mathrm{vec}(\mathbf{X}_{n(\ell)} - \mathbf{C}_{(\ell)})$.}  
Since  $\mathbf{W}_{n}$ is diagonal, the cost is equal to computing the Hadamard product of $(\mathbf{X}_{n(\ell)} - \mathbf{C}_{(\ell)})$ and $\mathbf{W}_{n(\ell)}$, both of size $P_\ell \times \prod_{i \neq \ell} P_i$, so it requires $O\!\left(P\right)$ operations.
\item \textbf{Compute $\mathcal{U}_n^{(-\ell)}$.} Since $\mathcal{U}_n^{(-\ell)}= \mathcal{U}_n \times_{-\ell} \lbrace \mathbf{V} \rbrace$, its computation is dominated by the cost for computing $\mathcal{U}_n \times \lbrace \mathbf{V} \rbrace$, which is $O\!\left(L K_L P\right)$ (as derived in step 1 for updating the center $\mathcal{C}$).
\item \textbf{Compute} $\left(\mathbf{U}_{n(\ell)}^{(-\ell)} \otimes \mathbf{I}_{P_\ell}\right) \mathbf{W}_{n}\mathrm{vec}(\mathbf{X}_{n(\ell)} - \mathbf{C}_{(\ell)})$ and $\left(\mathbf{U}_{n(\ell)}^{(-\ell)} \otimes \mathbf{I}_{P_\ell}\right)\mathbf{W}_{n}\left(\mathbf{U}_{n(\ell)}^{(-\ell)} \otimes \mathbf{I}_{P_\ell}\right)^{T}$.
Forming $\left(\mathbf{U}_{n(\ell)}^{(-\ell)} \otimes \mathbf{I}_{P_\ell}\right)$ costs $O\!\left(K_\ell P_\ell P\right)$. Multiplying a matrix of size $K_\ell P_\ell \times P$ with a vector of size $P$ also has $O\!\left(K_\ell P_\ell P\right)$. Since $\mathbf{W}_{n} \in \mathbb{R}^{P \times P}$ and $\left(\mathbf{U}_{n(\ell)}^{(-\ell)} \otimes \mathbf{I}_{P_\ell}\right)^{T} \in \mathbb{R}^{P \times K_\ell P_\ell}$, the total cost of the matrix multiplications is $O\!\left(K_\ell^2 P_\ell^2 P\right)$ operations. Steps 1-4 thus cost $O\!\left( P (L K_L + K_\ell^2 P_\ell^2) \right)$ for each $n$.
 
\item \textbf{Compute the pseudoinverse $\left(\Sumn \left(\mathbf{U}_{n(\ell)}^{(-\ell)} \otimes \mathbf{I}_{P_\ell}\right)\mathbf{W}_{n}\left(\mathbf{U}_{n(\ell)}^{(-\ell)} \otimes \mathbf{I}_{P_\ell}\right)^{T}\right)^{\dagger}$.} Computing the pseudoinverse of an $K_\ell P_\ell \times K_\ell P_\ell$ matrix via singular value decomposition requires $O\!\left(K_\ell^3 P_\ell^3\right)$ operations.
\item \textbf{Carry out the final multiplication.} Multiplying the $K_\ell P_\ell \times K_\ell P_\ell$ pseudoinverse by an $K_\ell P_\ell \times 1$ vector requires $O\!\left(K_\ell^2 P_\ell^2\right)$ operations.
\item \textbf{Orthogonalize via QR decomposition}. This requires $O\!\left(P_\ell K_\ell^2\right)$ computing time. 
\end{enumerate}

Since $N \geqslant K_\ell$, the time complexity for updating each $\mathbf{V}^{(\ell)}$ is $O\!\left( N P (L K_L + K_\ell^2 P_\ell^2) \right)$. Summing over all loading matrices gives a total time complexity of 
\begin{equation*}O\!\left( N P \left(L^2 K_L + \sum_{\ell=1}^L K_\ell^2 P_\ell^2\right) \right)
\end{equation*}

The time complexity for updating $ \mathcal{U}_n$ as in \eqref{eq:solU} is obtained as follows:

\begin{enumerate}
\item \textbf{Compute $\widetilde{\mathbf{W}}_n \mathrm{vec}\left(\mathcal{X}_n - \mathcal{C}\right)$.} The Hadamard product of $\left(\mathcal{X}_n - \mathcal{C}\right)$ and $\widetilde{\mathcal{W}}_{n}$ requires $O\!\left(P\right)$ operations.
\item \textbf{Compute $\left(\sbigotimes_{\ell=1}^{L} {\mathbf{V}^{(\ell)T}}\right)\mathrm{vec}\left(\left(\mathcal{X}_n - \mathcal{C}\right)\odot \widetilde{\mathcal{W}}_{n}\right)$}. 
Forming the matrix $\left(\sbigotimes_{\ell=1}^{L} \mathbf{V}^{(\ell)T}\right) \in \mathbb{R}^{K \times P}$ has cost $O\!\left( K P\right)$.  
The matrix–vector multiplication of $\left(\sbigotimes_{\ell=1}^{L} \mathbf{V}^{(\ell)T}\right)$ with $\text{vec}\!\left((\mathcal{X}_n - \mathcal{C}) \odot \widetilde{\mathcal{W}}_{n}\right) \in \mathbb{R}^{P}$ also requires $O\!\left( K P\right)$.  
\item \textbf{Compute $\left(\sbigotimes_{\ell=1}^{L} {\mathbf{V}^{(\ell)T}}\right)\widetilde{\mathbf{W}}_{n}\left(\sbigotimes_{\ell=1}^{L} {\mathbf{V}^{(\ell)}}\right)$.}
Since $\widetilde{\mathbf{W}}_{n}$ is diagonal, the matrix multiplication $\widetilde{\mathbf{W}}_{n}\left(\sbigotimes_{\ell=1}^{L} \mathbf{V}^{(\ell)}\right)$,  
with $\widetilde{\mathbf{W}}_{n} \in \mathbb{R}^{P \times P}$ and $\left(\sbigotimes_{\ell=1}^{L} \mathbf{V}^{(\ell)}\right) \in \mathbb{R}^{P \times K}$ requires $O\!\left(K P\right)$ operations. This is dominated by the multiplication of $\left(\sbigotimes_{\ell=1}^{L} \mathbf{V}^{(\ell)T}\right) \in \mathbb{R}^{K \times P}$ with $\widetilde{\mathbf{W}}_{n}\left(\sbigotimes_{\ell=1}^{L} \mathbf{V}^{(\ell)}\right) \in \mathbb{R}^{P \times K}$ which costs $O\!\left(K^2P\right)$.  
\item \textbf{Compute the pseudoinverse $\left(\left(\sbigotimes_{\ell=1}^{L} {\mathbf{V}_{q+1}^{(\ell)T}}\right)\widetilde{\mathbf{W}}_{n, q}\left(\sbigotimes_{\ell=1}^{L} {\mathbf{V}^{(\ell)}_{q+1}}\right)\right)^{\dagger}$.} 
Computing the pseudoinverse of an $K \times K$ matrix via a singular value decomposition requires $O\!\left(K^3\right)$ operations.
\item \textbf{Carry out the final multiplication.} 
Multiplying the $K \times K$ pseudoinverse by an $K \times 1$ vector requires $O\!\left(K^2\right)$ operations.
\end{enumerate}
Therefore, the time complexity for updating $\lbrace \mathcal{U}_n \rbrace$ is $O\!\left(N K^2 P\right)$.
Updating $\mathcal{C}$ as in \eqref{eq:solC} consists of the following steps:
\begin{enumerate}
\item \textbf{Compute $\mathcal{U}_{n} \times \lbrace \mathbf{V}\rbrace$.} 
To carry out a series of mode-$\ell$ products from the tensor $\mathcal{U}_n \in \mathbb{R}^{K_1 \times \cdots \times K_L}$ using the collection of matrices $\{\mathbf{V}^{(\ell)}\}$ with $\mathbf{V}^{(\ell)} \in \mathbb{R}^{P_\ell \times K_\ell}$ costs  
$O\!\left(\sum_{\ell=1}^L \left(\prod_{i \leqslant \ell} P_i\right)\!\cdot\!\left(\prod_{i \geqslant \ell} K_i\right)\right)$.  
For example, the product $\mathcal{U}_n \times_1 \mathbf{V}^{(1)}$ can be written as the matrix product $\mathbf{V}^{(1)} \mathbf{U}_{n(1)}$, with $\mathbf{V}^{(1)} \in \mathbb{R}^{P_1 \times K_1}$ and $\mathbf{U}_{n(1)} \in \mathbb{R}^{K_1 \times \prod_{i \neq 1} K_i}$, which requires $O(KP_1)$ operations.  
In general, step $\ell$ of the sequence of mode-$\ell$ products has complexity  
$O\!\left(\left(\prod_{i \leqslant \ell} P_i\right)\!\cdot\!\left(\prod_{i \geqslant \ell} K_i\right)\right)$, since at each step the dimension of the mode-$\ell$ of the core tensor changes.  
Repeating this process $L$ times yields the total cost of computing $\mathcal{U}_n \times \{\mathbf{V}\}$ as
\begin{equation*}
O\!\left( \sum_{\ell =1}^L \left(\prod_{i \leqslant \ell} P_i \right) \cdot \left(\prod_{i \geqslant \ell} K_i \right) \right) = O\!\left(L K_L P \right),
\end{equation*}
where the equality comes from the fact that $P_\ell \geqslant K_\ell$ with $\ell = 1, \ldots, L$, so  $O(L K_L P)$ serves as an upper bound.
\item \textbf{Compute $\Sumn \left({\mathcal{X}}_n  - \mathcal{U}_{n} \times \lbrace \mathbf{V}\rbrace\right) \odot \mathcal{W}_{n}$.} 
The dominant cost is the sequence of mode-$\ell$ products, 
hence, over $N$ samples the total cost is $O\!\left(NL K_LP \right)$. 
\item \textbf{Compute $\left(\Sumn \left({\mathcal{X}}_n  -\mathcal{U}_n \times \lbrace \mathbf{V}\rbrace\right) \odot \mathcal{W}_{n} \right) \odot \mathcal{H}$.} 
The computation of $\mathcal{H}$ requires $O\!\left(N P\right)$ operations. 
The elementwise product between $\left(\Sumn \left({\mathcal{X}}_n  -\mathcal{U} \times \lbrace \mathbf{V}\rbrace\right)\odot \mathcal{W}_{n} \right)$ and $\mathcal{H}$ costs $O\!\left(P\right)$. 
\end{enumerate}

Hence, the complexity for updating $\mathcal{C}$ is
$ O\!\left(N L K_LP \right)$.
Finally, updating $\mathcal{W}_n$ is an elementwise 
operation with complexity $O\!\left(NP \right)$.

Therefore, the total time complexity of the IRLS algorithm in Section~\ref{sec:algo} is
\begin{equation*}
O\!\left(NP \max \Big\lbrace K^2, L^2 K_L, \sum_{\ell=1}^L K_\ell^2 P_\ell^2 \Big\rbrace \right)
\end{equation*}

For the full ROMPCA method we need to add the time complexity of the initialisation described in Section~\ref{sec:init}, and the re-estimation of the center and core tensors given in Section~\ref{sec:restim}. FastDDC has a time complexity of $O\!\left(NP \log P\right)$.  Standard MPCA needs an additional $O\!\left(\sum_{\ell} K_\ell P_\ell^2 \right)$ for computing for each $\ell$ the $K_\ell$ leading eigenvectors of $\Phi^{(\ell)*} = \sum_{n = 1}^H \left( \mathbf{X}_{n(\ell)} - \widebar{\mathbf{X}}_{(\ell)} \right)\left( \mathbf{X}_{n(\ell)} - \widebar{\mathbf{X}}_{(\ell)} \right)^T$ that serve as initialisation for the IRLS algorithm with all weights fixed to 1. Constructing all the $\Phi^{(\ell)*}$ requires $O\!\left(N P \,\sum_{\ell} P_\ell \right)$ time. 
The computation of the second initialization candidate does not increase the overall time complexity as we use the IRLS algorithm starting from the first initialization candidate.
The cost of computing the $P$ robust M-scales $\scell$ is $O\!\left(NP\right)$ while it is $O(N)$ for the single M-scale $\hat{\sigma}_2$.

To re-estimate the center and the core tensors as
in Section~\ref{sec:restim}, the time complexity is derived as follows:
\begin{enumerate}
\item \textbf{Compute $\left(\Sumn \mathcal{X}_n \odot \mathcal{W}_n \right) \odot \mathcal{H}$.} This requires $O\!\left( NP \right)$ operations.
\item 
\textbf{Compute $\mathcal{C} + \left( \bar{\mathcal{X}}^{\mathcal{W}} - \mathcal{C}\right)\times \lbrace \mathbf{V}^T\rbrace\times \lbrace \mathbf{V}\rbrace$.}
The mode-$\ell$ products between $\left( \bar{\mathcal{X}}^{\mathcal{W}} - \mathcal{C}\right)$ and $\lbrace \mathbf{V}^{T} \rbrace$ cost $O\!\left(\sum_{\ell=1}^L \left(\prod_{i \leqslant \ell} K_i\right)\!\cdot\!\left(\prod_{i \geqslant \ell} P_i\right)\right) = O(L K_1 P)$. The mode-$\ell$ products between $\left( \bar{\mathcal{X}}^{\mathcal{W}} - \mathcal{C}\right)\times \lbrace \mathbf{V}^T\rbrace$ and $\lbrace \mathbf{V} \rbrace$ costs $O(L K_L P)$.
Therefore, the computational cost is $O\!\left(L(K_1+K_L) P \right)$, since the remaining operations are elementwise and require $O(P)$ time.
\item \textbf{Compute $\mathcal{U}_{n} - \left( \bar{\mathcal{X}}^{\mathcal{W}} - \mathcal{C}\right)\times \lbrace \mathbf{V}^T\rbrace$.} 
As this operation needs to be done for all $n = 1, \ldots, N$, the complexity is $O\!\left( NK \right)$.
\end{enumerate}

Since $K_1+K_L \leq 2N \leqslant NLK_L$, and $\log P = \sum_{\ell=1}^L \log P_\ell \leqslant \sum_{\ell=1}^L P_\ell$, the overall time complexity of ROMPCA is 
\begin{equation*}
O\!\left(NP \max \Big\lbrace K^2, L^2 K_L, \sum_{\ell=1}^L K_\ell^2 P_\ell^2,  \Big\rbrace \right)
\end{equation*}
If we further assume that $K_1 = K_2 = \cdots = K_L = \widetilde{K}$ and $P_1 = P_2 = \cdots = P_L = \widetilde{P}$, the time complexity simplifies to
\begin{equation*}
O\!\left(N\widetilde{P}^L \max \Big\lbrace \widetilde{K}^{2L}, L^2 \widetilde{K}, L \widetilde{K}^2 \widetilde{P}^2  \Big\rbrace \right).
\end{equation*}

For the space complexity of ROMPCA we need $O\!\left(NP+ \sum_{\ell=1}^{L}P_\ell K_\ell+ NK\right) = O\!\left(NP \right)$ for storing the original tensors, the weight matrices, the loading matrices and the core tensors. The space complexity for updating $\{\mathbf{V}^{(\ell)}\}$ as in \eqref{eq:solV} is dominated by the computation of $\left(\mathbf{U}_{n(\ell)}^{(-\ell)} \otimes \mathbf{I}_{P_\ell}\right)\mathbf{W}_{n}\left(\mathbf{U}_{n(\ell)}^{(-\ell)} \otimes \mathbf{I}_{P_\ell}\right)^{T}$. To store $\mathbf{U}_{n(\ell)}^{(-\ell)}$ we need only $O\!\left(P\right)$ additional storage since all projection matrices are already stored from previous steps and all $L-2$ intermediate matrices of $O\!\left(P\right)$ are discarded. 
To build the matrix $\left(\mathbf{U}_{n(\ell)}^{(-\ell)} \otimes \mathbf{I}_{P_\ell}\right) \in \mathbb{R}^{K_\ell P_\ell \times P}$, the space complexity is $O\!\left(K_\ell P_\ell P\right)$. 
To obtain the intermediate $\mathbf{W}_{n} \left(\mathbf{U}_{n(\ell)}^{(-\ell)} \otimes \mathbf{I}_{P_\ell}\right)^{T} \in \mathbb{R}^{P \times K_\ell P_\ell}$,  the  space complexity is  $O\!\left(K_\ell P_\ell P\right)$. 
The final matrix $\left(\mathbf{U}_{n(\ell)}^{(-\ell)} \otimes \mathbf{I}_{P_\ell}\right)\mathbf{W}_{n}\left(\mathbf{U}_{n(\ell)}^{(-\ell)} \otimes \mathbf{I}_{P_\ell}\right)^{T} \in \mathbb{R}^{K_\ell P_\ell \times K_\ell P_\ell}$ has space complexity $O\!\left(K_\ell^2 P_\ell^2\right)$. 
Hence, the additional space complexity for updating all $\{\mathbf{V}^{(\ell)}\}$ is
$O\!\left(\max_{1 \leqslant \ell \leqslant L }\!\left\{K_\ell P_\ell P,\ K_\ell^2 P_\ell^2\right\}\right)$ since the matrices computed for the update of each $\mathbf{V}^{(\ell)}$ don't need to be stored.  

The space complexity for updating $\{\mathcal{U}_n\}$ as in \eqref{eq:solU} is dominated by computing 
$\left(\sbigotimes_{\ell=1}^{L} \mathbf{V}^{(\ell)T}\right)\widetilde{\mathbf{W}}_{n}\left(\sbigotimes_{\ell=1}^{L} \mathbf{V}^{(\ell)}\right)$.
To build $\left(\sbigotimes_{\ell=1}^{L} \mathbf{V}^{(\ell)}\right)\in\mathbb{R}^{P\times K}$, the space complexity is $O(PK)$, while to obtain
$\left(\sbigotimes_{\ell=1}^{L} \mathbf{V}^{(\ell)T}\right)\widetilde{\mathbf{W}}_{n}\left(\sbigotimes_{\ell=1}^{L} \mathbf{V}^{(\ell)}\right)\in\mathbb{R}^{K\times K}$, it is  $O(KP+K^2)=O(KP)$.

The space complexity for updating $\mathcal{C}$ as in \eqref{eq:solC} is dominated by computing $\mathcal{U}_n \times \lbrace \mathbf{V} \rbrace$ which needs $O\!\left(P\right)$ additional storage. 

The initialization and re-estimation of the center and core tensors does not add space complexity as FastDDC requires $O\!\left(NP\right)$ space. The total space complexity of ROMPCA is thus
\begin{equation*}O\!\left(NP + PK+\max_{1 \leqslant \ell \leqslant L }\!\left\{K_\ell P_\ell P,\ K_\ell^2 P_\ell^2\right\}\right).
\end{equation*}
If we further assume that $K_1 = K_2 = \cdots = K_L = \widetilde{K}$ and $P_1 = P_2 = \cdots = P_L = \widetilde{P}$, the space complexity simplifies to
\begin{equation*}
O\!\left(N\widetilde{P}^L \;+\; \widetilde{P}^L \widetilde{K}^L + \widetilde{K} \widetilde{P} \widetilde{P}^L \right) = O\!\left(\widetilde{P}^L  \max\left\{N, \widetilde{K}^L,\widetilde{P} \widetilde{K} \right\} \right)
\end{equation*}

\newpage
\section{The Rank Determination Method}
\label{app:addrank}
\begin{figure}[!ht]
\begin{tabular}{cc}
\includegraphics[width=0.45\textwidth]{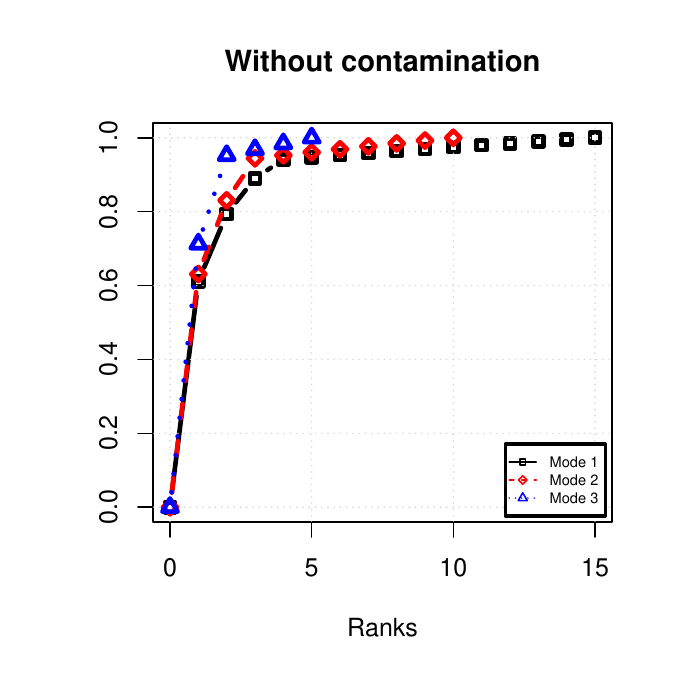} &
\includegraphics[width=0.45\textwidth]{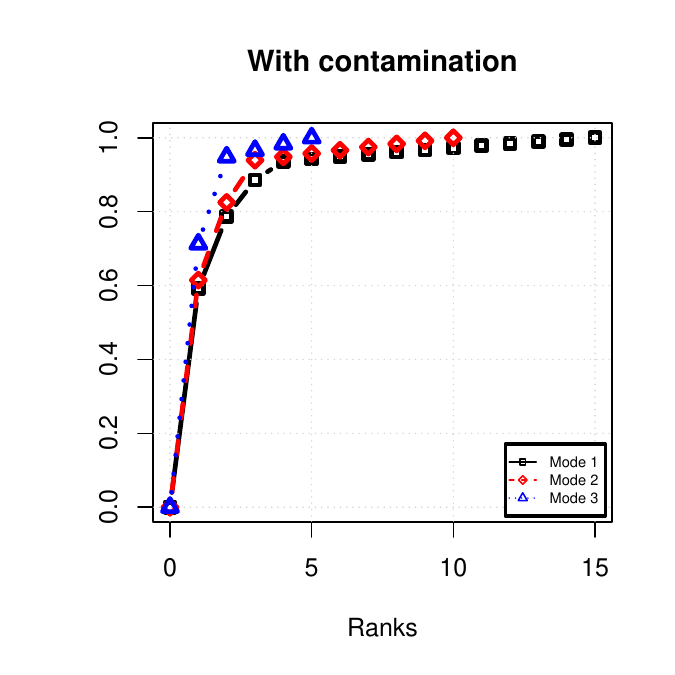} %
\end{tabular}
\vspace{-0.5cm}
\caption{Cumulative eigenvalues at an uncontaminated dataset (left) and at a contaminated dataset (right).}
\label{fig:Simulation_rankdet}
\end{figure}

To illustrate the performance of the rank determination method proposed in Section~\ref{sec:rank}, we consider a dataset generated as in Section~\ref{sec:simu}  with $(P_1, P_2, P_3) = (15, 10, 5)$ and $(K_1,K_2,K_3) = (4,3,2)$. The cumulative eigenvalues obtained from $\{ \mathcal{X}_n \}$ are plotted in the left panel of Figure~\ref{fig:Simulation_rankdet}. We clearly see that the cumulative variance in each mode stabilizes at the corresponding dimensions of the core tensor. 

Next we contaminate the data with cellwise and casewise outliers according to the procedure outlined in Section~\ref{sec:rank}, with $\gamma_{\text{cell}}=5$. The cumulative eigenvalues derived from $\{\mathcal{X}^{\text{\tiny DDC}}_h\}$ are shown in the right panel of Figure~\ref{fig:Simulation_rankdet} and are almost identical to the those of the left panel. 
This illustrates the effectiveness of our proposed rank determination method.

\newpage
\section{Additional Simulation Results}
\label{app:addsim}
\begin{figure}[!ht]
    \centering
    \includegraphics[width=.32\textwidth]{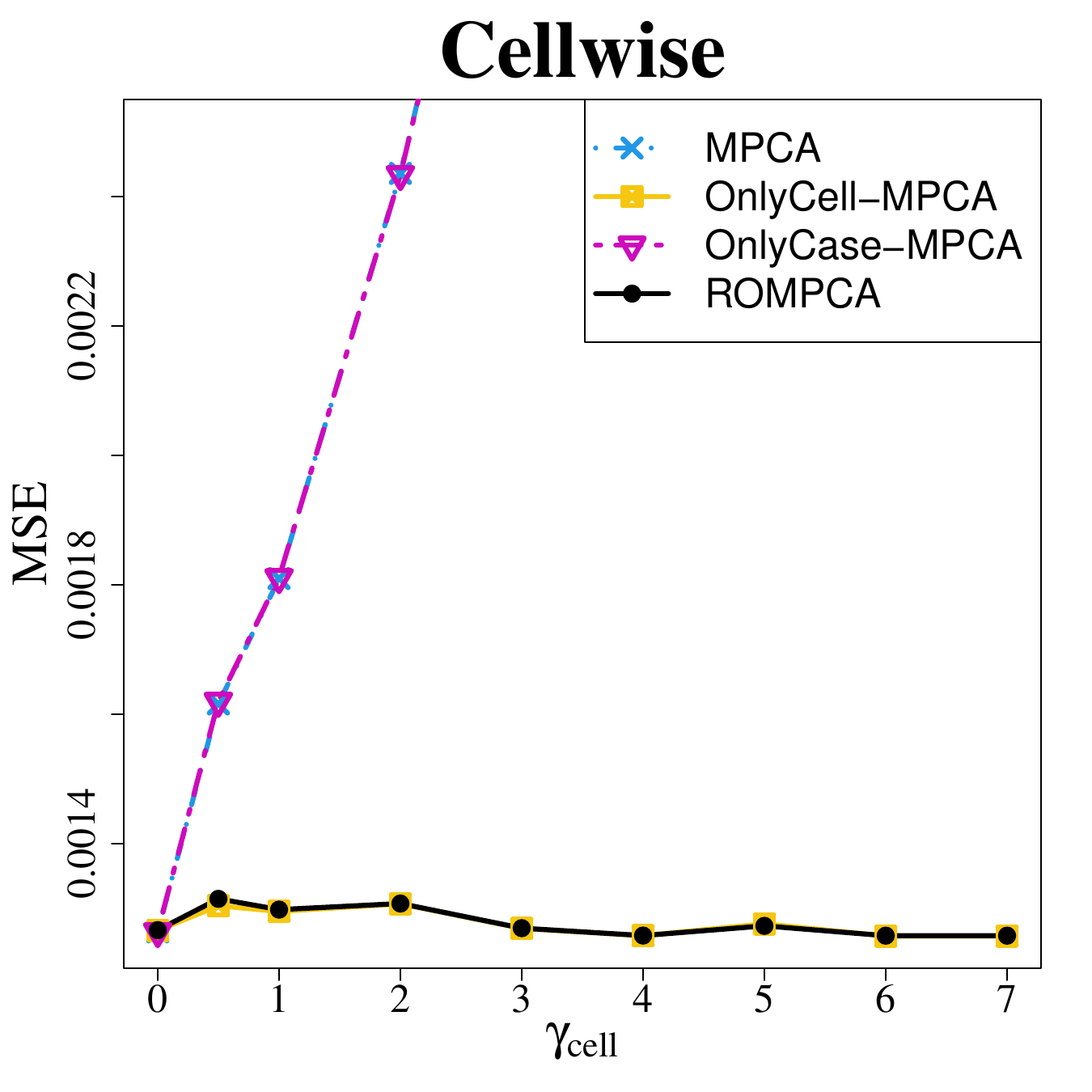} %
    \includegraphics[width=.32\textwidth]{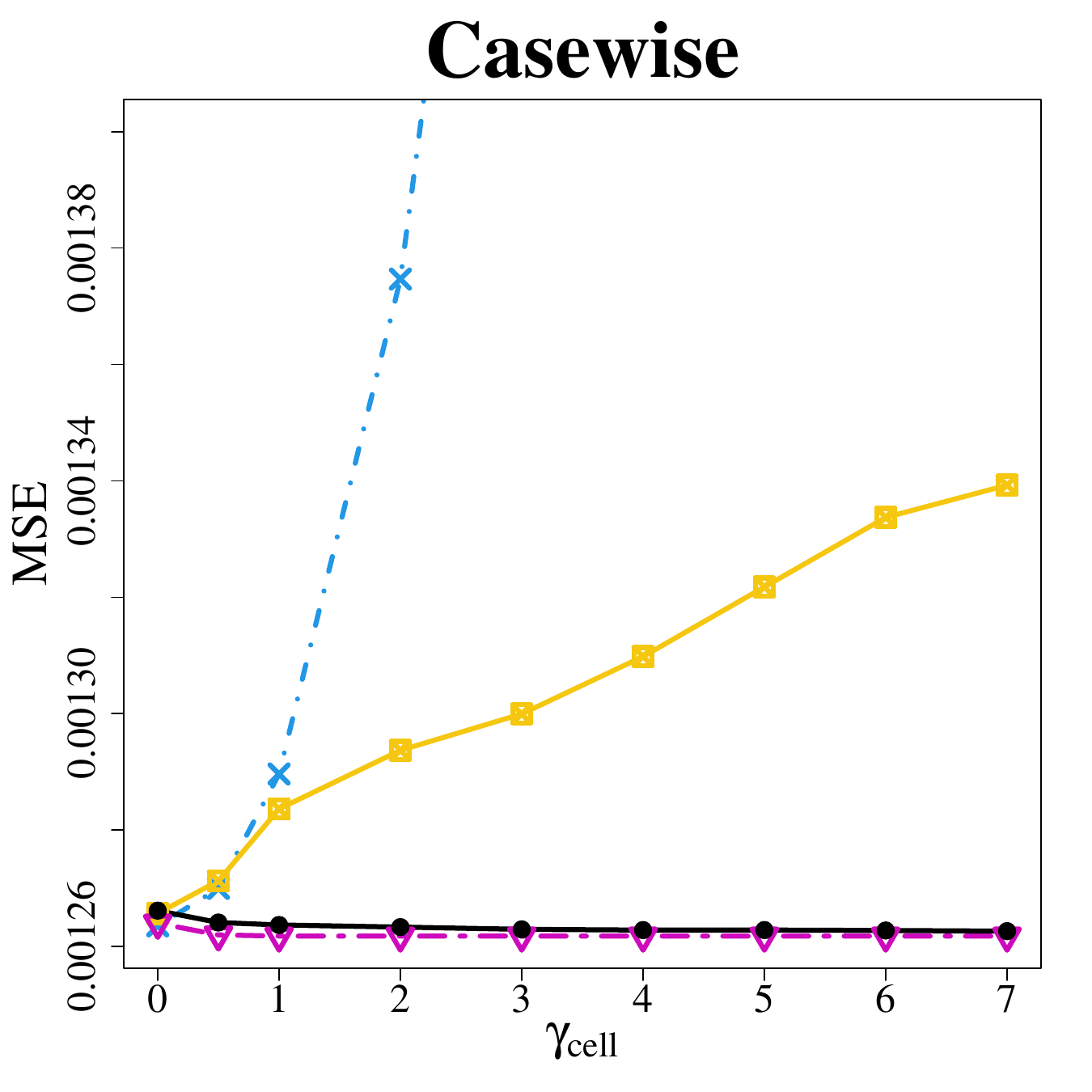} %
    \includegraphics[width=.32\textwidth]{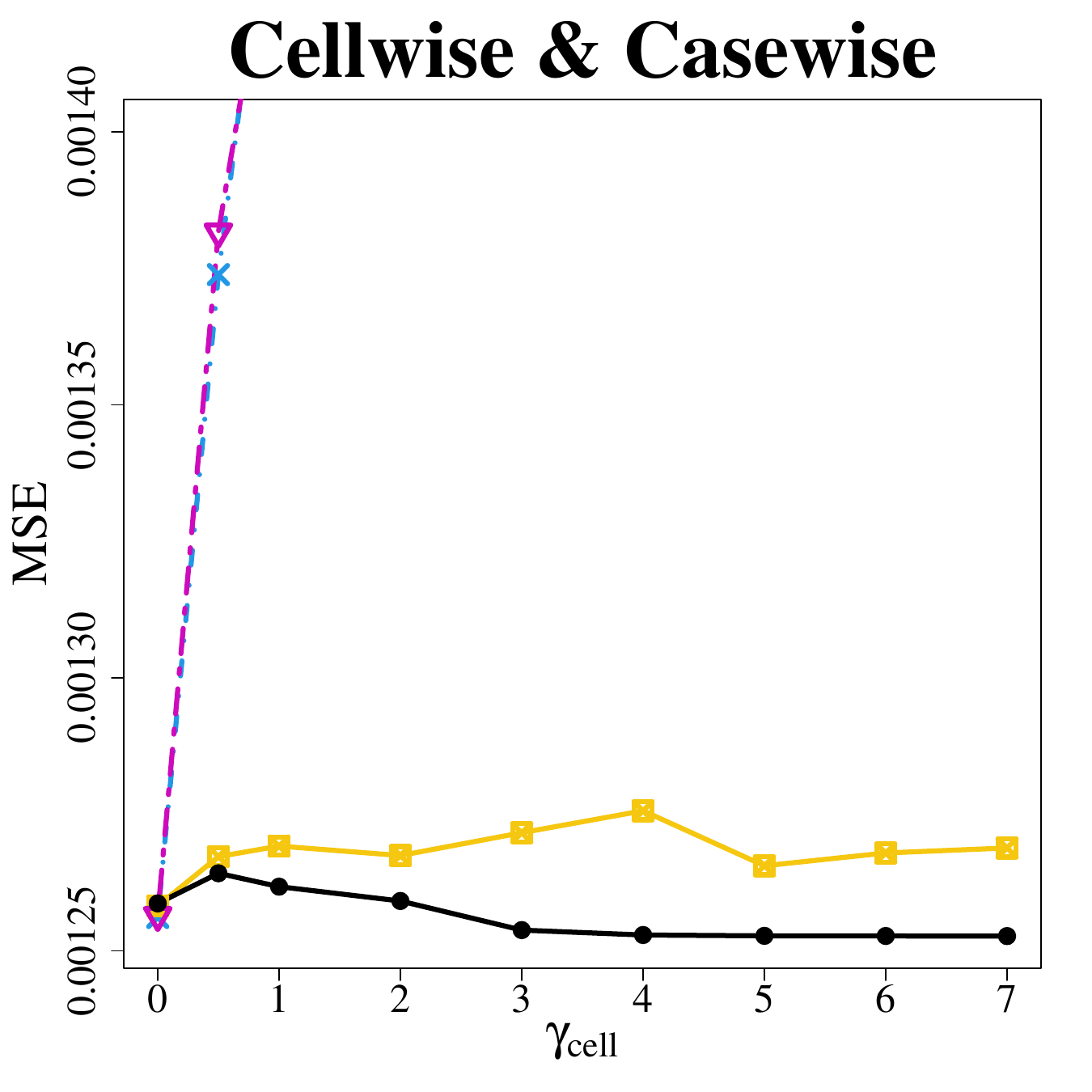} %
    \caption{Mean MSE attained by MPCA, OnlyCase-MPCA, OnlyCell-MPCA, and ROMPCA  for the setting $(P_1, P_2, P_3) = (15, 10, 5)$ and $(K_1, K_2, K_3) = (4, 3, 2)$ under cellwise contamination, casewise contamination or both, in function of $\gamma_{\text{cell}}$ for data without missing values.}%
    \label{fig:simulMSE_A09_low}%
\end{figure}

\begin{figure}[!ht]
    \centering
    \includegraphics[width=.32\textwidth]{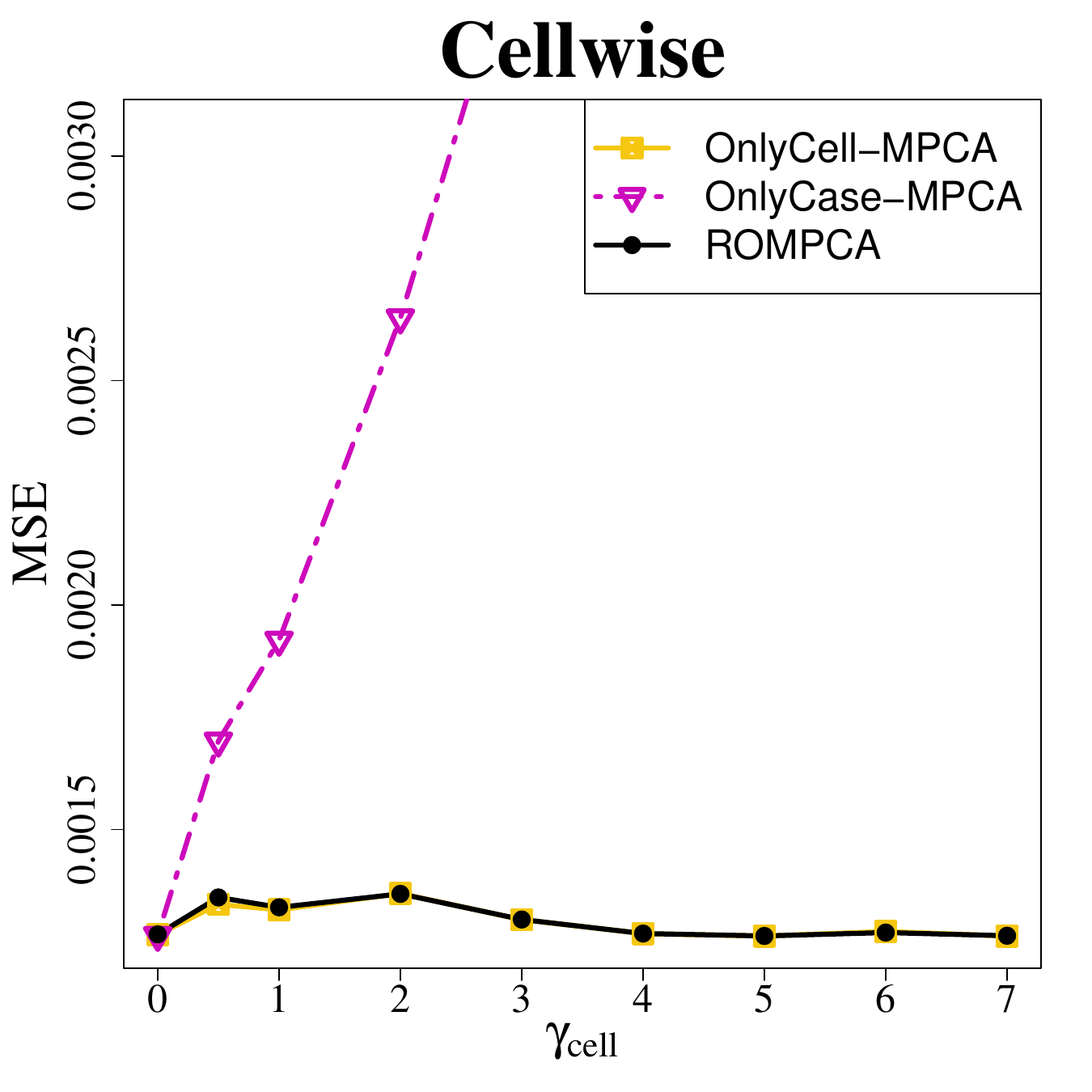} %
    \includegraphics[width=.32\textwidth]{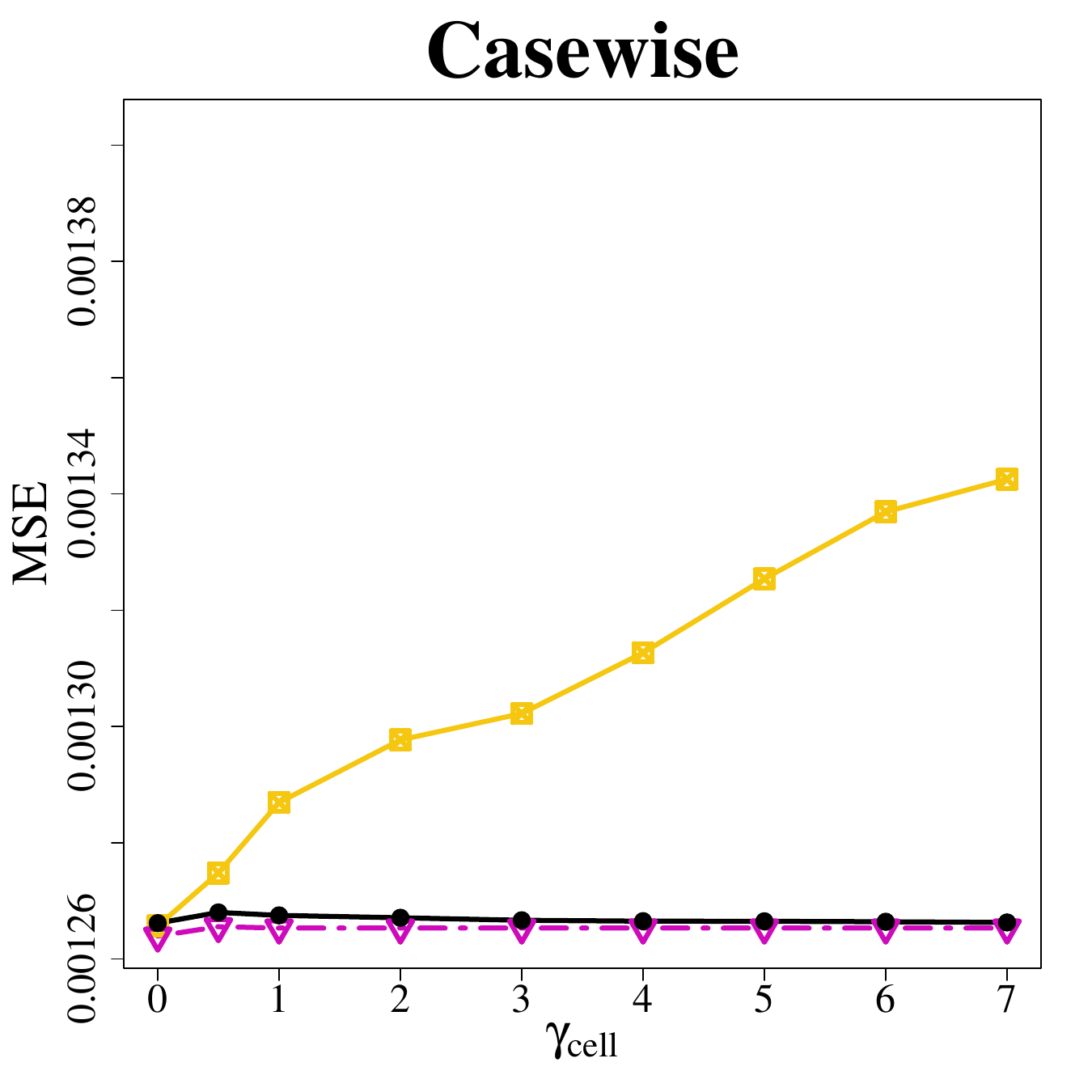} %
    \includegraphics[width=.32\textwidth]{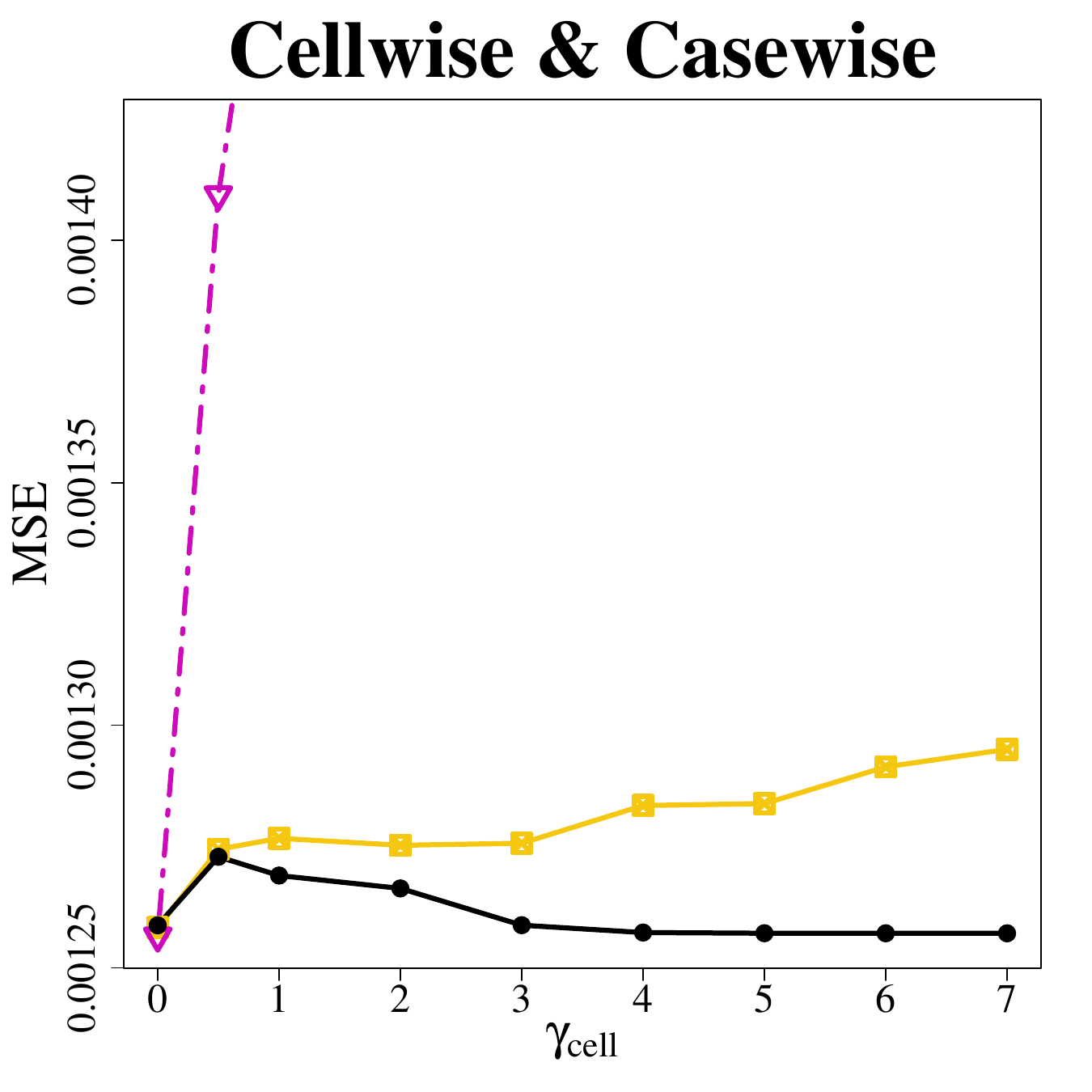} %
    \caption{Mean MSE attained by OnlyCase-MPCA, OnlyCell-MPCA, and ROMPCA  for the setting $(P_1, P_2, P_3) = (15, 10, 5)$ and $(K_1, K_2, K_3) = (4, 3, 2)$ under cellwise contamination, casewise contamination or both, in function of $\gamma_{\text{cell}}$ for data with $10\%$ of missing values.}%
    \label{fig:simulMSE_A09_NA_low}%
\end{figure}

Here, we provide the results for the lower dimensional setting 
for $(P_1, P_2, P_3) = (15, 10, 5)$ and $(K_1, K_2, K_3) = (4, 3, 2)$.
Figures~\ref{fig:simulMSE_A09_low} and~\ref{fig:simulMSE_A09_NA_low} show the mean MSE for the three contamination scenarios in absence and presence of missing values, respectively. 
The resulting curves are very similar to the results in Section~\ref{sec:simu} of the paper, leading to the same conclusions.

\begin{figure}[!ht]
    \centering
    \includegraphics[width=.4\textwidth]{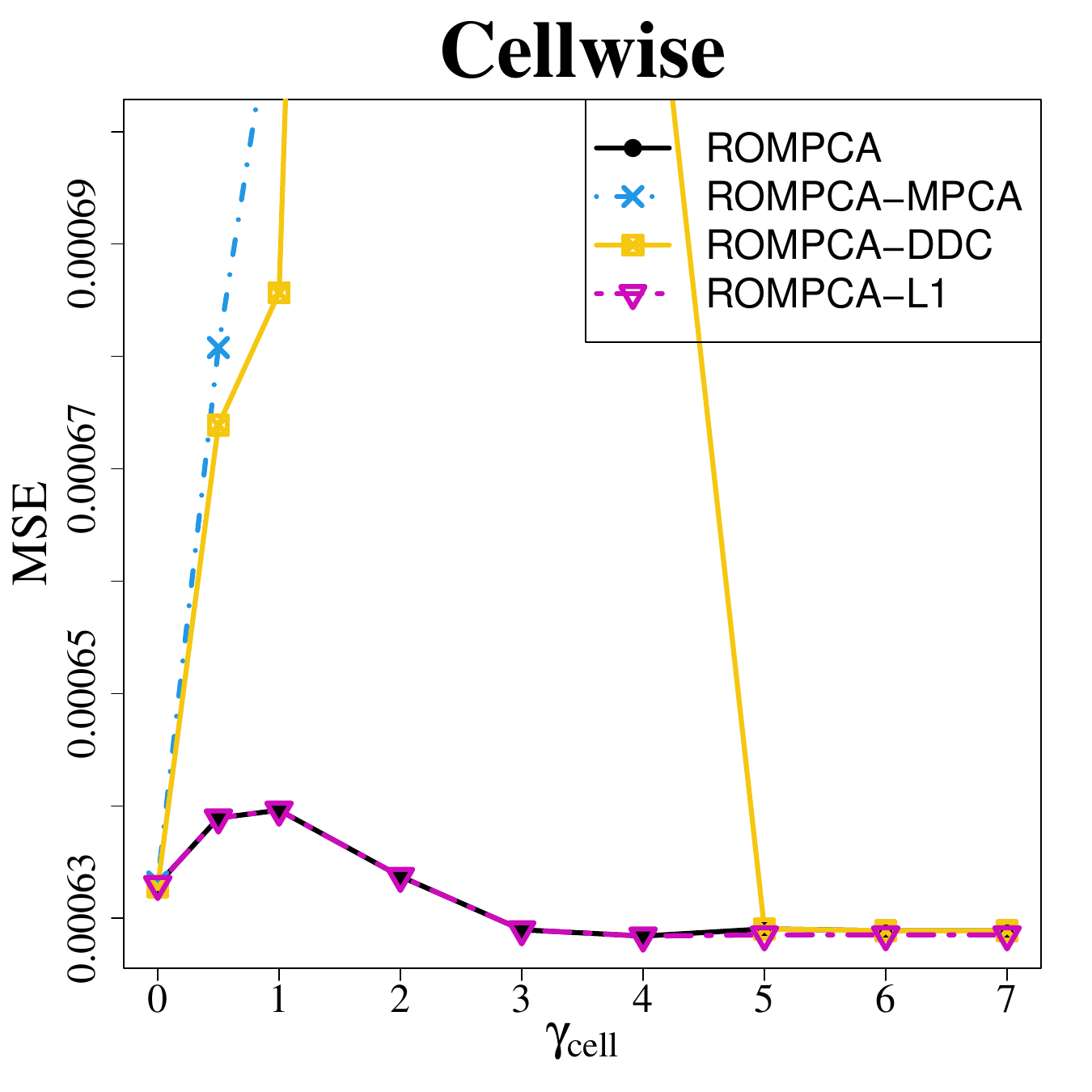} %
    \includegraphics[width=.4\textwidth]{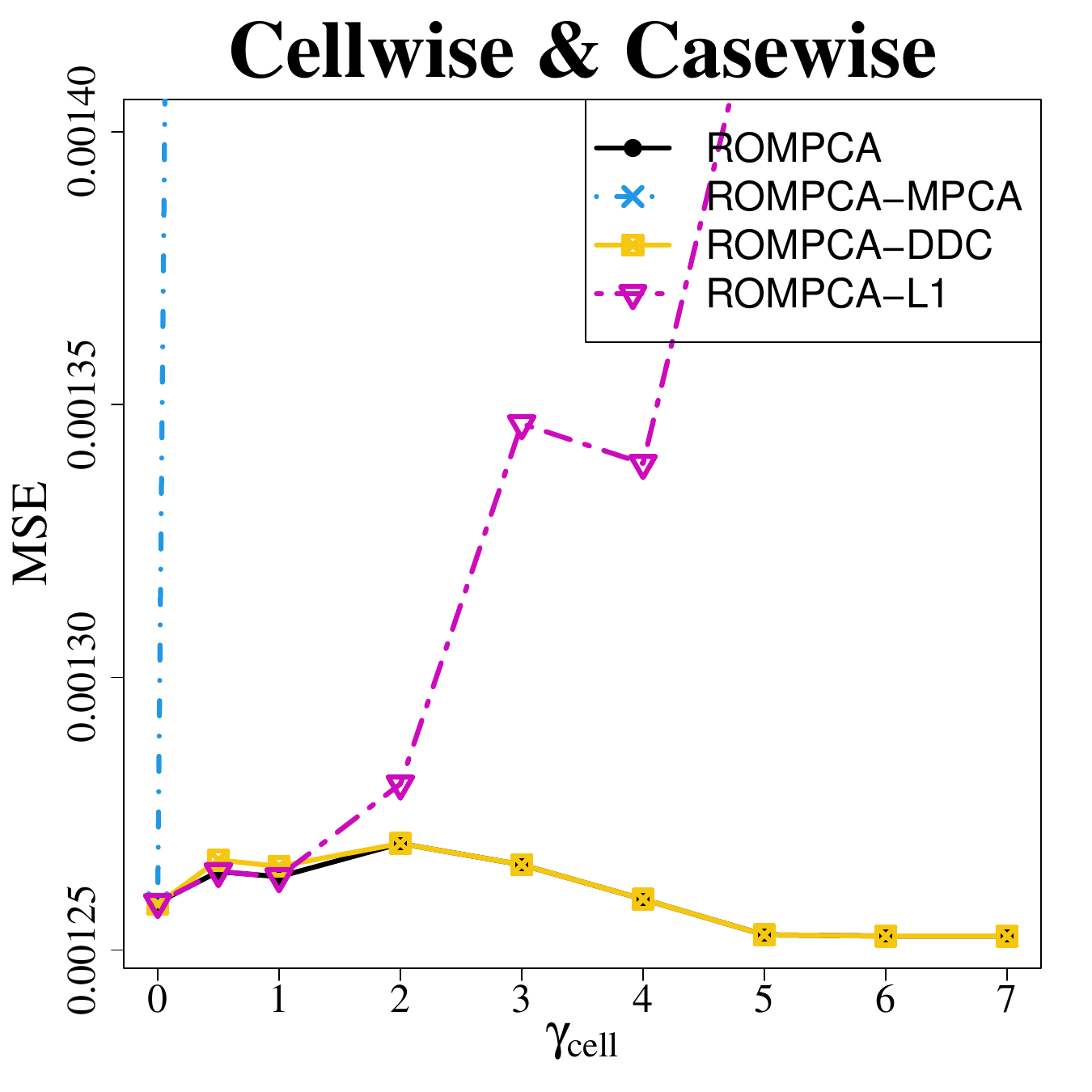} %
    \vspace{-3mm} 
    \caption{Mean MSE obtained by ROMPCA, ROMPCA-MPCA, ROMPCA-DDC, and ROMPCA-L1 for the settings $(P_1, P_2, P_3) = (30, 20, 5)$ and $(K_1, K_2, K_3) = (8, 3, 2)$ under cellwise contamination, and $(P_1, P_2, P_3) = (15, 10, 5)$ and $(K_1, K_2, K_3) = (4, 3, 2)$ in the presence of both casewise and cellwise outliers, as a function of $\gamma_{\text{cell}}$, for data without missing values.}

    \label{fig:simulMSE_ccase10_init_low}%
\end{figure}

Figure \ref{fig:simulMSE_ccase10_init_low} illustrates the sensitivity of the proposed ROMPCA method to the initialization strategy. The plots display the mean MSE under cellwise contamination and in the presence of both casewise and cellwise outliers, without missing values, for ROMPCA initialized in several ways. ROMPCA-MPCA refers to ROMPCA initialized with standard MPCA, while ROMPCA-DDC and ROMPCA-L1 correspond to using as initializations the first and second candidate procedures, respectively, as described in Section~\ref{sec:init}. 
As expected, ROMPCA initialized with standard MPCA performs the worst, since this initialization is not robust to outliers. Under cellwise contamination, the second initialization tends to perform the best, and ROMPCA typically selects it. When both outlier types are present, the best performance is obtained with the first candidate initialization. The original ROMPCA performs best overall, as it adaptively selects the most suitable initialization among the first and second candidates.

\begin{figure}[!ht]
    \centering
    \includegraphics[width=.4\textwidth]{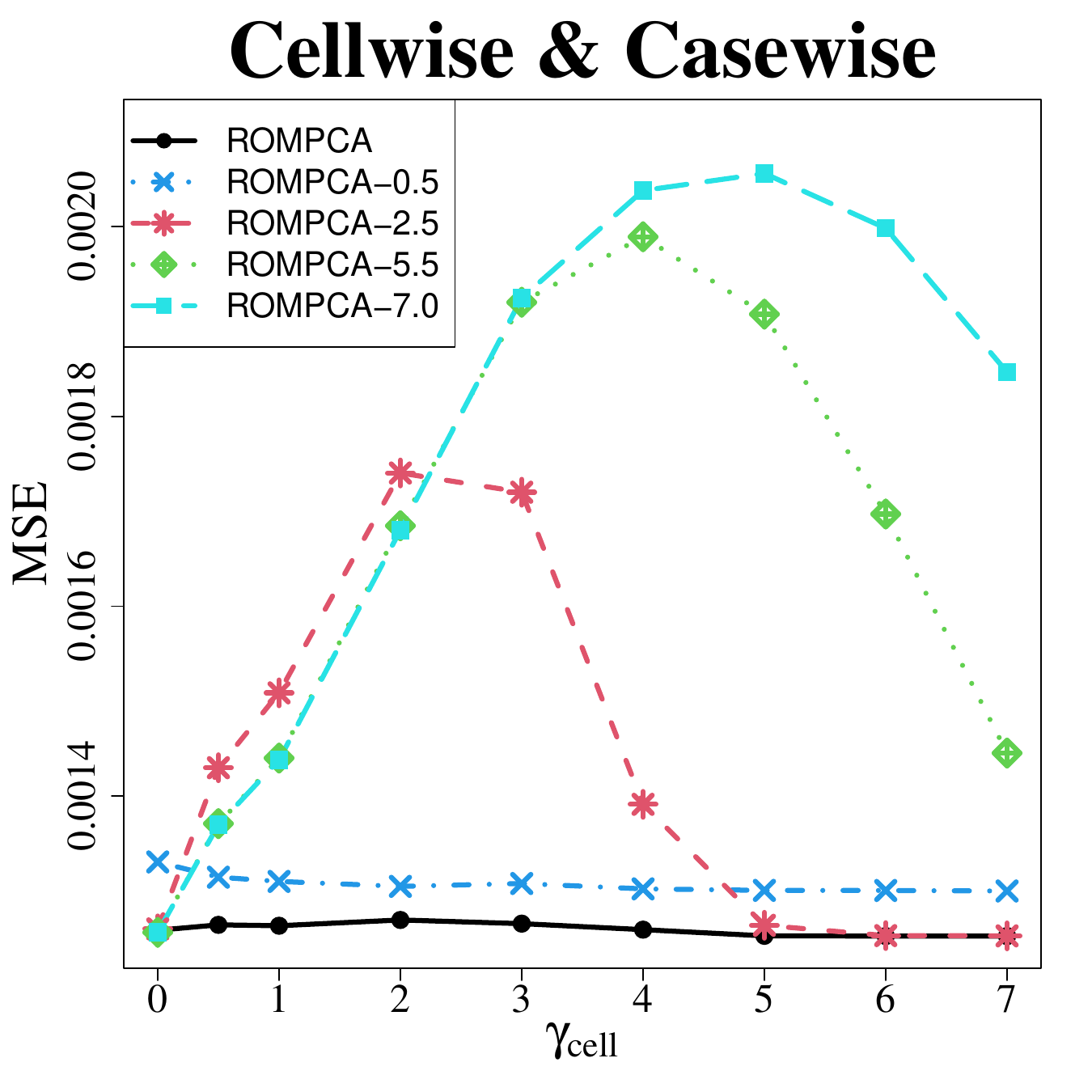} %
    \caption{Mean MSE attained by ROMPCA, ROMPCA-0.5, ROMPCA-2.5, ROMPCA-5.5  and ROMPCA-7.0  for the setting $(P_1, P_2, P_3) = (15, 10, 5)$ and $(K_1, K_2, K_3) = (4, 3, 2)$ under casewise and cellwise contamination, as a function of $\gamma_{\text{cell}}$, for data without  missing values.}%
    \label{fig:simulMSE_tune_low}%
\end{figure}

\newpage
Figure~\ref{fig:simulMSE_tune_low} illustrates the sensitivity of the proposed ROMPCA method to the choice of tuning constants $(b,c)$ in the hyperbolic tangent $\rho$-functions used in~\eqref{eq:obj}. The plots report the mean MSE in the presence of casewise and cellwise outliers, without missing values, for ROMPCA under different choices of the tuning constants. To assess the impact of these parameters, we compare the default tuning constants $b = 1.5$ and $c = 4$ with alternative values. To facilitate the comparison, we set $c = b \cdot (4/1.5)$, so that only $b$ needs to be adjusted.  ROMPCA-0.5 corresponds to $b = 0.5$, ROMPCA-2.5 to $b = 2.5$, ROMPCA-5.5 to $b = 5.5$, and ROMPCA-7.0 to $b = 7.0$.

ROMPCA achieves the best performance overall, as its choice of tuning constants provides an effective balance between robustness and efficiency. ROMPCA-0.5 also performs well due to its high robustness. For this choice of tuning constants, a larger number of cells and cases are downweighted in the algorithm. However, this comes at the cost of efficiency at uncontaminated data. In contrast, the other variants become more efficient as the parameters $b$ and $c$ increase, since a greater proportion of cells and cases are treated as regular. This gain in efficiency, however, is accompanied by a loss of robustness, as the MSE decreases only for higher values of $\gamma_{\text{cell}}$, while remaining relatively large for smaller contamination levels.

\begin{figure}[!ht]
    \centering
    \includegraphics[width=.4\textwidth]{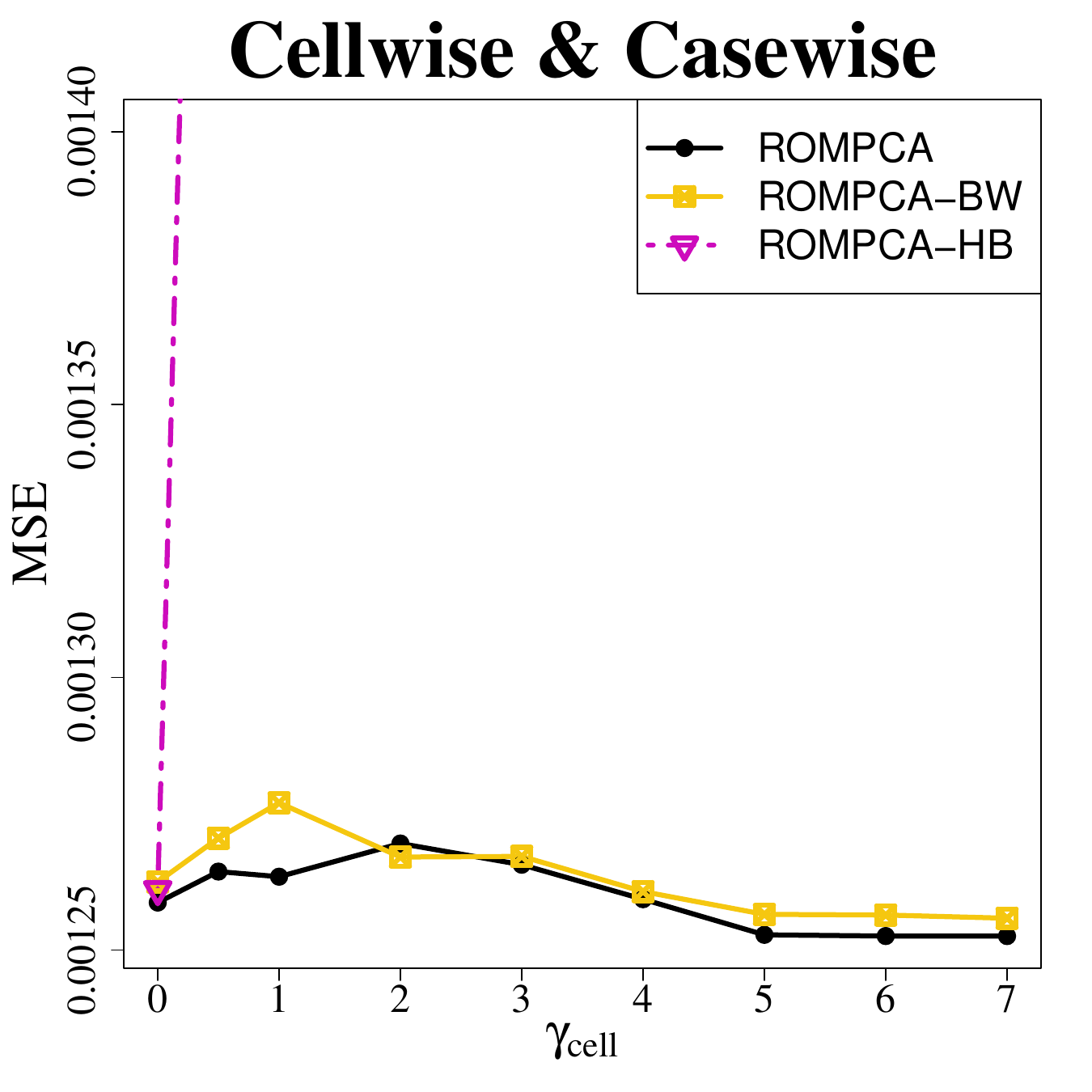} %
    \vspace{-3mm}
    \caption{Mean MSE attained by ROMPCA, ROMPCA-BW, and ROMPCA-HB for the setting $(P_1, P_2, P_3) = (15, 10, 5)$ and $(K_1, K_2, K_3) = (4, 3, 2)$ under cellwise and casewise contamination, as a function of $\gamma_{\text{cell}}$, for data without  missing values.}%
    \label{fig:simulMSE_rho_low}%
\end{figure}

Figure~\ref{fig:simulMSE_rho_low} assesses the performance of ROMPCA under different choices of the $\rho$-function in~\eqref{eq:obj}. ROMPCA uses the hyperbolic tangent $\rho$-function.
ROMPCA-HB uses the \textit{Huber} $\rho$-function \citep{Huber:RobStat},
\begin{equation*}
\rho_{\text{HB}}(x) = 
\begin{cases}
x^2/2 & \quad \text{if } |x| \leqslant c,\\[4pt]
c(|x| - c/2) & \quad \text{if } |x| > c,
\end{cases}
\end{equation*}
with $c=0.732$, while
 ROMPCA-BW uses  \textit{Tukey's biweight} $\rho$-function \citep{maronna2019robust},
\begin{equation*}
\rho_{\text{BW}}(x) = 
\begin{cases}
d^2/6\!\left[1 - \left(1 - \left(x/d\right)^2\right)^{\!3}\right] & \quad \text{if } |x| \leqslant d,\\[4pt]
d^2/6 & \quad \text{if } |x| > d
\end{cases}
\end{equation*}
with $d=3.44$, both attaining 85\% efficiency at the Gaussian distribution.

The Huber $\rho$-function is not redescending, meaning that although it reduces the influence of moderate outliers, extreme outliers can still affect the estimation, as shown in Figure~\ref{fig:simulMSE_rho_low}. 
In contrast, Tukey's biweight $\rho$-function is redescending and therefore behaves similarly to the hyperbolic tangent $\rho$-function. 
Although Tukey's biweight and hyperbolic tangent $\rho$-functions yield similar performance, the latter is preferred because it is theoretically optimal among redescending M-estimators. 
It is specifically designed to maximize efficiency for a given level of robustness, according to the change-of-variance curve criterion introduced by \cite{Hampel:CVC}.

\clearpage

\section{Additional Results for the Dog Walker Data}
\label{app:addDogRes}
\begin{figure}[!ht]
\begin{tabular}{cc}
\includegraphics[width=0.45\textwidth]{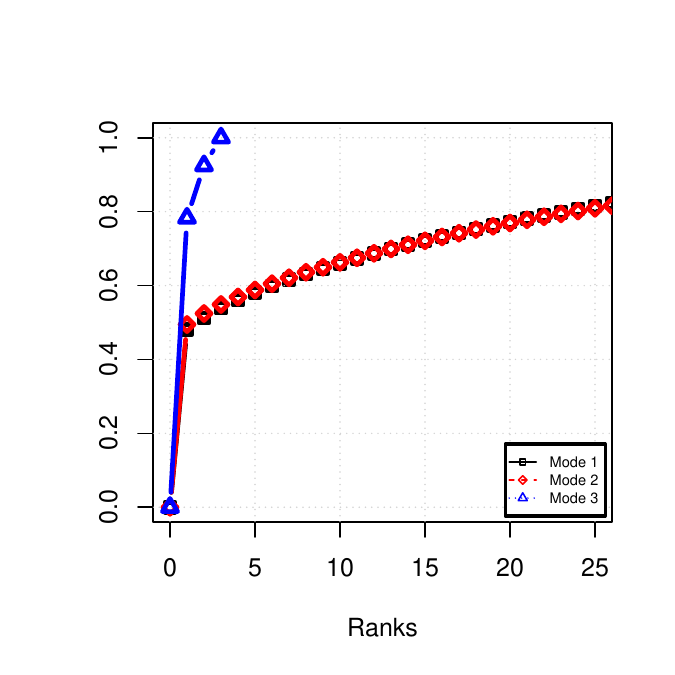} &
\includegraphics[width=0.45\textwidth]{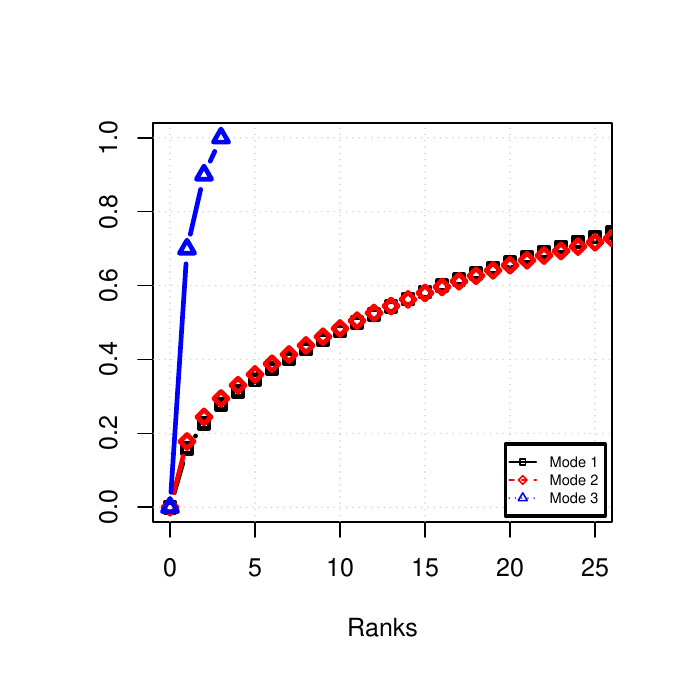} %
\end{tabular}
\vspace{-0.5cm}
\caption{Comparison of the cumulative eigenvalues for the Dog Walker data using ROMPCA (left) and MPCA (right).}
\label{fig:DogWalker_EigenCumDist}
\end{figure}

To reduce the dimensionality of the Dog Walker data, the rank of the projection matrices must first be determined. The ranks for ROMPCA are identified as $(1,1,1)$ based on the plot of the cumulative  eigenvalues for each mode in the left frame of Figure~\ref{fig:DogWalker_EigenCumDist}. The ranks $(1,1,1)$ suggest that the center already explains a large portion of the data variability. However, additional variability remains, which is captured in the core tensors $\lbrace \mathcal{U}_n \rbrace$. As shown in Figure~\ref{fig:DogWalker_Scores_ROMPCA} the set of core tensors reveal a non-constant pattern. This indicates the presence of variability that cannot be fully explained using only the center.

\vspace{-1cm}

\begin{figure}[!ht]
\begin{center}
    \includegraphics[width=1\textwidth]{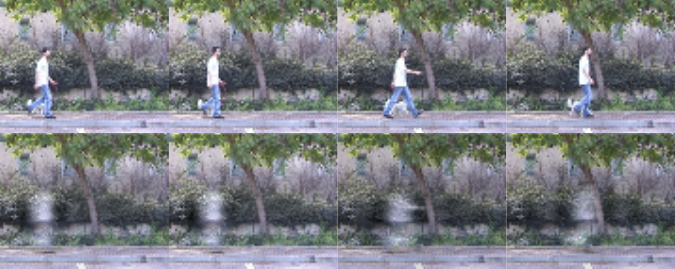}
\end{center}
\vspace{-3mm}
\caption{Frames from the Dog Walker data (upper row) and reconstructed data frames using MPCA (lower row).}
\label{fig:DogWalker_Reconst_DataFrames_MPCA}
\end{figure}

The right frame of Figure~\ref{fig:DogWalker_EigenCumDist} shows the cumulative eigenvalues of MPCA. We choose the ranks as $(9,9,1)$ to match the variance explained from the robust procedure, and apply MPCA. Figure~\ref{fig:DogWalker_Reconst_DataFrames_MPCA} compares some of the original frames with the reconstructed frames from MPCA. We clearly see that the background is not well separated from the walking man.

\begin{figure}[!ht]
    \centering
     \includegraphics[width=.9\textwidth]{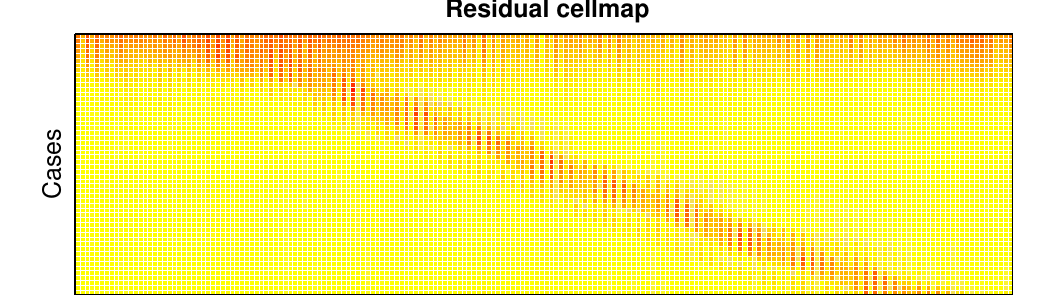}
      \includegraphics[width=.9\textwidth]{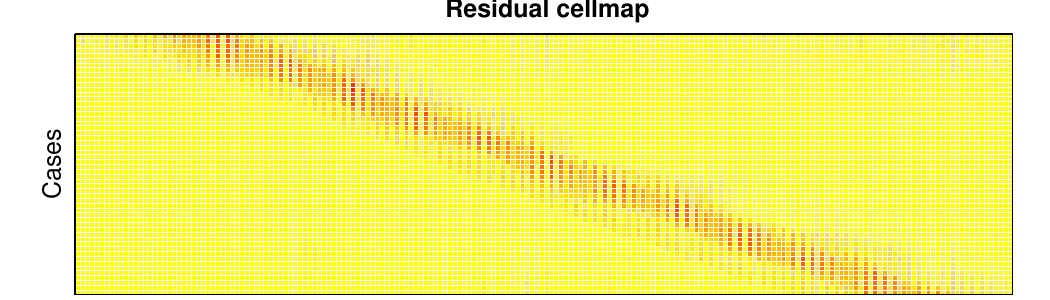}
    \caption{Residual cellmap from DDC (top) and ROMPCA (bottom).}
    \label{fig:DogWalker_cellmap_comp}
\end{figure}

The dog walker example allows us to illustrate the added value of using a multilinear approach. Figure~\ref{fig:DogWalker_cellmap_comp} shows two residual cellmaps using $100$ aggregated columns. The top panel is the residual cellmap obtained from DDC applied to the vectorized frames, while the bottom panel is the ROMPCA residual cellmap. In contrast to ROMPCA, the multivariate DDC method flags nearly all cells in the first frames as outliers. It is thus not able to capture the variation in the illumination. 

\clearpage

\section{Additional Results for the Dorrit Data}
\label{app:addDorRes}
\begin{figure}[!ht]
\begin{center}
\begin{tabular}{cc}
\includegraphics[width=0.5\textwidth]{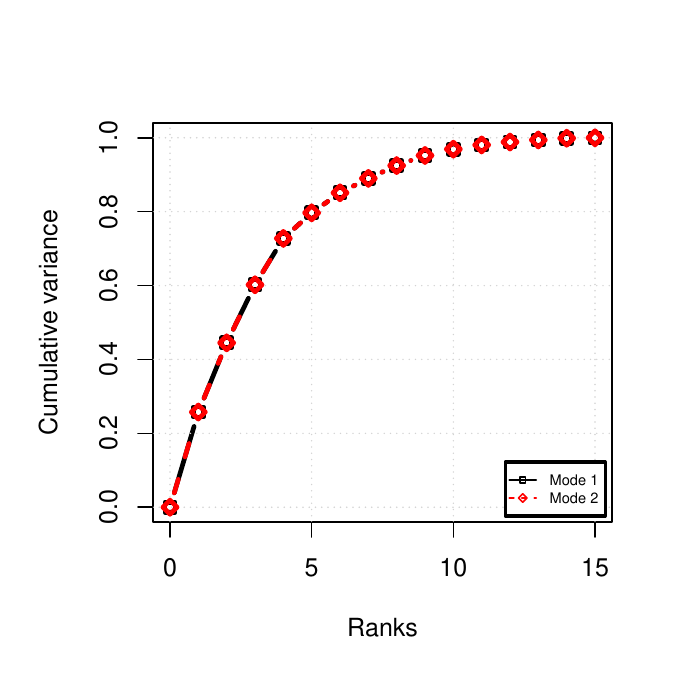} &
\includegraphics[width=0.5\textwidth]{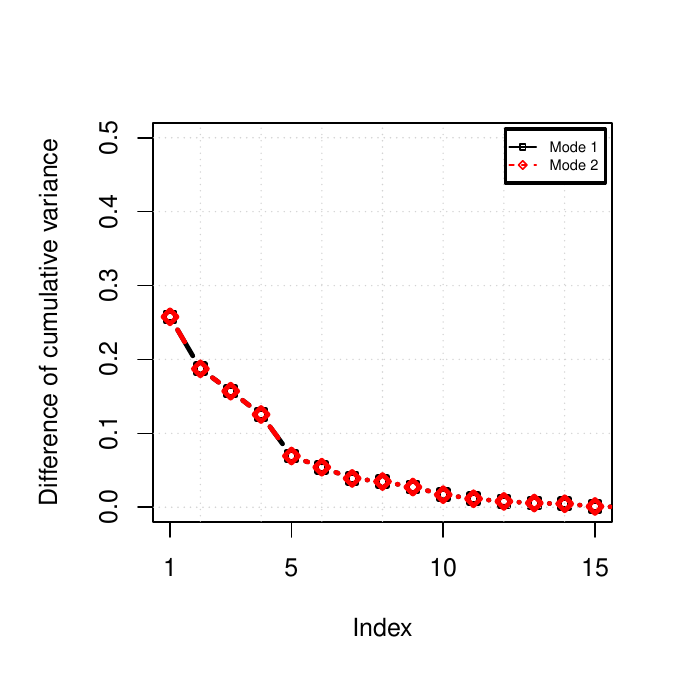} %
\end{tabular}
\end{center}
\vspace{-5mm}
\caption{Cumulative eigenvalues (left) and eigenvalues (right) using DDC imputation on the Dorrit data.}
\label{fig:Dorrit_EigenCum}
\end{figure}

Applying the elbow rule to the plot of the cumulative eigenvalues in the left panel of Figure~\ref{fig:Dorrit_EigenCum} suggests selecting the ranks \((K_1, K_2) = (4,4)\). This choice is motivated by the fact that the increase in cumulative variance is most pronounced for the first four ranks, after which the gains become substantially smaller. To further support this choice, the right panel of Figure~\ref{fig:Dorrit_EigenCum} displays the differences in cumulative variance (i.e.\ the eigenvalues) explained by successive ranks. These differences are relatively large for the first four components and decay from the fifth component onward. This is in line with the ranks used in \cite{Heng:RobustTensor} in their data application for the Dorrit data.

\newpage
\section{Discussion About Temporal Dependence and Computational Scalability}
\label{app:disc}
In some applications, tensor data exhibit an inherent temporal structure. In such cases, the assumption that the $N$ tensors are independent may be too restrictive. For example, in Section~\ref{sec:realdata}, it is reasonable to assume that the frames from the Dog Walker data are temporally correlated. 
Although our method performed well in this setting, it could be further adapted to directly address datasets exhibiting temporal dependencies.
Indeed, such dependences could lead to inferential issues, particularly in the anomaly detection stage, where some cases or cells might be wrongly flagged as outliers or incorrectly considered as regular observations.

In the context of MPCA, datasets with a temporal mode have also been studied \citep{Fartash:SpeechMPCA, Wu:MPCAnet}. For instance, in \cite{Lu:MPCA}, within their gait recognition application, the data are structured such that one mode corresponds to time. Each gait sequence is modeled as a third-order tensor whose three modes represent the spatial column, spatial row, and temporal dimensions.
The idea is to segment the walking sequence into individual cycles and treat each cycle as one sample. By doing so, the periodic motion inherent in walking is removed from the sample mode, allowing the samples to be treated as independent. The MPCA algorithm then performs dimensionality reduction along all three modes while keeping the set of samples (half gait cycles) intact.

A more appropriate strategy for datasets in which the temporal dimension represents the sample mode is to use a tensor decomposition that explicitly compresses this dimension. Among common tensor decompositions, PARAFAC and Tucker models can incorporate a temporal mode as the sample mode, which is then compressed along that dimension \citep{Latchoumane:TuckerEEG, Acar:CPDEEG, Cong:TensEEG}. 
For instance, in the analysis of electroencephalography (EEG) signals for brain activity studies, this compression is meaningful because temporal modes in such data are typically highly correlated: walking sequences exhibit periodic motion, and EEG rhythms display oscillatory patterns. Consequently, the temporal mode can often be represented by a small number of dominant eigenvectors or principal temporal components.

Although our proposed method already shows good computational performance in both simulations and real-data case studies, it is important to note that tensor dimensions can grow rapidly as additional modes are introduced. Therefore, computational scalability becomes a key consideration for high-dimensional tensors.

Conceptually, the ROMPCA method is scalable for tensors of any order. This is because the projection matrices ${\mathbf{V}^{(\ell)}}$ are updated sequentially, so the number of modes does not affect their storage requirements. Likewise, the storage of the core tensors is comparable to that of standard MPCA, meaning that the scalability limit of ROMPCA is essentially the same as that of MPCA, apart from the additional computational cost. The main computational bottleneck in the ROMPCA algorithm lies in the Kronecker product. To address this, instead of explicitly forming the Kronecker product, the series of Kronecker products appearing, for example, in \eqref{eq:solU}, can be reformulated as a sequence of matrix–matrix multiplications, as described in \cite{Fackler:KronAlgo}. This approach could substantially reduce both memory and computational requirements, thus improving scalability to higher-dimensional tensors.

To further enhance computational efficiency, the algorithm can be implemented in C++, which allows for low-level memory management and optimized numerical operations compared to high-level interpreted languages \citep{Rolinger:PerfTensDecomp}. Moreover, the updates of the projection matrices ${\mathbf{V}^{(\ell)}}$ in \eqref{eq:solV} and the core tensors ${\mathcal{U}_{n}}$ in \eqref{eq:solU} can be parallelized wherever feasible \citep{Cichocki:TensNet}, thereby exploiting multi-core architectures to further reduce execution time.

\renewcommand{\refname}{Additional References}

\end{document}